\documentclass[12pt,a4paper]{article}
\usepackage{jheppub}  
\usepackage{amssymb} 
\usepackage{amsmath}
\usepackage{mathtools}
\usepackage{amsfonts}    
\usepackage{dsfont}
\usepackage{pdfpages}
\usepackage{verbatim}
\usepackage{tensor}
\usepackage{mathrsfs}
\usepackage{textgreek}
\usepackage{braket}
\usepackage{diagbox}
\usepackage{tikz}
\usetikzlibrary{decorations.pathreplacing, decorations.markings,calc,shapes.misc,decorations.pathmorphing,patterns.meta, math}
\usepackage{listings}
\usepackage[most]{tcolorbox}

\definecolor{codegreen}{rgb}{0,0.6,0}
\definecolor{codegray}{rgb}{0.5,0.5,0.5}
\definecolor{codepurple}{rgb}{0.58,0,0.82}
\definecolor{backcolour}{rgb}{0.95,0.95,0.92}

\lstdefinestyle{mystyle}{
    backgroundcolor=\color{backcolour},   
    commentstyle=\color{codegreen},
    keywordstyle=\color{magenta},
    numberstyle=\tiny\color{codegray},
    stringstyle=\color{codepurple},
    basicstyle=\ttfamily\footnotesize,
    breakatwhitespace=false,         
    breaklines=true,                 
    captionpos=b,                    
    keepspaces=true,                 
    numbers=left,                    
    numbersep=5pt,                  
    showspaces=false,                
    showstringspaces=false,
    showtabs=false,                  
    tabsize=2
}
\newcommand{\ba}{\begin{align}}

\newcommand{\be}{\begin{equation}}
\newcommand{\ee}{\end{equation}}
\def\bd{\begin{tikzpicture}}
\def\ed{\end{tikzpicture}}

\DeclareMathOperator\tr{tr}
\DeclareMathOperator\diag{diag}

\renewcommand\Re{\mathop{\text{Re}}}

\allowdisplaybreaks[1]

\newcommand\PSL{\text{PSL}}

\newcommand\SL{\text{SL}}

\newcommand\Map{\text{Map}}

\newcommand\vol{\text{vol}}
\newcommand\Li{\text{Li}}

\newcommand\CC{\mathbb{C}}
\newcommand\ZZ{\mathbb{Z}}
\newcommand\RR{\mathbb{R}}

\renewcommand\d{\text{d}}

\newcommand{\id}{\mathds{1}}

\title{3d gravity from Virasoro TQFT:\\  Holography, wormholes and knots}

\author[a,b]{Scott Collier}\emailAdd{sac@mit.edu}
\author[c,d]{\!\!, Lorenz Eberhardt}\emailAdd{l.eberhardt@uva.nl}
\author[e]{and Mengyang Zhang}\emailAdd{mengyang@princeton.edu}
\affiliation[a]{Princeton Center for Theoretical Science, Princeton University, Princeton, NJ 08544,
USA}
\affiliation[b]{Center for Theoretical Physics, Massachusetts Insitute of Technology, Cambridge, MA 02139,
USA}
\affiliation[c]{Institute for Advanced Study, Einstein Drive, Princeton, NJ 08540, USA}
\affiliation[d]{Institute for Theoretical Physics, University of Amsterdam, PO Box 94485, 1090 GL Amsterdam, The Netherlands}
\affiliation[e]{Joseph Henry Laboratories, Princeton University, Princeton, NJ 08544, USA}

\abstract{
We further develop the description of three-dimensional quantum gravity with negative cosmological constant in terms of Virasoro TQFT formulated in our previous paper \cite{Collier:2023fwi}.
We compare the partition functions computed in the Virasoro TQFT formalism to the semiclassical evaluation of Euclidean gravity partition functions. This matching is highly non-trivial, but can be checked directly in some examples.
We then showcase the formalism in action, by computing the gravity partition functions of many relevant topologies. For holographic applications, we focus on the partition functions of Euclidean multi-boundary wormholes  with three-punctured spheres as boundaries. This precisely quantifies the higher moments of the structure constants in the proposed ensemble boundary dual and subjects the proposal to thorough checks.
Finally, we investigate in detail the example of the figure eight knot complement as a hyperbolic 3-manifold. We show that the Virasoro TQFT partition function is identical to the partition function computed in Teichm\"uller theory, thus giving strong evidence for the equivalence of these TQFTs. We also show how to produce a large class of manifolds via Dehn surgery on the figure eight knot.
}

\begin{document}

\begin{flushright}
\hfill{\tt MIT-CTP/5691}
\end{flushright}

\maketitle

\makeatletter
\g@addto@macro\bfseries{\boldmath}
\makeatother

\section{Introduction}
Three-dimensional quantum gravity with negative cosmological constant has proven to be one of the most interesting and productive toy models of quantum gravity. The major outstanding problems are to fully solve the theory from first principles and firmly establish a holographic correspondence for the theory. There has been major progress on both fronts over the last few years; see \cite{Yin:2007gv, Giombi:2008vd, Cotler:2018zff, Maxfield:2020ale, Cotler:2020ugk, Eberhardt:2022wlc, Collier:2023fwi} and \cite{Belin:2020hea, Schlenker:2022dyo, Chandra:2022bqq, Belin:2023efa, DiUbaldo:2023qli, Collier:2023cyw, deBoer:2023vsm}, respectively.
\medskip

In our previous paper \cite{Collier:2023fwi}, we developed a formalism that computes the gravity partition function algorithmically on a background of fixed (on-shell) topology. This fixes the contributions of hyperbolic three-manifolds to the gravitational path integral and represents a large step towards a complete solution of the theory directly from the bulk. The next step would involve performing the sum over all three-dimensional topologies that appear in the gravitational path integral, which may also require suitable non-perturbative or off-shell contributions. While our previous paper developed the formalism in terms of the Virasoro TQFT, this paper gives several interesting applications that exemplify its practical utility and should be viewed as a natural continuation of \cite{Collier:2023fwi}.

On the holographic side, a consistent picture is emerging that the gravitational path integral computes certain universal statistical features in a putative ensemble of holographic 2d CFTs. While the full non-perturbative definition of such an ensemble is still not settled, this perspective makes very concrete predictions that can be quantitatively matched between the bulk and boundary. In this paper we will study partition functions of Virasoro TQFT on multi-boundary wormholes to exemplify the extent to which the gravitational path integral precisely captures universal statistics of CFT data, transcending the Gaussian approximation of \cite{Chandra:2022bqq}.
\medskip

We will assume that the reader is acquainted with the concepts introduced in \cite{Collier:2023fwi}, but now recall some key features. As suggested by holography, the Hilbert space of 3d quantum gravity is spanned by the left- and right-moving Virasoro conformal blocks on the spatial surface $\Sigma$. This factorization of the Hilbert space allows one to consider, say, only the left-movers as a fundamental building block.\footnote{The left- and right-movers are entangled only by the sum over topologies in the gravitational path integral.} A key ingredient in the proposal of \cite{Collier:2023fwi} is an explicit form of the inner product on this conformal block Hilbert space. Important for the consistency of this structure is the fact that the conformal blocks transform among each other under crossing transformations and as such the Hilbert space carries a unitary action under crossing transformations. There are remarkably explicit expressions for the crossing transformation in terms of the Ponsot-Teschner fusion kernel $\mathbb{F}$ and the modular crossing kernel $\mathbb{S}$ \cite{Ponsot:1999uf, Ponsot:2000mt, Teschner:2012em, Teschner:2013tqy}. Since the Hamiltonian in gravity vanishes, the theory can be viewed as a TQFT on a background topology. This data completely specifies the TQFT that we called Virasoro TQFT in \cite{Collier:2023fwi}. The TQFT partition function on a fixed topology can be computed via surgery techniques similarly to Chern-Simons theory. The Virasoro TQFT partition function then immediately leads to the full 3d gravity partition function via the following formula, valid for all hyperbolic three-manifolds
\be 
Z_\text{grav}(M)=\sum_{\gamma \in \Map(\partial M)/\Map(M,\partial M)} |Z_\text{Vir}(M^\gamma)|^2\ . \label{eq:sum over topologies}
\ee
Here we sum over all images of the manifold $M$ under the boundary mapping class group $\Map(\partial M)$, which is part of the sum over topologies, modulo the bulk mapping class group $\Map(M,\partial M)$, which is gauged in gravity.
\bigskip

We start in Section~\ref{sec:structural properties} by analyzing the gravity partition functions as computed in Virasoro TQFT and their relation to the semiclassical evaluation of the gravitational path integral. Comparing the two expressions leads to the (refined) volume conjecture that we already mentioned in \cite{Collier:2023fwi} and discuss further here. We also discuss the existence of non-isomorphic hyperbolic manifolds with identical Virasoro TQFT partition functions. This in particular implies that the gravitational path integral is not powerful enough to detect the topology of hyperbolic manifolds.

We then discuss examples that are relevant for the holographic description of 3d gravity in Section~\ref{sec:holography}. We focus on a class of manifolds obtained by removing three-punctured spheres from $\mathrm{S}^3$ and connecting the boundaries appropriately with Wilson lines. They compute holographically the higher moments of the structure constants in the proposed ensemble description of the boundary dual. We find that the partition functions may be computed using diagrammatic rules that are simply the $q$-deformations of the rules for the computation of disk partition functions in JT gravity + matter \cite{Mertens:2017mtv, Lam:2018pvp, Jafferis:2022wez}. When projecting the Wilson lines on a disk, one associates a Virasoro 6j-symbol to every crossing of lines and one integral to every loop formed by the internal lines, see eqs.~\eqref{eq:trivalent vertex} and 
\eqref{eq:quartic vertex} for the precise formulae. We also use Virasoro TQFT to compute the gravity partition function on a class of contributions to the single-boundary gravitational path integral that are not handlebodies. These non-handlebody instantons are formed by quotients of the two-boundary Euclidean wormhole and we find that the gravity partition function is related to the partition function of Liouville CFT on a particular non-orientable surface.

Finally, we consider the example of the figure eight knot complement in Section~\ref{sec:figure eight knot}, which constitutes one of the simplest examples of a hyperbolic manifold with no asymptotic boundary. We compute its Virasoro TQFT partition function from a variety of perspectives and demonstrate that it agrees with the partition function computed in an a priori different TQFT known as Teichm\"uller TQFT. This lends strong credence to the equivalence of the two theories, even though Virasoro TQFT provides a far more convenient framework for holographic applications. We also illustrate the procedure of Dehn surgery on the figure eight knot, which leads to the gravity partition function on a whole family of hyperbolic three-manifolds whose volume accumulates to that of the figure eight knot complement.

\section{Structural properties of Virasoro TQFT} \label{sec:structural properties}
We start by discussing the relation of the formulation of gravity in terms of Virasoro TQFT and the semiclassical gravity path integral. Comparing the two leads to the volume conjecture and we discuss various consequences for the volumes of hyperbolic 3-manifolds, conformal blocks and one-loop determinants. 
We then also explain some of the consistency conditions of the Virasoro TQFT. Such consistency conditions are all implied by the consistency of the mapping class representation on the initial value surface, but often the three-dimensional viewpoint is much more powerful. 
\subsection{Volume conjecture}
We already stated the (refined) volume conjecture in \cite{Collier:2023fwi}, but it will play a much more prominent role in the present paper. By comparing the usual metric approach of 3d gravity and the Virasoro TQFT approach, one obtains the following prediction for the semiclassical expansion of partition functions:
\be 
|Z_\text{Vir}(M)|^2=\mathrm{e}^{-\frac{c}{6\pi} \vol(M)}\left[\prod_{\gamma\in \mathcal{P}} \prod_{m=2}^\infty \frac{1}{|1-q_\gamma^m|^2} +\mathcal{O}(c^{-1})\right]\ . \label{eq:refined volume conjecture}
\ee
Here we used that the gravity tree-level action is $\frac{c}{6\pi} \, \vol(M)$, where $\vol(M)$ is the volume of the hyperbolic manifold.
We also used the explicit form of the one-loop determinant as computed in \cite{Giombi:2008vd}. This explicit formula for the one-loop determinant is valid for hyperbolic manifolds without defects that can be written as $\mathbb{H}^3/\Gamma$ for a so-called Kleinian group $\Gamma$. In case $M$ has defects, the volume conjecture should still hold, but there is no known general formula for the one-loop determinant. We recall that $\mathcal{P}$ denotes the set of all primitive geodesics on the three-manifold in question. Alternatively, we can think of $\mathcal{P}$ as the set of primitive conjugacy classes in the Kleinian group $\Gamma$ (i.e.\ conjugacy classes that are not powers of other conjugacy classes) and also identify the conjugacy class of $\gamma$ with the conjugacy class of $\gamma^{-1}$, since this corresponds to orientation reversal of the corresponding geodesic.
We could of course extend the matching to higher loop order, but will restrict here to the tree-level and one-loop piece.

We refer to this equation as the refined volume conjecture, since the classical volume conjecture is the corresponding statement for the tree-level term in the $\frac{1}{c}$-expansion \cite{Kashaev:1996kc}. The relation \eqref{eq:refined volume conjecture} should also hold in the presence of boundaries, in which case the volume of the hyperbolic manifold is the renormalized volume \cite{Henningson:1998gx}.

\subsection{The volume of hyperbolic tetrahedra}
Let us explain one of the simplest non-trivial instances of the volume conjecture in more detail. Consider a single hyperbolic tetrahedron as in Figure~\ref{fig:hyperbolic tetrahedron} with dihedral angles $\theta_i$ specifying the angle between the two faces meeting at the edge.

\begin{figure}[ht]
    \centering
    \begin{tikzpicture}
    \draw[very thick] (0,0) to node[below] {$\theta_6$} (4,0);
    \draw[very thick, densely dashed] (0,0) to node[right, shift={(0,-.1)}] {$\theta_5$} (2.8,1.5) to node[left] {$\theta_4$} (4,0);
    \draw[very thick, densely dashed] (2.8,1.5) to node[left, shift={(.1,-.2)}] {$\theta_3$} (2.5,3.5);
    \draw[very thick] (0,0) to node[left] {$\theta_1$} (2.5,3.5) to node[right] {$\theta_2$} (4,0);
\end{tikzpicture}
    \caption{Tetrahedron with dihedral angles specified.}
    \label{fig:hyperbolic tetrahedron}
\end{figure}
The dihedral angles have to satisfy rather complicated conditions for such a hyperbolic tetrahedron to exist. We can take two identical such hyperbolic tetrahedra and identify them along the corresponding faces. This leads to a topological three-sphere with conical defects running in the form of the tetrahedron through it. The hyperbolic tetrahedron is specified by the dihedral angles $\theta_j \in (0,\pi)$ spanned by the two faces meeting at an edge. Upon gluing two tetrahedra, the conical defect angle becomes $2\pi-2\theta_j$. Semiclassically, the relation between defect angles $\alpha_j$ and Liouville momentum reads\footnote{Here we are adopting the standard notation from Liouville theory for the central charge and conformal weights:
\begin{equation}
    c = 1+6(b+b^{-1})^2 = 1 + 6 Q^2, \quad \Delta_j = \frac{c-1}{24}+P_j^2.
\end{equation}} \cite{Teschner:2003em}
\be 
P_j=\frac{i \alpha_j}{4\pi b}\sim \frac{iQ}{2}-\frac{i\theta_j}{2\pi b}
\ee
This is the hyperbolic three-manifold that we use for the volume conjecture.
Some extra care is necessary to correctly normalize the vertices. We observe that the renormalized volume of the Euclidean wormhole of the form $\Sigma_{0,3} \times I$ exactly vanishes. Since it evaluates to the Liouville structure constant $C_0(P_1,P_2,P_3)$ in the Virasoro TQFT, this means that for the purposes of the volume conjecture, we should define a juncture with a normalization constant $C_0(P_1,P_2,P_3)^{-\frac{1}{2}}$ as follows
\be 
\frac{1}{\sqrt{C_0(P_1,P_2,P_3)}} \times 
\begin{tikzpicture}[baseline={([yshift=-.5ex]current bounding box.center)}]
\draw[very thick, red] (.25,.43) to (.5,.87) node[right, black] {$P_1$};
\draw[very thick, red] (.25,-.43) to (.5,-.87) node[right, black] {$P_3$};
\draw[very thick, red] (-.5,0) to (-1,0) node[left, black] {$P_2$};
\draw[very thick] (0,0) circle (.5);
\draw[very thick, densely dashed] (.5,0) arc (0:180:.5 and .2);
\draw[very thick] (.5,0) arc (0:-180:.5 and .2);
\end{tikzpicture}
\ . 
\ee

We compute the TQFT partition function $Z_\text{Vir}$ of this tetrahedral configuration in Section~\ref{subsec:four-boundary wormhole}. The result is given by
\begin{align} 
Z_\text{Vir}\left(\ \begin{tikzpicture}[baseline={([yshift=-.5ex]current bounding box.center)}, scale=.4]
    \draw[very thick, red] (0,0) to (4,0);
    \draw[very thick, densely dashed, red] (0,0) to  (2.8,1.5) to  (4,0);
    \draw[very thick, densely dashed, red] (2.8,1.5) to (2.5,3.5);
    \draw[very thick, red] (0,0) to (2.5,3.5) to (4,0);
    \fill[white] (0,0) circle (.5);
    \draw[very thick] (0,0) circle (.5);
    \draw[very thick, densely dashed] (.5,0) arc (0:180:.5 and .2);
    \draw[very thick] (.5,0) arc (0:-180:.5 and .2);
    \begin{scope}[shift={(4,0)}]
        \fill[white] (0,0) circle (.5);
        \draw[very thick] (0,0) circle (.5);
        \draw[very thick, densely dashed] (.5,0) arc (0:180:.5 and .2);
        \draw[very thick] (.5,0) arc (0:-180:.5 and .2);
    \end{scope}
    \begin{scope}[shift={(2.8,1.5)}]
        \fill[white] (0,0) circle (.5);
        \draw[very thick] (0,0) circle (.5);
        \draw[very thick, densely dashed] (.5,0) arc (0:180:.5 and .2);
        \draw[very thick] (.5,0) arc (0:-180:.5 and .2);
    \end{scope}    
    \begin{scope}[shift={(2.5,3.5)}]
        \fill[white] (0,0) circle (.5);
        \draw[very thick] (0,0) circle (.5);
        \draw[very thick, densely dashed] (.5,0) arc (0:180:.5 and .2);
        \draw[very thick] (.5,0) arc (0:-180:.5 and .2);
    \end{scope}
\end{tikzpicture}\ \right)&=\rho_0(P_6)^{-1}\, C_0(P_1,P_2,P_3)C_0(P_3,P_4,P_5) \, \mathbb{F}_{P_3,P_6} \! \begin{bmatrix}
    P_4 & P_2 \\ P_5 & P_1
\end{bmatrix}\\
&=\sqrt{C_0(P_1,P_2,P_3) C_0(P_1,P_5,P_6) C_0(P_2,P_4,P_6) C_0(P_3,P_4,P_5)} \nonumber\\
&\qquad\times\begin{Bmatrix} P_1 & P_2 & P_3 \\ P_4 & P_5 & P_6 \end{Bmatrix}\ .
\end{align}
The symbol $\{\begin{smallmatrix}  P_1 & P_2 & P_3 \\ P_4 & P_5 & P_6\end{smallmatrix}\}$ is the crossing kernel for sphere four-point function conformal blocks in the Racah-Wigner normalization \cite{Teschner:2012em}, which we also call the Virasoro 6j-symbol. It has the correct tetrahedral symmetry as required by the picture, where the vertices of the tetrahedron are formed by $(P_1,P_2,P_3)$, $(P_1,P_5,P_6)$, $(P_2,P_4,P_6)$ and $(P_3,P_4,P_5)$.

Thus the prediction of the volume conjecture is now that
\be 
2\, \vol(\eta_1,\eta_2,\eta_3,\eta_4,\eta_5,\eta_6)=-2\pi \lim_{b \to 0} b^2\log \begin{Bmatrix} \frac{iQ}{2}-\frac{i\eta_1}{2\pi b} &  \frac{iQ}{2}-\frac{i\eta_2}{2\pi b} &  \frac{iQ}{2}-\frac{i\eta_3}{2\pi b} \\  \frac{iQ}{2}-\frac{i\eta_4}{2\pi b} &  \frac{iQ}{2}-\frac{i\eta_5}{2\pi b} &  \frac{iQ}{2}-\frac{i\eta_6}{2\pi b} \end{Bmatrix}\ ,
\ee
where the volume on the left hand side is the volume of the hyperbolic tetrahedron specified by the dihedral angles $\eta_j$. 

One can evaluate the integral in the defining formula for the crossing kernel via saddle-point approximation in this limit and confirm that it agrees with the volume formula for a hyperbolic tetrahedron. This was done in \cite{Teschner:2012em}, but the volume conjecture gives a conceptual derivation of that fact.

We also mention that the Virasoro crossing kernel has the following Regge symmetry \cite{Apresyan:2022erh, Eberhardt:2023mrq}
\be 
\mathbb{F}_{P_3,P_6} \! \begin{bmatrix}
    P_4 & P_2 \\ P_5 & P_1
\end{bmatrix}=\mathbb{F}_{P_3,P_6} \! \begin{bmatrix}
    \frac{1}{2}(P_2+P_4+P_5-P_1) & \frac{1}{2}(P_1+P_2+P_4-P_5) \\ \frac{1}{2}(P_1+P_4+P_5-P_2) & \frac{1}{2}(P_1+P_2+P_5-P_4)
\end{bmatrix}\ .
\ee
This implies via the volume conjecture that the volume of a hyperbolic tetrahedron is invariant under the replacement
\begin{align} 
\theta_1 &\to \tfrac{1}{2}(\theta_1+\theta_2+\theta_4-\theta_5)\ , & \theta_5 &\to \tfrac{1}{2}(-\theta_1+\theta_2+\theta_4+\theta_5)\ , \\
\theta_2 &\to \tfrac{1}{2}(\theta_1+\theta_2+\theta_5-\theta_4)\ , & \theta_4 &\to \tfrac{1}{2}(\theta_1-\theta_2+\theta_4+\theta_5)\ ,
\end{align} 
with $\theta_3$ and $\theta_6$ unchanged.
This property is very non-trivial to see geometrically and giving a direct proof of it is rather hard.

\subsection{Volume conjecture for handlebodies}
\paragraph{Semiclassical vacuum blocks.} Let us apply the volume conjecture in the form \eqref{eq:refined volume conjecture} to a handlebody. Recall that the Virasoro TQFT partition function on a genus-$g$ handlebody $\mathsf{S}\Sigma_g$ evaluates to the vacuum Virasoro conformal block,
\be 
Z_\text{Vir}(\mathsf{S}\Sigma_g)= \begin{tikzpicture}[baseline={([yshift=-.5ex]current bounding box.center)}, xscale=.9]
       \begin{scope}
        \draw[very thick, red, out=90, in=150] (-2.4,0) to (-.5,0);
        \draw[very thick, red, out=90, in=30] (2.4,0) to (.5,0);
        \draw[very thick, red, out=-90, in=210] (-2.4,0) to (-.5,0);
        \draw[very thick, red, out=-90, in=-30] (2.4,0) to (.5,0);
        \draw[very thick, red] (-.5,0) to node[above, black] {$\id$} (.5,0);
        \fill[red] (-.5,0) circle (.07);
        \fill[red] (.5,0) circle (.07);
        \node at (1.6,.6) {$\id$};
        \node at (-1.6,.6) {$\id$};
        \draw[very thick, in=180, out=0] (-1.5,1) to (0,.5) to (1.5,1);
        \draw[very thick, in=180, out=0] (-1.5,-1) to (0,-.5) to (1.5,-1);
        \draw[very thick, in=180, out=180, looseness=2] (-1.5,1) to (-1.5,-1);
        \draw[very thick, in=0, out=0, looseness=2] (1.5,1) to (1.5,-1);
        \draw[very thick, bend right=30] (-2,0) to (-1,0);
        \draw[very thick, bend left=30] (-1.9,-.05) to (-1.1,-.05);
        \draw[very thick, bend left=30] (2,0) to (1,0);
        \draw[very thick, bend right=30] (1.9,-.05) to (1.1,-.05);
        \end{scope}
    \end{tikzpicture}\ ,
\ee
where we drew a genus-2 surface for concreteness. As such the volume conjecture \eqref{eq:refined volume conjecture} gives the semiclassical expansion of vacuum blocks, 
\be 
\begin{tikzpicture}[baseline={([yshift=-.5ex]current bounding box.center)}, xscale=.9]
       \begin{scope}
        \draw[very thick, red, out=90, in=150] (-2.4,0) to (-.5,0);
        \draw[very thick, red, out=90, in=30] (2.4,0) to (.5,0);
        \draw[very thick, red, out=-90, in=210] (-2.4,0) to (-.5,0);
        \draw[very thick, red, out=-90, in=-30] (2.4,0) to (.5,0);
        \draw[very thick, red] (-.5,0) to node[above, black] {$\id$} (.5,0);
        \fill[red] (-.5,0) circle (.07);
        \fill[red] (.5,0) circle (.07);
        \node at (1.6,.6) {$\id$};
        \node at (-1.6,.6) {$\id$};
        \draw[very thick, in=180, out=0] (-1.5,1) to (0,.5) to (1.5,1);
        \draw[very thick, in=180, out=0] (-1.5,-1) to (0,-.5) to (1.5,-1);
        \draw[very thick, in=180, out=180, looseness=2] (-1.5,1) to (-1.5,-1);
        \draw[very thick, in=0, out=0, looseness=2] (1.5,1) to (1.5,-1);
        \draw[very thick, bend right=30] (-2,0) to (-1,0);
        \draw[very thick, bend left=30] (-1.9,-.05) to (-1.1,-.05);
        \draw[very thick, bend left=30] (2,0) to (1,0);
        \draw[very thick, bend right=30] (1.9,-.05) to (1.1,-.05);
        \end{scope}
    \end{tikzpicture} \sim \mathrm{e}^{-\frac{c}{12\pi}\vol(\mathsf{S}\Sigma_g)} \prod_{\gamma\in \mathcal{P}(\Gamma_g)} \prod_{m=2}^\infty \frac{1}{1-q_\gamma^m}\ . \label{eq:vacuum conformal block semiclassical expansion}
\ee
Here, $\vol(\mathsf{S}\Sigma_g)$ is the in general complex volume of the handlebody.\footnote{In general to write this formula only for a chiral half (which goes beyond the volume conjecture \eqref{eq:refined volume conjecture}), we also need to assign an imaginary part to the volume which is known as the Chern-Simons invariant.}
Such a semiclassical expansion of the conformal blocks is familiar from 2d CFT, where the leading term is called the semiclassical conformal block \cite{Belavin:1984vu, Zamolodchikov:1984eqp, Besken:2019jyw}, but to our knowledge there is no general CFT derivation of the one-loop determinant, and even direct derivations of the leading term are somewhat limited. The group $\Gamma_g \subset \PSL(2,\CC)$ that appears in the one-loop determinant is the Schottky group of the corresponding handlebody.

The Virasoro TQFT approach gives a simple derivation of this fact. It also shows that the semiclassical block is nothing else than the volume of the corresponding handlebody. It was shown in \cite{Krasnov:2000zq} that this volume is identified with the on-shell value of the Liouville action as defined by Takhtajan and Zograf \cite{ZografTakhtajan},
\be 
S_\text{L}(\Sigma_g)=-4 \Re \vol(\mathsf{S}\Sigma_g)\ .
\ee
Defining the on-shell Liouville action requires one to pick a conformal block channel.
The Virasoro TQFT also makes a prediction about the order one term in the semiclassical expansion.

\paragraph{One-loop determinant.}
Let us recall the formula derived in \cite{McIntyre:2004xs} for the holomorphic factorization of the Laplacian on a Riemann surface.
We have
\be 
\frac{\det' \Delta_2}{\det N_2}=c_g\,  \mathrm{e}^{-\frac{13}{12\pi} \, S_\text{L}(\Sigma_g)} \bigg|(1-q_1)^2(1-q_2)\prod_{\gamma \in \mathcal{P}(\Gamma_g)}\prod_{m=2}^\infty (1-q_\gamma^m)^2\bigg|^2\ .
\ee
for some constant $c_g$ independent of the moduli. It depends on the renormalization scheme used to define the determinant $\det' \Delta_2$. Here $\Delta_2$ is the Laplacian acting on holomorphic quadratic differentials on the surface $\Sigma$ and the prime indicates that we removed the zero modes. $\det N_2$ is the determinant of $\langle \varphi_j \, |\,  \varphi_k \rangle$ and $\{\varphi_j\}_{j=1,\dots,3g-3}$ is a natural basis of holomorphic quadratic differentials as defined in \cite{McIntyre:2004xs}. We also denoted $q_j=q_{\gamma_j}$ with $\gamma_1,\dots,\gamma_g$ the $g$ free generators of the Schottky group. The perhaps unnatural seeming factor $(1-q_1)^2(1-q_2)$ appears because of the specific way in which $\varphi_j$ is defined and is a result of fixing the $\PSL(2,\CC)$ conjugacy freedom for the Schottky group.
Thus we have 
\be\label{eq:volume conjecture for handlebodies}
\left|\begin{tikzpicture}[baseline={([yshift=-.5ex]current bounding box.center)}, xscale=.9]
       \begin{scope}
        \draw[very thick, red, out=90, in=150] (-2.4,0) to (-.5,0);
        \draw[very thick, red, out=90, in=30] (2.4,0) to (.5,0);
        \draw[very thick, red, out=-90, in=210] (-2.4,0) to (-.5,0);
        \draw[very thick, red, out=-90, in=-30] (2.4,0) to (.5,0);
        \draw[very thick, red] (-.5,0) to node[above, black] {$\id$} (.5,0);
        \fill[red] (-.5,0) circle (.07);
        \fill[red] (.5,0) circle (.07);
        \node at (1.6,.6) {$\id$};
        \node at (-1.6,.6) {$\id$};
        \draw[very thick, in=180, out=0] (-1.5,1) to (0,.5) to (1.5,1);
        \draw[very thick, in=180, out=0] (-1.5,-1) to (0,-.5) to (1.5,-1);
        \draw[very thick, in=180, out=180, looseness=2] (-1.5,1) to (-1.5,-1);
        \draw[very thick, in=0, out=0, looseness=2] (1.5,1) to (1.5,-1);
        \draw[very thick, bend right=30] (-2,0) to (-1,0);
        \draw[very thick, bend left=30] (-1.9,-.05) to (-1.1,-.05);
        \draw[very thick, bend left=30] (2,0) to (1,0);
        \draw[very thick, bend right=30] (1.9,-.05) to (1.1,-.05);
        \end{scope}
    \end{tikzpicture}\right|^2 \sim c_g' \frac{\mathrm{e}^{-\frac{c-13}{6\pi}\vol(\mathsf{S}\Sigma_g)}}{\sqrt{\det' \Delta_2}}  \times |1-q_1|^2|1-q_2| \sqrt{\det N_2}\ .
\ee
This tells us that the one-loop partition function is exactly the inverse square root of the partition function of a $bc$-ghost system with a particular choice of ghost insertions.

\paragraph{Explicit check.}
Here we explicitly check the one-loop refinement of the volume conjecture for handlebodies in a simple example, perturbatively in the moduli of the Riemann surface in an expansion about a pinching limit. Consider for concreteness a genus-two Riemann surface formed by plumbing two two-holed disks $D_1$ and $D_2$:
\begin{subequations}
\begin{align}
        D_1 &= \{z_1\in\mathbb{C}\,  | \,  r_1 < |z_1| < r_3,\,  |z_1-1| > r_2\}\ ,\\
        D_2 &= \{ z_2 \in \mathbb{C} \, | \, \tilde r_1 < |z_2| <\tilde r_3,\,  |z_2-1|> \tilde r_2\}\ .
\end{align}
\end{subequations}
Gluing the boundaries of the disks according to the following inversion map prepares a disk with three holes
\begin{align}
    |z_2| &= \tilde r_3\, : \quad z_2 \sim \frac{1}{p_1 z_1}\, , & \text{with}\ |p_1| &= \frac{1}{r_3\tilde r_3}\ .
\end{align}
The remaining identifications are
\begin{subequations} \label{eq:genus two identifications}
    \begin{align}
        |z_2| &= \tilde r_1\, : \quad z_2 \sim \frac{p_3}{z_1}\, , & \text{with}\  |p_3| &= r_1\tilde r_1\ ,\\
        |z_2-1| &= \tilde r_3\, : \quad z_2-1\sim \frac{p_2}{z_1-1}\,,& \text{with}\ |p_2| &= r_2 \tilde r_2\ .
    \end{align}
\end{subequations}
The complex plumbing parameters $p_i$ parameterize the moduli of the Riemann surface, with the $p_i\to 0$ limit a pinching locus in which the surface is realized by gluing two spheres along long narrow tubes. The corresponding Virasoro conformal blocks may then straightforwardly be computed as an expansion in powers of the plumbing parameters $p_i$, see for example \cite{Cho:2017oxl} for details.

This parameterization of the genus-two Riemann surface is clearly equivalent to the Schottky parameterization, in which one realizes the Riemann surface $\Sigma_g$  as a quotient of the form
\begin{equation}
    \Sigma_g = (\mathbb{C}\cup\{\infty\}-\Lambda)/\Gamma\, .
\end{equation}
Here $\Gamma = \langle\gamma_1,\ldots,\gamma_g\rangle$ is the Schottky group, which is a free group generated by the loxodromic elements $\gamma_1,\ldots,\gamma_g$ of $\PSL(2,\mathbb{C})$, and $\Lambda$ is the limit set of the action of $\Gamma$. The generators $\gamma_i$ act on the Riemann sphere by M\"obius transformation. In our example of the genus-two Riemann surface formed by plumbing two-holed disks as above, the generators of the Schottky group may be taken to be
\begin{equation}
    \gamma_1(z) = p_1 p_3 z\ , \qquad
    \gamma_2(z) = \frac{(1-p_2)z-1/p_1}{z-1/p_1} \ .
\end{equation}
Each generator $\gamma$ is conjugate to $\diag(q_\gamma^{1/2},q_\gamma^{-1/2})$, with $|q_\gamma|<1$. Here we have
\begin{equation}
        q_{\gamma_1} = p_1 p_3 \ , \qquad
        q_{\gamma_2} = \frac{1-p_1 + p_1 p_2 - \sqrt{1-2p_1(1+p_2)+p_1^2(1-p_2)^2}}{1-p_1 + p_1 p_2 + \sqrt{1-2p_1(1+p_2)+p_1^2(1-p_2)^2}} \ .
\end{equation}

We are now in a position to directly compare the perturbative expansion of the $c\to\infty$ limit of the genus-two Virasoro vacuum block as parameterized in the plumbing frame above\footnote{As explained in \cite{Cho:2017oxl}, in the plumbing frame the $c\to\infty$ limit of the Virasoro blocks is actually finite; in other words, the corresponding Liouville action vanishes. The $c\to\infty$ limit of the vacuum block as computed in (\ref{eq:vacuum conformal block semiclassical expansion}) plays an important role in determining the seed of the recursive representation of arbitrary Virasoro blocks.} with that of the gravity one-loop determinant on the genus-two handlebody (\ref{eq:vacuum conformal block semiclassical expansion}). We find
\begin{align}
    &\left.
    \begin{tikzpicture}[baseline={([yshift=-.5ex]current bounding box.center)}, xscale=.9]
       \begin{scope}
        \draw[very thick, red, out=90, in=150] (-2.4,0) to (-.5,0);
        \draw[very thick, red, out=90, in=30] (2.4,0) to (.5,0);
        \draw[very thick, red, out=-90, in=210] (-2.4,0) to (-.5,0);
        \draw[very thick, red, out=-90, in=-30] (2.4,0) to (.5,0);
        \draw[very thick, red] (-.5,0) to node[above, black] {$\id$} (.5,0);
        \fill[red] (-.5,0) circle (.07);
        \fill[red] (.5,0) circle (.07);
        \node at (1.6,.6) {$\id$};
        \node at (-1.6,.6) {$\id$};
        \draw[very thick, in=180, out=0] (-1.5,1) to (0,.5) to (1.5,1);
        \draw[very thick, in=180, out=0] (-1.5,-1) to (0,-.5) to (1.5,-1);
        \draw[very thick, in=180, out=180, looseness=2] (-1.5,1) to (-1.5,-1);
        \draw[very thick, in=0, out=0, looseness=2] (1.5,1) to (1.5,-1);
        \draw[very thick, bend right=30] (-2,0) to (-1,0);
        \draw[very thick, bend left=30] (-1.9,-.05) to (-1.1,-.05);
        \draw[very thick, bend left=30] (2,0) to (1,0);
        \draw[very thick, bend right=30] (1.9,-.05) to (1.1,-.05);
        \end{scope}
    \end{tikzpicture}
    \right|_{c\to\infty} =
    \prod_{\gamma\in \mathcal{P}(\Gamma)} \prod_{m=2}^\infty \frac{1}{1-q_\gamma^m}\\
    &\qquad\qquad= 1 + p_1^2p_2^2 + p_2^2p_3^2+p_3^2p_1^2 + 4(p_1^3p_2^2 + p_2^3p_3^2 + p_2^2 p_3^3) + \ldots 
\end{align}
On the left-hand side we evaluate the genus-two identity block in the plumbing frame perturbatively in the moduli by brute force, and on the right-hand side we evaluate the one-loop determinant by taking the product over primitive conjuguacy classes of the Schottky group. We have verified the agreement between these two expressions up to total degree 12 in the expansion in the plumbing parameters. 

\subsection{Mutations of hyperbolic manifolds}
One may ask whether $Z_\text{Vir}$ is a perfect invariant of a hyperbolic three-manifold, or, in other words, is $Z_\text{Vir}$ powerful enough to distinguish any two hyperbolic manifolds?
As in Chern-Simons theory, the answer to this question is negative. There exist non-isometric hyperbolic three-manifolds $M_1$ and $M_2$ with $Z_\text{Vir}(M_1)=Z_\text{Vir}(M_2)$. The reason for this is a general operation known as mutation.

There are different kinds of mutations, these are all relatively subtle operations that go undetected by most knot invariants, including the Virasoro TQFT partition function $Z_\text{Vir}$.\footnote{In Virasoro TQFT, we think of a knot as a defect inserted in the three-sphere $\mathrm{S}^3$ that is knotted appropriately.} Let us first explain the classical example of knot mutations.
\begin{figure}[ht]
    \centering
    \begin{tikzpicture}
        \begin{scope}
        \begin{scope}
        \clip (2,0) rectangle (4,2);
        \draw[very thick, out=-90, in=0, red] (3.2,1.5) to (-1,1.5);
        \end{scope}
        \fill[white] (2.96,.9) circle (.07);
        \draw[very thick, out=0, in=90, red] (0,3) to (3,.5);
        \fill[white] (2.43,1.99) circle (.07);
        \begin{scope}
        \clip (1.5,-1) rectangle (3.2,2);
        \draw[very thick, out=-90, in=-90, looseness=2, red] (3,.5) to (1,.7);
        \end{scope}
        \draw[very thick, out=90, in=180, red] (1,.7) to (2.5,2);
        \fill[white] (1.05,1.07) circle (.07);
        \fill[white] (2.04,-.58) circle (.07);
        \draw[very thick, out=0, in=90, red] (2.5,2) to (3.2,1.5);
        \draw[very thick, out=180, in=90, red] (0,3) to (-1.5,1.6);
        \fill[white] (-1.37,2.17) circle (.07);
        \begin{scope}
        \clip (-1,-.5) rectangle (2.1,2);
        \draw[very thick, out=-90, in=120, red] (-1.5,1.6) to (2,-.3);
        \end{scope}
        \fill[white] (-.33,.61) circle (.07);
        \draw[very thick, out=-60, in=0, red] (2,-.3) to (0,-1);
        \draw[very thick, out=-90, in=180, red] (-2.5,.7) to (-1.2,0);
        \fill[white] (-1.66,0.03) circle (.07);
        \draw[very thick, out=180, in=-90, red] (0,-1) to (-1.9,1);
        \draw[very thick, out=90, in=180, red] (-1.9,1) to (-1.2,2.2);
        \fill[white] (-1.87,1.39) circle (.07);
        \draw[very thick, out=180, in=90, red] (-1,1.5) to (-2.5,.7);
        \fill[white] (-1.5,1.46) circle (.07);
        \begin{scope}
        \clip (-1.6,-.5) rectangle (-.9,2);
        \draw[very thick, out=-90, in=120, red] (-1.5,1.6) to (2,-.3);
        \end{scope}
        \draw[very thick, out=0, in=0, looseness=1.5, red] (-1.2,0) to (-1.2,2.2);
        \fill[white] (-.28,1.41) circle (.07);
        \begin{scope}
        \clip (2.1,2) rectangle (-1.2,0);
        \draw[very thick, out=-90, in=0, red] (3.2,1.5) to (-1,1.5);
        \end{scope}
        \fill[white] (1.03,.38) circle (.07);
        \begin{scope}
        \clip (.5,-1) rectangle (1.6,2);
        \draw[very thick, out=-90, in=-90, looseness=2, red] (3,.5) to (1,.7);
        \end{scope}
        \draw[very thick, blue] (2.2,.6) circle (1.7); 
        \begin{scope}[shift={(2.2,.6)}]
        \draw[very thick, blue,<->] {(-1,0)++(-30:2.9)} arc (-30:30:2.9);
        \end{scope}
        \end{scope}

        \begin{scope}[shift={(8,0)}]
        \draw[very thick, out=0, in=90, red] (0,3) to (2.3,1.8);
       \fill[white] (2.28,1.98) circle (.07);
        \draw[very thick, out=90, in=180, red] (1,.7) to (2.5,2);
       \fill[white] (1.09,1.16) circle (.07);
        \begin{scope}
        \clip (1.9,-.4) rectangle (3,0);
        \draw[very thick, out=0, in=0, looseness=3, red] (2.15,-.3) to (2,.5);
        \end{scope}
       \fill[white] (2.25,-.29) circle (.09);
        \draw[very thick, out=0, in=90, red] (2.5,2) to (3.2,1);
        \draw[very thick, out=180, in=90, red] (0,3) to (-1.5,1.6);
        \fill[white] (-1.37,2.17) circle (.07);
        \begin{scope}
        \clip (-1,-.5) rectangle (2.1,2);
        \draw[very thick, out=-90, in=180, red] (-1.5,1.6) to (2,.5);
        \end{scope}
        \fill[white] (1.03,.48) circle (.07);
        \draw[very thick, out=-90, in=180, red] (1,.7) to (2.15,-.3);
        \fill[white] (-.37,.52) circle (.07);
        \draw[very thick, out=220, in=0, red] (2.5,-.1) to (0,-1);
        \draw[very thick, out=40, in=-90, red] (2.5,-.1) to (3.2,1);
        \draw[very thick, out=-90, in=180, red] (-2.5,.7) to (-1.2,0);
        \fill[white] (-1.66,0.03) circle (.07);
        \draw[very thick, out=180, in=-90, red] (0,-1) to (-1.9,1);
        \draw[very thick, out=90, in=180, red] (-1.9,1) to (-1.2,2.2);
        \fill[white] (-1.87,1.39) circle (.07);
        \draw[very thick, out=180, in=90, red] (-1,1.5) to (-2.5,.7);
        \fill[white] (-1.5,1.46) circle (.07);
        \begin{scope}
        \clip (-1.6,-.5) rectangle (-.9,2);
        \draw[very thick, out=-90, in=180, red] (-1.5,1.6) to (2,.5);
        \end{scope}
        \draw[very thick, out=0, in=0, looseness=1.5, red] (-1.2,0) to (-1.2,2.2);
       \fill[white] (2.78,.15) circle (.07);
        \begin{scope}
        \clip (1.9,1) rectangle (3,0);
        \draw[very thick, out=0, in=0, looseness=3, red] (2.15,-.3) to (2,.5);
        \end{scope}
        \draw[very thick, blue] (2.2,.6) circle (1.7); 
        \fill[white] (-.28,1.41) circle (.07);
        \draw[very thick, out=-90, in=0, red] (2.3,1.8) to (-1,1.5);
        \end{scope}
    \end{tikzpicture}
    \caption{The Conway knot and its mutant, the Kinoshita-Teresaka knot. They are not equivalent, but the value of $Z_\text{Vir}$ is the same.}
    \label{fig:Conway knot mutation}
\end{figure}
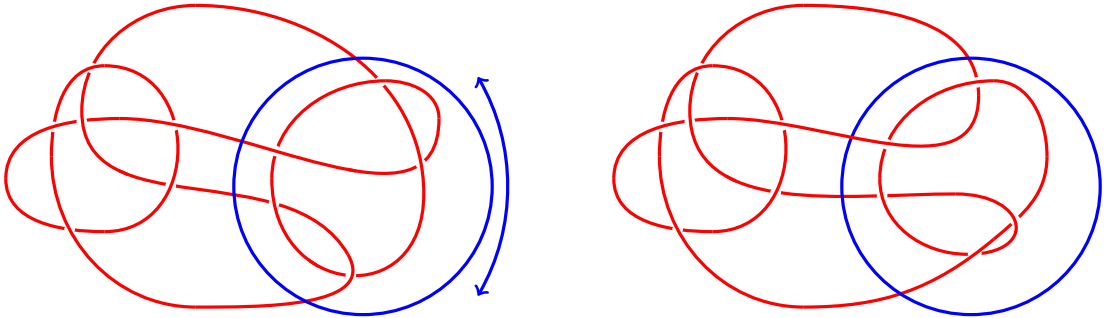
Consider a region of a knot in which two strands enter and two strands exit. The path integral over this region gives a state in the Hilbert space of the four-punctured sphere where all the four labels are identical (since we considering a knot associated to a single Virasoro representation). Thus, whatever the resulting state is, it can be expanded in terms of $s$-channel conformal blocks. However, any $s$-channel conformal block with identical external labels is invariant under a $\ZZ_2 \times \ZZ_2$ symmetry group generated by rotations around the $x$ and $y$ axis as follows:
\be 
\begin{tikzpicture}[baseline={([yshift=-.5ex]current bounding box.center)}]
    \draw[thick, densely dashed] (0,-.8) to (0,.8);
    \draw[thick, densely dashed] (-2.2,0) to (2.2,0);
    \draw[very thick, red] (-1,.5) node[left, black] {$P_0$} to (-.5,0);    
    \draw[very thick, red] (-1,-.5) node[left, black] {$P_0$} to (-.5,0);
    \draw[very thick, red] (1,.5) node[right, black] {$P_0$} to (.5,0);    
    \draw[very thick, red] (1,-.5) node[right, black] {$P_0$} to (.5,0);
    \draw[very thick, red] (-.5,0) to (.5,0);
    \draw[very thick,->, bend right=60] (1.9,-.5) to (1.9,.5);
    \draw[very thick,->, bend right=60] (-.5,-.5) to (.5,-.5);
    \node at (.22,.23) {$P$};
    \fill[red] (-.5,0) circle (.07);
    \fill[red] (.5,0) circle (.07);
\end{tikzpicture}\ ,
\ee
the composition of which yields a rotation by $180$ degrees. It thus follows that the Virasoro TQFT partition function on the excised four-punctured sphere is invariant under the same symmetry operations. In particular, this means that one can cut the four-punctured sphere with a tangle inside, apply one of these symmetry operations, and then reglue the tangle. This leads in general to an inequivalent knot, but the difference is not detectable by computing $Z_\text{Vir}$. A famous example of a mutant hyperbolic knot pair is the Conway knot and the Kinoshita-Terasaka knot shown in Figure~\ref{fig:Conway knot mutation}. In particular, since for this example, the mapping class group of both knot complements is trivial, the gravitational path integral on the Conway knot and the Kinoshita-Terasaka knot is exactly the same and the gravitational path integral is hence not a sufficiently refined observable to be able to detect the topology of all hyperbolic three-manifolds.

Via the volume conjecture \eqref{eq:refined volume conjecture}, this implies in particular that mutant knot complements have the same hyperbolic volume. This result is known in the math literature \cite{Mutation_volume}, but the present discussion makes it tautological. More surprisingly, the refined volume conjecture \eqref{eq:refined volume conjecture} also implies that the corresponding manifolds have the same one-loop determinants.

Using the same techniques of Virasoro TQFT, one can also show that the geodesics fully inside or outside the cutting surface have the same length.\footnote{This is based on the observation that inserting Wilson lines with degenerate Virasoro representations measure the geodesic length in the classical limit \cite{Eberhardt:2023mrq}.} Thus the length spectra of two mutant manifolds partially coincide. However, the length spectrum in general differs as one can see by an explicit computation using the software \texttt{SnapPy} \cite{SnapPy}. We display in Table~\ref{tab:length spectrum Conway knot} the low-lying length spectrum on the Conway knot and the Kinoshita-Teresaka knot.
Thus even though the geodesic length spectrum determines the one-loop determinant (we have $q_\gamma=\mathrm{e}^{-\ell_\gamma}$) and the one-loop determinants agree, the length spectrum is in general different. 

\begin{table}
    \centering
    \begin{tabular}{c|c}
    Conway & Kinoshita-Teresaka \\
    \hline
 ${\color{codegreen} 1.044 +2.327 i}$ & ${\color{codegreen} 1.044 +2.327 i}$ \\
 ${\color{codegreen} 1.152 -2.266 i}$ & ${\color{codegreen} 1.152 -2.266 i}$ \\
 ${\color{codegreen} 1.384 +2.840 i}$ & ${\color{codegreen} 1.384 +0.508 i}$ \\
 ${\color{red} 1.756 -2.011 i}$ & ${\color{red}1.530 +2.037 i}$ \\
 ${\color{codegreen} 1.831 -0.095 i}$ & ${\color{codegreen} 1.831 -0.095 i}$ \\
 ${\color{codegreen} 1.907 +2.521 i}$ & ${\color{codegreen} 1.907 +2.521 i}$ \\
 ${\color{codegreen} 1.938 -2.402 i}$ & ${\color{codegreen} 1.938 -2.402 i}$ \\
 ${\color{red}2.011 +0.738 i}$ & ${\color{red}2.031 +2.934 i}$ \\
 ${\color{red}2.184 -1.327 i}$ & ${\color{red}2.097 +2.938 i}$ \\
 ${\color{red}2.230 -1.770 i}$ & ${\color{red}2.183 -1.425 i}$ \\
 ${\color{codegreen} 2.233 -1.893 i}$ & ${\color{codegreen} 2.233 -1.893 i}$ 
    \end{tabular}
    \caption{The low-lying length-spectrum of primitive geodesics on the complement of the Conway and the Kinoshita-Teresaka knot. The imaginary part encodes the holonomy around the geodesic. The length spectrum partially agrees. All geodesics have multiplicity 1 since the two manifolds have trivial isometry groups.}
    \label{tab:length spectrum Conway knot}
\end{table}

There are other versions of mutations. We can consider any embedded surface in $M$ with a special symmetry such as the four-punctured sphere above. Cutting $M$ along such a surface, applying the symmetry and regluing leads to a mutated manifold. For example, we can cut along a genus 2 surface without punctures and use the $\ZZ_2$ hyperelliptic involution. Every genus 2 conformal block is invariant under the corresponding $\ZZ_2$ symmetry acting by a rotation around the $x$ axis as follows:\footnote{This would fail at genus 3 since we have in general two different Liouville momenta on the bottom and top of the middle loop, which get exchanged by this operation. Correspondingly, not every genus 3 surface is hyperelliptic.} 
\be 
    \begin{tikzpicture}[baseline={([yshift=-.5ex]current bounding box.center)}]
    \begin{scope}
        \draw[densely dashed, thick] (-3.2,0) to (3.2,0);
        \draw[very thick, red] (-1.5,0) circle (.8 and .5);
        \draw[very thick, red] (1.5,0) circle (.8 and .5);
        \draw[very thick, red] (-.7,0) to (.7,0);
        \node at (-1.5,.75) {$P_1$};
        \node at (0,.25) {$P_2$};
        \node at (1.5,.75) {$P_3$};
        \fill[red] (-.7,0) circle (.07);
        \fill[red] (.7,0) circle (.07);
        \draw[very thick, in=180, out=0] (-1.5,1) to (0,.5) to (1.5,1);
        \draw[very thick, in=180, out=0] (-1.5,-1) to (0,-.5) to (1.5,-1);
        \draw[very thick, in=180, out=180, looseness=2] (-1.5,1) to (-1.5,-1);
        \draw[very thick, in=0, out=0, looseness=2] (1.5,1) to (1.5,-1);
        \draw[very thick, bend right=30] (-2,0.05) to (-1,0.05);
        \draw[very thick, bend left=30] (-1.9,0) to (-1.1,0);
        \draw[very thick, bend left=30] (2,0.05) to (1,0.05);
        \draw[very thick, bend right=30] (1.9,0) to (1.1,0);
        \draw[very thick, bend right=60,->] (2.8,-.5) to (2.8,.5);
        \end{scope}
    \end{tikzpicture}\ .
\ee
Thus any partition function of Virasoro TQFT on a hyperbolic three-manifold with only a genus 2 boundary must have the same property. In particular, we can produce two hyperbolic three-manifolds by cutting along a genus 2 surface and applying such a rotation. This yields in general non-equivalent hyperbolic three-manifolds, but with the same value of $Z_\text{Vir}$. Via the volume conjecture \eqref{eq:refined volume conjecture}, this implies again that such a pair of manifolds has the same hyperbolic volume and the same value of the infinite product appearing as the one-loop determinant in \eqref{eq:refined volume conjecture}. 

As a concrete example that is perhaps more familiar and directly relevant to holography, consider the Euclidean wormhole of the form $\Sigma_2 \times [0,1]$. Since the genus 2 surface is hyperelliptic, we can perform the hyperelliptic involution on one side, which formally leads to a different manifold, but with identical partition function. 
As explained in \cite{Collier:2023fwi}, the Virasoro TQFT partition function on the Euclidean wormhole is simply given by the partition function of Liouville CFT $Z_\text{Liouville}(\Sigma_{2,0}|\mathbf{m}_1,\mathbf{m}_2)$, where the left-moving moduli $\mathbf{m}_1$ are associated to the left boundary and the right-moving moduli $\mathbf{m}_2$ to the right-moving boundary. 
However, the Liouville partition function is already invariant when we apply the hyperelliptic involution to only $\mathbf{m}_1$, which means that the partition function of this twisted wormhole also equals the Liouville partition function.

\subsection{Consistency conditions on the crossing kernels}\label{subsec:consistency conditions}
The crossing kernels $\mathbb{F}$ and $\mathbb{S}$ on the four-punctured sphere and the once-punctured torus, respectively, are subject to a number of constraints known as the Moore-Seiberg consistency conditions \cite{Moore:1988qv}. We listed them in the Appendix of \cite{Collier:2023fwi}, see also \cite{Teschner:2013tqy}. They express consistency of the projective representation of the 2d mapping class group on the space of conformal blocks. For example, the modular crossing kernel $\mathbb{S}$ has to satisfy the $\SL(2,\ZZ)$ relations together with the Dehn twist $\mathbb{T}$ on the once-punctured torus.\footnote{More precisely, the crossing kernels give rise to a projective representation of $\SL(2,\ZZ)$.}

These relations are also necessary for the consistency of the three-dimensional theory. However, they can often be seen much easier from the three-dimensional perspective. We explain here one simple example that shows that $\mathbb{S}$ can be fully expressed in terms of $\mathbb{F}$ that we also use later in the paper. We should mention that this construction is standard in the context of modular tensor categories which can be viewed as the rational counterpart of Virasoro TQFT \cite{Kitaev:2005hzj}.\footnote{We thank Sahand Seifnashri for explaining the MTC computation to us.} 

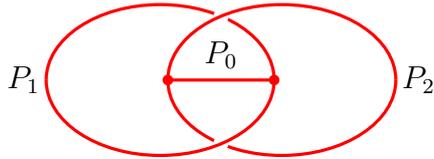
\begin{figure}
    \centering
    \begin{tikzpicture}
        \draw[very thick, red] (-.7,0) arc (-180:0:1.5 and 1);
        \draw[very thick, red] (.7,0) arc (0:180:1.5 and 1);
        \fill[white] (0,.84) circle (.1);
        \fill[white] (0,-.84) circle (.1);
        \draw[very thick, red] (-.7,0) arc (180:0:1.5 and 1);
        \draw[very thick, red] (.7,0) arc (0:-180:1.5 and 1);
        \draw[very thick, red] (-.7,0) to node[above, black] {$P_0$} (.7,0);
        \fill[red] (-.7,0) circle (.07);
        \fill[red] (.7,0) circle (.07);
        \node at (-2.6,0) {$P_1$};
        \node at (2.6,0) {$P_2$};
    \end{tikzpicture}
    \caption{The network of Wilson lines used to derive the relation between the modular crossing kernel $\mathbb{S}$ and the sphere crossing kernel $\mathbb{F}$.}
    \label{fig:Wilson lines crossing kernel}
\end{figure}

Consider $\mathrm{S}^3$ with a network of Wilson lines as in Figure~\ref{fig:Wilson lines crossing kernel}. We recall that a juncture of Wilson lines was defined as follows in \cite{Collier:2023fwi}:\footnote{$C_0(P_1,P_2,P_3)$ is the universal Liouville three-point function, see \cite[eq.~(2.17)]{Collier:2023fwi}.}
\be 
\begin{tikzpicture}[baseline={([yshift=-.5ex]current bounding box.center)}]
\draw[very thick, red] (0,0) to (.5,.87) node[right, black] {$P_1$};
\draw[very thick, red] (0,0) to (.5,-.87) node[right, black] {$P_3$};
\draw[very thick, red] (0,0) to (-1,0) node[left, black] {$P_2$};
\fill[red] (0,0) circle (0.07);
\end{tikzpicture}\equiv
\frac{1}{C_0(P_1,P_2,P_3)} \times 
\begin{tikzpicture}[baseline={([yshift=-.5ex]current bounding box.center)}]
\draw[very thick, red] (.25,.43) to (.5,.87) node[right, black] {$P_1$};
\draw[very thick, red] (.25,-.43) to (.5,-.87) node[right, black] {$P_3$};
\draw[very thick, red] (-.5,0) to (-1,0) node[left, black] {$P_2$};
\draw[very thick] (0,0) circle (.5);
\draw[very thick, densely dashed] (.5,0) arc (0:180:.5 and .2);
\draw[very thick] (.5,0) arc (0:-180:.5 and .2);
\end{tikzpicture}
\ . \label{eq:definition juncture}
\ee
On the right hand side, we excise a spherical boundary around the puncture. The path integral then creates a state in the boundary Hilbert space, which is one-dimensional and hence can be canonically identified with $\mathbb{C}$ by fixing the standard normalization of the three-point function on the sphere.

The main point is now that the value of the partition function on the network of Wilson lines in Figure~\ref{fig:Wilson lines crossing kernel} can be computed in two different ways as follows.

Let us first consider the Heegaard splitting into two once-punctured tori. The two once-punctured tori are homeomorphic to tubular neighborhoods of the Wilson lines $P_1$ and $P_2$, respectively. The normalization of the juncture in \eqref{eq:definition juncture} is chosen such that the Virasoro TQFT path integral on the once-punctured tori leads precisely to the respective conformal blocks on the boundary torus. The two once-punctured tori are interlocking and hence we have to apply an S-modular transformation. Being more careful about the definition of the S-modular transformation actually shows that we need the inverse of the modular crossing kernel $\mathbb{S}$. Since $\mathbb{S}$ squares to $\mathrm{e}^{\pi i \Delta_0}$,\footnote{See \cite[eq.~(A.5a)]{Collier:2023fwi}.} the inverse of $\mathbb{S}$ differs from $\mathbb{S}$ only by the phase $\mathrm{e}^{-\pi i \Delta_0}$. In the end, we obtain for the partition function of the Wilson line network $M$,
\begin{align} 
Z_\text{Vir}(M)&=\mathrm{e}^{-\pi i \Delta_0} \int \mathrm{d} P_1' \ \mathbb{S}_{P_1,P_1'}[P_0]\,  \bigg \langle \begin{tikzpicture}[baseline={([yshift=-.5ex]current bounding box.center)}]
\draw[very thick, red] (-.6,0) node[above, black] {$P_0$} to (0,0);
\draw[very thick, red] (.6,0) circle (.6);
\node at (.9,0) {$P_2$};
\fill[red] (0,0) circle (.07);
\end{tikzpicture} \bigg| \begin{tikzpicture}[baseline={([yshift=-.5ex]current bounding box.center)}]
\draw[very thick, red] (-.6,0) node[above, black] {$P_0$} to (0,0);
\draw[very thick, red] (.6,0) circle (.6);
\node at (.9,0) {$P_1'$};
\fill[red] (0,0) circle (.07);
\end{tikzpicture}\bigg \rangle \\
&=\frac{\mathrm{e}^{-\pi i \Delta_0}\, \mathbb{S}_{P_1,P_2}[P_0]}{\rho_0(P_2) C_0(P_0,P_2,P_2)}\ ,\label{eq:FS relation first evaluation ZVir}
\end{align}
where we applied \cite[eq.~(2.21)]{Collier:2023fwi} for the evaluation of the inner product between conformal blocks. It reads
\be 
\langle \mathcal{F}_{g,n}^{\mathcal{C}} (\vec P) \, | \, \mathcal{F}_{g,n}^{\mathcal{C}}(\vec P') \rangle= \frac{\delta^{3g-3+n}(\vec P-\vec P')}{\prod_{\text{cuffs a}} \rho_0(P_a) \prod_{\text{pair of pants }(i,j,k)} C_0(P_i,P_j,P_k)}\ .
\ee
Here, $\mathcal{F}^{\mathcal{C}}_{g,n}$ are the genus-$g$ $n$-point Virasoro blocks in a particular OPE channel $\mathcal{C}$, $C_0(P_1,P_2,P_3)$ is the Liouville three-point function, and $\rho_0(P)$ the inverse of the two-point function.
In other words, conformal blocks in a channel $\mathcal{C}$ are orthogonal with a density given by the inverse OPE density of Liouville theory. See \cite{Collier:2023fwi} for our conventions for the Liouville structure constants.

We can alternatively compute the partition function by a Heegaard splitting along a four-punctured sphere containing the Wilson line $P_0$ and the stubs of the Wilson lines $P_1$ and $P_2$. We have in hopefully obvious notation
\begin{align}
    Z_\text{Vir}(M)&=\int \d P\ \mathbb{F}_{P_0,P}\begin{bmatrix} P_2 & P_1 \\ P_2 & P_1
    \end{bmatrix} Z_\text{Vir} \Bigg( \begin{tikzpicture}[baseline={([yshift=-.5ex]current bounding box.center)}]
    \draw[very thick, out=60, in=90, looseness=3, red] (0,.3) to (-.7,0);
    \fill[white] (0,.77) circle (.09);
    \draw[very thick, out=120, in=90, looseness=3, red] (0,.3) to (.7,0);
    \draw[very thick, out=-120, in=-90, looseness=3, red] (0,-.3) to (.7,0);
    \fill[white] (0,-.77) circle (.09);
    \draw[very thick, out=-60, in=-90, looseness=3, red] (0,-.3) to (-.7,0);
    \draw[very thick, red] (0,-.3) to (0,.3);
    \node at (.2,0) {$P$};
    \node at (-.43,0) {$P_1$};
    \node at (.95,0) {$P_2$};
    \fill[red] (0,.3) circle (.06);
    \fill[red] (0,-.3) circle (.06);
\end{tikzpicture}\!\!\Bigg) \\
&=\int \d P\ \mathbb{F}_{P_0,P}\begin{bmatrix} P_2 & P_1 \\ P_2 & P_1
    \end{bmatrix} \, \mathrm{e}^{2\pi i (\Delta-\Delta_1-\Delta_2)} Z_\text{Vir} \Bigg( \begin{tikzpicture}[baseline={([yshift=-.5ex]current bounding box.center)}]
    \draw[very thick, red] (0,0) circle (.7 and .6);
    \draw[very thick, red] (0,-.6) to (0,.6);
    \node at (.2,0) {$P$};
    \node at (-.43,0) {$P_1$};
    \node at (.95,0) {$P_2$};
    \fill[red] (0,.6) circle (.06);
    \fill[red] (0,-.6) circle (.06);
\end{tikzpicture}\!\!\Bigg) \\
&=\int \d P\ \mathbb{F}_{P_0,P}\begin{bmatrix} P_2 & P_1 \\ P_2 & P_1
    \end{bmatrix} \, \frac{\mathrm{e}^{2\pi i (\Delta-\Delta_1-\Delta_2)}}{C_0(P,P_1,P_2)}\ , \label{eq:FS relation second evaluation ZVir}
\end{align} 
where we used the braiding move twice in the second line. In the last line we recognize the Euclidean wormhole with two three-punctured sphere on both ends, which evaluates to the Liouville three-point function. Taking into account the normalization in eq.~\eqref{eq:definition juncture}, we get an inverse structure constant.

Comparing \eqref{eq:FS relation first evaluation ZVir} and \eqref{eq:FS relation second evaluation ZVir} then expresses the modular crossing kernel fully in terms of the sphere crossing kernel,
\be 
\label{eq:FS relation}
\mathbb{S}_{P_1,P_2}[P_0]=\int \d P\ \frac{\rho_0(P_2) C_0(P_0,P_2,P_2)}{C_0(P,P_1,P_2)}\, \mathrm{e}^{\pi i (2\Delta+\Delta_0-2\Delta_1-2\Delta_2)}\, \mathbb{F}_{P_0,P}\begin{bmatrix} P_2 & P_1 \\ P_2 & P_1
    \end{bmatrix} \ .
\ee
Using the explicit form of the sphere crossing kernel given e.g.\ in eq.~(2.42a) of \cite{Collier:2023fwi}, one can use various known identities of the involved integrals of special functions to derive the known expression of the modular fusion kernel from this integral formula \cite{Teschner:2013tqy, Eberhardt:2023mrq}.

This identity can also be derived from a two-dimensional point of view by requiring the consistency of the representation of the mapping class group on the space of conformal blocks on the two-punctured torus. It is in fact a special case of the corresponding Moore-Seiberg relation. However, the corresponding derivation is much more complicated and subtle than the three-dimensional point of view.

\section{Holographic examples} \label{sec:holography}
We now move on to holographic applications of the Virasoro TQFT formalism. 
We will mostly focus on multi-boundary wormholes that have direct implications for the description of the holographic dual of 3d gravity in terms of an ensemble of CFT data. In order to set the stage for this discussion, let us briefly recapitulate the ensemble description of AdS$_3$ gravity. 

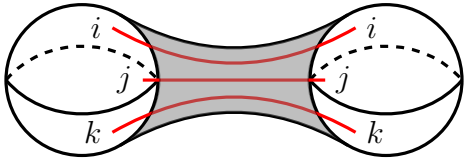
\begin{figure}[ht]
    \centering
    \begin{tikzpicture}
        \draw[very thick, dashed, out=50, in=130] (-3,0) to (-1,0);

        \draw[very thick, dashed, out=50, in=130] (1,0) to (3,0);

        \draw[very thick] (-2,0) ellipse (1 and 1);
        \draw[very thick, out=-50, in=230] (-3,0) to (-1,0);
        \draw[very thick] (2,0) ellipse (1 and 1);
        \draw[very thick, out=-50, in=230] (1,0) to (3,0);

        \draw[very thick, out = 330, in = 210] ({-2+cos(60)},{sin(60)}) to ({{2+cos(120)}},{sin(120)});
        \draw[very thick, out = 30, in = 150] ({-2+cos(-60)},{sin(-60)}) to ({{2+cos(240)}},{sin(240)});

        \draw[very thick, red, out = 330, in = 210] ({-2+4/5*cos(60)},{4/5*sin(60)}) to ({{2+4/5*cos(120)}},{4/5*sin(120)});
        \draw[very thick, red, out = 0, in = 180] ({-2+4/5},{0}) to ({2-4/5},{0});
        \draw[very thick, red, out = 30, in = 150] ({-2+4/5*cos(-60)},{4/5*sin(-60)}) to ({{2+4/5*cos(240)}},{4/5*sin(240)});

        \draw[fill=gray,opacity=.5] ({-2+cos(60)},{sin(60)}) to[out = 330, in = 210] ({{2+cos(120)}},{sin(120)}) to[out=210, in = 150,looseness=1.2] ({{2+cos(240)}},{sin(240)}) to[out=150, in =30] ({{-2+cos(-60)}},{sin(-60)}) to[out=30,in=330,looseness=1.2] ({{-2+cos(60)}},{sin(60)});

        \node[left] at ({-2+4/5*cos(60)},{4/5*sin(60)}) {$i$};
        \node[left] at ({-2+4/5},{0}) {$j$};
        \node[left] at ({-2+4/5*cos(-60)},{4/5*sin(-60)}) {$k$};
        
        \node[right] at ({2+4/5*cos(120)},{4/5*sin(120)}) {$i$};
        \node[right] at ({2-4/5},{0}) {$j$};
        \node[right] at ({2+4/5*cos(240)},{4/5*sin(240)}) {$k$};
    \end{tikzpicture}
    \caption{The Euclidean wormhole with the topology of a three-punctured sphere times an interval that contributes to the variance of the structure constants in the ensemble description of the dual of AdS$_3$ gravity.}\label{fig:variance wormhole}
\end{figure}

In \cite{Chandra:2022bqq} it was shown that averaged products of CFT observables in a Gaussian ensemble for the CFT data defined by
\begin{subequations}
    \begin{align}
        \overline{c_{ijk}} &= 0\ ,\\
        \overline{c_{ijk}c^*_{\ell m n}} &= C_0(P_i,P_j,P_k)C_0(\bar P_i, \bar P_j, \bar P_k)\nonumber\\
    &\qquad\times\left(\delta_{i\ell}\delta_{j m}\delta_{kn} + (-1)^{\ell_i+\ell_j+\ell_k}\delta_{i\ell}\delta_{j n}\delta_{k m} + 4\text{ permutations}\right) ,
    \end{align} \label{eq:ensemble definition}%
\end{subequations}
together with a Cardy spectrum of heavy states, agree with the on-shell actions (and in certain cases, the one-loop determinants) of suitable Euclidean wormholes in semiclassical AdS$_3$ gravity coupled to massive point particles. Here $\ell_i = P_i^2-\bar P_i^2$ is the spin of the corresponding primary. The averaged CFT quantities are computed by performing a simultaneous conformal block decomposition of the observables and computing Wick contractions of the structure constants using (\ref{eq:ensemble definition}). The gravity computations were mostly restricted to two-boundary wormholes with topology $\Sigma\times [0,1]$, with $\Sigma$ a (possibly punctured) Riemann surface, corresponding to two-copy averaged observables on the CFT ensemble side. Indeed one may view (\ref{eq:ensemble definition}) as being determined by an explicit computation of the 3d gravity partition function on a Euclidean wormhole with the topology of a three-punctured sphere times an interval \cite{Chandra:2022bqq}; see figure \ref{fig:variance wormhole}.

In \cite{Collier:2023fwi} the correspondence between two-boundary Euclidean wormhole partition functions and averaged products of CFT observables was extended to finite central charge using Virasoro TQFT. In particular the TQFT partition function on the Euclidean wormhole was computed, with the result
\begin{equation}\label{eq:VTQFT Euclidean wormhole}
    Z_{\text{Vir}}(\Sigma\times [0,1]|\mathbf{m}_1,\mathbf{m}_2) = Z_{\text{Liouville}}(\Sigma|\mathbf{m}_1,\mathbf{m}_2),
\end{equation}
where $\mathbf{m}_1,\mathbf{m}_2$ collectively denote the moduli of the Riemann surfaces at the two boundaries, and $Z_{\text{Liouville}}(\Sigma)$ is the correlation function on $\Sigma$ in Liouville CFT. $|Z_{\text{Vir}}(\Sigma\times[0,1])|^2$ agrees with the the averaged CFT computations performed in the Gaussian ensemble (\ref{eq:ensemble definition}), and its large-$c$ expansion agrees with the semiclassical gravity saddle-point computations in \cite{Chandra:2022bqq}. 

Except in certain very special cases it is not clear how to compute the gravity path integral on configurations with more than two asymptotic boundaries in the metric formalism. Such configurations in particular encode non-Gaussian corrections to the ensemble formulation of the boundary theory defined in (\ref{eq:ensemble definition}), which are known to be needed for the internal consistency of the ensemble description from a variety of points of view \cite{Belin:2021ryy, Jafferis:2022wez,Belin:2023efa}. For example, the existence of a Gaussian contraction often depends on the specific choice of channel in the conformal block decomposition of the CFT observables; crossing symmetry then requires non-Gaussian statistics in the dual channel in order to reproduce the result in the channel where the Gaussian contraction exists.  Hence non-Gaussian corrections, which are necessary for an internally consistent description of the boundary ensemble, are not presently accessible in the metric formulation of AdS$_3$ gravity. 

In the remainder of this section we will study Euclidean wormholes in AdS$_3$ gravity with more than two asymptotic boundaries using Virasoro TQFT. We will mostly focus on wormholes with more than two three-punctured sphere boundaries, since these determine the leading contributions to higher moments of the structure constants in the ensemble description of the holographic dual. We will see in some examples that the resulting non-Gaussian statistics precisely affirm the consistency of the results computed in the Gaussian ensemble.

There may also be non-Gaussian corrections to (\ref{eq:ensemble definition}) associated to Euclidean wormholes with two three-punctured sphere boundaries but with higher topology in the bulk. We will not study such corrections here, but let us briefly mention that we have already encountered such a correction associated with a higher-topology wormhole. In section \ref{subsec:consistency conditions} we studied a configuration of Wilson lines equivalent to the following two-boundary wormhole with linked Wilson lines
\begin{equation}
    M = 
    \vcenter{\hbox{
    \begin{tikzpicture}[scale=.75]
        \draw[very thick, dashed, out=50, in=130] (-3,0) to (-1,0);

        \draw[very thick, dashed, out=50, in=130] (1,0) to (3,0);

        \draw[very thick] (-2,0) ellipse (1 and 1);
        \draw[very thick, out=-50, in=230] (-3,0) to (-1,0);
        \draw[very thick] (2,0) ellipse (1 and 1);
        \draw[very thick, out=-50, in=230] (1,0) to (3,0);

        \draw[very thick, out = 330, in = 210] ({-2+cos(60)},{sin(60)}) to ({{2+cos(120)}},{sin(120)});
        \draw[very thick, out = 30, in = 150] ({-2+cos(-60)},{sin(-60)}) to ({{2+cos(240)}},{sin(240)});

        \draw[very thick, red, out = 330, in = 210] ({-2+4/5*cos(60)},{4/5*sin(60)}) to ({{2+4/5*cos(120)}},{4/5*sin(120)});
        \draw[very thick, red, looseness=.2] ({-2+4/5*cos(-60)},{4/5*sin(-60)}) to[out=60, in = -90] (1/4,-.15) to[out=90, in = 0]  ({-2+4/5},{0});
        \draw[fill=white, draw=white] (0,-.05) circle (1/10);
        \draw[very thick, red, looseness=.2] ({{2+4/5*cos(240)}},{4/5*sin(240)}) to[out=120, in = -90] (-1/4,-.15) to [out=90, in = 180] ({2-4/5},{0});
        \draw[fill=white,draw=white] (0,-.27) circle (1/10);
        \draw[very thick, red] ({.11*cos(196.5)},{-.27+.11*sin(196.5)}) to ({.11*cos(16.5)},{-.27+.11*sin(16.5)});

        \draw[fill=gray,opacity=.5] ({-2+cos(60)},{sin(60)}) to[out = 330, in = 210] ({{2+cos(120)}},{sin(120)}) to[out=210, in = 150,looseness=1.2] ({{2+cos(240)}},{sin(240)}) to[out=150, in =30] ({{-2+cos(-60)}},{sin(-60)}) to[out=30,in=330,looseness=1.2] ({{-2+cos(60)}},{sin(60)});

        \node[left] at ({-2+4/5*cos(60)},{4/5*sin(60)}) {$i$};
        \node[left] at ({-2+4/5},{0}) {$j$};
        \node[left] at ({-2+4/5*cos(-60)},{4/5*sin(-60)}) {$j$};
        
        \node[right] at ({2+4/5*cos(120)},{4/5*sin(120)}) {$i$};
        \node[right] at ({2-4/5},{0}) {$k$};
        \node[right] at ({2+4/5*cos(240)},{4/5*sin(240)}) {$k$};
    \end{tikzpicture}
    }}\,.
\end{equation}
The TQFT partition function on this wormhole may be computed as described in section \ref{subsec:consistency conditions}. One finds the following for the wormhole partition function
\begin{equation}
    Z_{\text{Vir}}(M) = \frac{C_0(P_i,P_j,P_j)\mathbb{S}_{P_jP_k}[P_i]}{\rho_0(P_k)}\, ,
\end{equation}
corresponding to the following averaged product of structure constants that would otherwise vanish (in the case that $j\ne k$) in the Gaussian ensemble\footnote{Strictly speaking, we get a different bulk manifold when we exchange the two ends of the Wilson line labelled by $k$. Exchanging them leads to a braiding phase $\mathrm{e}^{\pi i \Delta_j}$. Summing over both possibilities imposes that the spin of $i$ has to be even.}
\begin{equation}
    \overline{c_{ijj}c_{ikk}^*} = |Z_{\text{Vir}}(M)|^2 = \left|\frac{C_0(P_i,P_j,P_j)\mathbb{S}_{P_jP_k}[P_i]}{\rho_0(P_k)}\right|^2\, .
\end{equation}

\subsection{Cyclic defect wormholes}\label{subsec:symmetric wormholes}

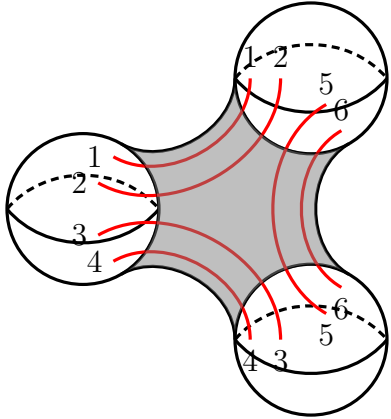
\begin{figure}[ht]
	\centering
	\begin{tikzpicture}
		\begin{scope}[shift={(-2,-3)}]
			\draw[very thick] (0,3) ellipse (1 and 1);
	        \draw[very thick, out=-50, in=230] (-1,3) to (1,3);
	        \draw[very thick, densely dashed, out=50, in=130] (-1,3) to (1,3);
        \end{scope}
        \begin{scope}[shift={({0+2*cos(60)},{-3+2*sin(60)})}]
			\draw[very thick] (0,3) ellipse (1 and 1);
	        \draw[very thick, out=-50, in=230] (-1,3) to (1,3);
	        \draw[very thick, densely dashed, out=50, in=130] (-1,3) to (1,3);
        \end{scope}
        \begin{scope}[shift={({0+2*cos(-60)},{-3+2*sin(-60)})}]
			\draw[very thick] (0,3) ellipse (1 and 1);
	        \draw[very thick, out=-50, in=230] (-1,3) to (1,3);
	        \draw[very thick, densely dashed, out=50, in=130] (-1,3) to (1,3);
        \end{scope}
        \draw[very thick, out = 330, in = 270] ({-2+cos(60)},{sin(60)}) to ({{2*cos(60)-1}},{2*sin(60)});
        \draw[very thick, out = 30, in = 90] ({-2+cos(-60)},{sin(-60)}) to ({{2*cos(-60)-1}},{2*sin(-60)});
        \draw[very thick, out = 210, in = 150] ({2*cos(60)+cos(-60)},{2*sin(60)+sin(-60)}) to ({{2*cos(-60)+cos(60)}},{2*sin(-60)+sin(60)});

        \draw[very thick, red, out = 330, in = 270] ({-2+4/5*cos(60)},{4/5*sin(60)}) to ({{2*cos(60)-4/5}},{2*sin(60)});
        \draw[very thick, red, out = 330, in = 270] ({-2+2/5*cos(60)},{2/5*sin(60)}) to ({{2*cos(60)-2/5}},{2*sin(60)});

        \draw[very thick, red, out =30, in = 90] ({-2+4/5*cos(-60)},{4/5*sin(-60)}) to ({{2*cos(-60)-4/5}},{2*sin(-60)});
        \draw[very thick, red, out =30, in = 90] ({-2+2/5*cos(-60)},{2/5*sin(-60)}) to ({{2*cos(-60)-2/5}},{2*sin(-60)});

        \draw[very thick, red, out = 210, in = 150] ({2*cos(60)+4/5*cos(-60)},{2*sin(60)+4/5*sin(-60)}) to ({{2*cos(-60)+4/5*cos(60)}},{2*sin(-60)+4/5*sin(60)});
        \draw[very thick, red, out = 210, in = 150] ({2*cos(60)+2/5*cos(-60)},{2*sin(60)+2/5*sin(-60)}) to ({{2*cos(-60)+2/5*cos(60)}},{2*sin(-60)+2/5*sin(60)});

        \node[left] at ({-2+4/5*cos(60)},{4/5*sin(60)}) {$1$};
        \node[left] at ({-2+2/5*cos(60)},{2/5*sin(60)}) {$2$};
        \node[left] at ({-2+4/5*cos(-60)},{4/5*sin(-60)}) {$4$};
        \node[left] at ({-2+2/5*cos(-60)},{2/5*sin(-60)}) {$3$};

        \node[above] at ({{2*cos(60)-4/5}},{2*sin(60)}) {$1$};
        \node[above] at ({{2*cos(60)-2/5}},{2*sin(60)}) {$2$};
        \node[above] at ({2*cos(60)+4/5*cos(-60)},{2*sin(60)+4/5*sin(-60)}) {$6$};
        \node[above] at ({2*cos(60)+2/5*cos(-60)},{2*sin(60)+2/5*sin(-60)}) {$5$};

        \node[below] at ({{2*cos(-60)-4/5}},{2*sin(-60)}) {$4$};
        \node[below] at ({{2*cos(-60)-2/5}},{2*sin(-60)}) {$3$};
        \node[below] at ({{2*cos(-60)+4/5*cos(60)}},{2*sin(-60)+4/5*sin(60)}) {$6$};
        \node[below] at ({{2*cos(-60)+2/5*cos(60)}},{2*sin(-60)+2/5*sin(60)}) {$5$};

        \draw[fill=gray,opacity=.5] ({-2+cos(60)},{sin(60)}) to[out = 330, in = 270] ({{2*cos(60)-1}},{2*sin(60)}) to[out=270, in =210,looseness=1.1] ({{2*cos(60)+cos(-60)}},{2*sin(60)+sin(-60)}) to[out=210, in =150] ({{2*cos(-60)+cos(60)}},{2*sin(-60)+sin(60)}) to[out=150, in = 90,looseness=1.1] ({{2*cos(-60)-1}},{2*sin(-60)}) to[out=90, in =30] ({{-2+cos(-60)}},{sin(-60)}) to[out=30, in = 330,looseness=1.1] ({{-2+cos(60)}},{sin(60)});

	\end{tikzpicture}
	\caption{The three-boundary sphere four-point wormhole $M_3$.}\label{fig:three boundary sphere four-point wormhole}
\end{figure}

A simple class of examples that demonstrate the practical utility of the TQFT reformulation of 3d gravity is provided by multi-boundary wormholes with defects connecting the sphere boundaries. For concreteness, consider the case where each boundary is a four-punctured sphere with defects connected in a pairwise cyclic way. See figure \ref{fig:three boundary sphere four-point wormhole} for a depiction of such a wormhole with three boundaries. In \cite{Chandra:2022bqq} it was argued that such on-shell wormholes contribute to the following averaged product of four-point functions
\begin{equation}
\overline{\langle\mathcal{O}_1\mathcal{O}_2\mathcal{O}_3\mathcal{O}_4\rangle\langle\mathcal{O}_3\mathcal{O}_4\mathcal{O}_5\mathcal{O}_6\rangle\cdots \langle\mathcal{O}_{2k-1}\mathcal{O}_{2k}\mathcal{O}_1\mathcal{O}_2\rangle}
\end{equation}
in the ensemble of CFT data dual to semiclassical 3d gravity.\footnote{Wormholes of this sort also contribute to the Renyi entropies of certain coarse-grained states in 2d CFT \cite{Chandra:2023dgq}.} The Gaussian ensemble hence makes a specific prediction for the gravitational partition function of these wormholes \cite{Chandra:2022bqq}: 
\begin{multline}
    Z_{\rm grav}(M_k) \stackrel{?}{=} \Bigg|\int_0^\infty \d P \ \rho_0(P) C_0(P_1,P_2,P) C_0(P_3,P_4,P)\cdots C_0(P_{2k-1},P_{2k}, P)\\
    \times  \begin{tikzpicture}[baseline={([yshift=-.5ex]current bounding box.center)}]
            \draw[very thick,red] (-3/4,1/2) to (-3/8,0);
            \draw[very thick,red] (-3/4,-1/2) to (-3/8,0);
            \draw[very thick,red] (-3/8,0) to (3/8,0);
            \draw[very thick,red] (3/8,0) to (3/4,1/2);
            \draw[very thick,red] (3/8,0) to (3/4,-1/2);
            \node[left] at (-3/4,1/2) {$P_1$}; 
            \node[left] at (-3/4,-1/2) {$P_2$};
            \node[right] at (3/4,1/2) {$P_4$};
            \node[right] at (3/4,-1/2) {$P_3$};
            \node[above] at (0,0) {$P$};
            \draw[draw=red, fill=red] (-3/8,0) circle (.07);
            \draw[draw=red, fill=red] (3/8,0) circle (.07);
        \end{tikzpicture}\!\!(z_1)
        \begin{tikzpicture}[baseline={([yshift=-.5ex]current bounding box.center)}]
            \draw[very thick,red] (-3/4,1/2) to (-3/8,0);
            \draw[very thick,red] (-3/4,-1/2) to (-3/8,0);
            \draw[very thick,red] (-3/8,0) to (3/8,0);
            \draw[very thick,red] (3/8,0) to (3/4,1/2);
            \draw[very thick,red] (3/8,0) to (3/4,-1/2);
            \node[left] at (-3/4,1/2) {$P_4$}; 
            \node[left] at (-3/4,-1/2) {$P_3$};
            \node[right] at (3/4,1/2) {$P_5$};
            \node[right] at (3/4,-1/2) {$P_6$};
            \node[above] at (0,0) {$P$};
            \draw[draw=red, fill=red] (-3/8,0) circle (.07);
            \draw[draw=red, fill=red] (3/8,0) circle (.07);
        \end{tikzpicture}\!\!
        (z_2)\ 
        \cdots
        \begin{tikzpicture}[baseline={([yshift=-.5ex]current bounding box.center)}]
            \draw[very thick,red] (-3/4,1/2) to (-3/8,0);
            \draw[very thick,red] (-3/4,-1/2) to (-3/8,0);
            \draw[very thick,red] (-3/8,0) to (3/8,0);
            \draw[very thick,red] (3/8,0) to (3/4,1/2);
            \draw[very thick,red] (3/8,0) to (3/4,-1/2);
            \node[left] at (-3/4,1/2) {$P_{2k}$}; 
            \node[left] at (-3/4,-1/2) {$P_{2k-1}$};
            \node[right] at (3/4,1/2) {$P_1$};
            \node[right] at (3/4,-1/2) {$P_2$};
            \node[above] at (0,0) {$P$};
            \draw[draw=red, fill=red] (-3/8,0) circle (.07);
            \draw[draw=red, fill=red] (3/8,0) circle (.07);
        \end{tikzpicture}\!\!
        (z_k)
\Bigg|^2\, . \label{eq:k-boundary four-point wormhole prediction}
\end{multline}
Here we have used the notation $M_k$ to refer to the $k$-boundary sphere four-point wormhole, the stick diagrams are shorthand for the conformal blocks as usual, and $z_i$ refers to the cross-ratio of the defect insertions on the $i^\text{th}$ boundary. The effect of the Gaussian average is to set all the internal weights equal in this particular conformal block decomposition of the wormhole partition function.

It is not at all clear how to compute the wormhole partition function of the $k$-boundary sphere four-point wormhole in the metric formalism of 3d gravity, even in the semiclassical limit. Here we will describe how this wormhole partition function may be straightforwardly computed in the Virasoro TQFT, reproducing the expectation from the Gaussian ensemble (\ref{eq:k-boundary four-point wormhole prediction}).

\begin{figure}[ht]
    \centering
    \begin{tikzpicture}
        \draw[very thick, densely dashed,out=50, in=130] (-3/2,0) to (3/2,0);

        \draw[fill=gray, opacity=.5, draw=gray] (0,0) ellipse (3/2 and 3/2);
        \draw[fill=white, draw=white] (-3/4,0) ellipse (1/2 and 1/2);
        \draw[fill=white, draw=white] (3/4,0) ellipse (1/2 and 1/2);

        \draw[very thick, fill=white] (-3/4,0) ellipse (1/2 and 1/2);
        \draw[very thick, out=-50, in=230] (-3/4-1/2,0) to (-3/4+1/2,0);
        \draw[very thick, densely dashed, out=50, in=130] (-3/4-1/2,0) to (-3/4+1/2,0);

        \draw[very thick, fill=white] (3/4,0) ellipse (1/2 and 1/2);
        \draw[very thick, out=-50, in=230] (3/4-1/2,0) to (3/4+1/2,0);
        \draw[very thick, densely dashed, out=50, in=130] (3/4-1/2,0) to (3/4+1/2,0);

        \draw[very thick, red] (-3/4+1/2-1/5,1/5) to (3/4-1/2+1/5,1/5);
        \draw[very thick, red] (-3/4+1/2-1/5,-1/5) to (3/4-1/2+1/5,-1/5);
        \draw[very thick, red] ({3/4+7/16*cos(45)},{7/16*sin(45)}) to ({3/2*cos(30)},{3/2*sin(30)});
        \draw[very thick, red] ({3/4+7/16*cos(45)},{-7/16*sin(45)}) to ({3/2*cos(30)},{-3/2*sin(30)});
        \draw[very thick, red] ({-3/4-7/16*cos(45)},{7/16*sin(45)}) to ({-3/2*cos(30)},{3/2*sin(30)});
        \draw[very thick, red] ({-3/4-7/16*cos(45)},{-7/16*sin(45)}) to ({-3/2*cos(30)},{-3/2*sin(30)});

        \draw[very thick] (0,0) ellipse (3/2 and 3/2);
        \draw[very thick, out=-50, in=230] (-3/2,0) to (3/2,0);

        \node[right] at ({3/2*cos(30)},{3/2*sin(30)}) {$6$};
        \node[right] at ({3/2*cos(30)},{-3/2*sin(30)}) {$5$};
        \node[left] at ({-3/2*cos(30)},{3/2*sin(30)}) {$1$};
        \node[left] at ({-3/2*cos(30)},{-3/2*sin(30)}) {$2$};
        \node[above] at (0,1/5) {$3$};
        \node[below] at (0,-1/5) {$4$};
    \end{tikzpicture}
    \caption{The three-boundary sphere four-point wormhole as a compression body.}\label{fig:three-boundary sphere four-point wormhole compression body}
\end{figure}
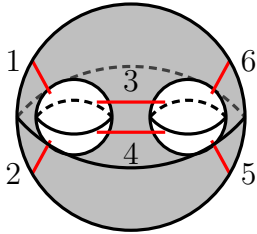

For concreteness and brevity of the equations we consider here the case $k=3$, but emphasize that the generalization to higher $k$ is completely straightforward. The idea is to view the wormhole as a compression body as indicated in figure \ref{fig:three-boundary sphere four-point wormhole compression body}. To compute the partition function on the corresponding compression body we insert a complete set of states in the Hilbert space of the inner boundaries. This produces a particular state in the Hilbert space of the outer boundary. Proceeding in this way we have
\begin{multline}
    Z_{\text{Vir}}(M_3) = \int_0^\infty \d P_a \, \d P_b \ \rho_0(P_a)\rho_0(P_b) C_0(P_1,P_2,P_a)C_0(P_3,P_4,P_a)C_0(P_3,P_4,P_b)\\
    \times
    C_0(P_5,P_6,P_b)\begin{tikzpicture}[baseline={([yshift=-.5ex]current bounding box.center)},scale=.9]
            \draw[very thick, densely dashed,out=50, in=130] (-3/2,0) to (3/2,0);
            \draw[very thick, red] ({3/2*cos(30)},{3/2*sin(30)}) to (1,0);
            \draw[very thick, red] ({3/2*cos(-30)},{3/2*sin(-30)}) to (1,0);
            \draw[very thick, red] (1,0) to (-1,0);
            \draw[very thick, red] ({3/2*cos(150)},{3/2*sin(150)}) to (-1,0);
            \draw[very thick, red] ({3/2*cos(210)},{3/2*sin(210)}) to (-1,0);
            \draw[very thick] (0,0) ellipse (3/2 and 3/2);
            \draw[very thick, out=-50, in=230] (-3/2,0) to (3/2,0); 
            \node[left] at ({3/2*cos(150)},{3/2*sin(150)}) {$P_1$};
            \node[left] at ({3/2*cos(210)},{3/2*sin(210)}) {$P_2$};
            \node[right] at ({3/2*cos(30)},{3/2*sin(30)}) {$P_4$};
            \node[right] at ({3/2*cos(-30)},{3/2*sin(-30)}) {$P_3$};
            \node[above] at (0,0) {$P_a$};
            \draw[draw=red,fill=red] (-1,0) circle (.07);
            \draw[draw=red,fill=red] (1,0) circle (.07);
            \begin{scope}[shift={(4,0)}]
                \draw[very thick, densely dashed,out=50, in=130] (-3/2,0) to (3/2,0);
                \draw[very thick, red] ({3/2*cos(30)},{3/2*sin(30)}) to (1,0);
                \draw[very thick, red] ({3/2*cos(-30)},{3/2*sin(-30)}) to (1,0);
                \draw[very thick, red] (1,0) to (-1,0);
                \draw[very thick, red] ({3/2*cos(150)},{3/2*sin(150)}) to (-1,0);
                \draw[very thick, red] ({3/2*cos(210)},{3/2*sin(210)}) to (-1,0);
                \draw[very thick] (0,0) ellipse (3/2 and 3/2);
                \draw[very thick, out=-50, in=230] (-3/2,0) to (3/2,0); 
                \node[left] at ({3/2*cos(150)},{3/2*sin(150)}) {$P_4$};
                \node[left] at ({3/2*cos(210)},{3/2*sin(210)}) {$P_3$};
                \node[right] at ({3/2*cos(30)},{3/2*sin(30)}) {$P_5$};
                \node[right] at ({3/2*cos(-30)},{3/2*sin(-30)}) {$P_6$};
                \node[above] at (0,0) {$P_b$};
                \draw[draw=red,fill=red] (-1,0) circle (.07);
                \draw[draw=red,fill=red] (1,0) circle (.07);
            \end{scope}
            \begin{scope}[shift={(8,0)},scale=1]
                \draw[very thick, densely dashed,out=50, in=130] (-3/2,0) to (3/2,0);
                \draw[very thick, red] ({3/2*cos(30)},{3/2*sin(30)}) to (1,0);
                \draw[very thick, red] ({3/2*cos(-30)},{3/2*sin(-30)}) to (1,0);
                \draw[very thick, red] (1,0) to (1/2,0);
                \draw[very thick, red] (0,0) circle (1/2);
                \draw[very thick, red] (-1/2,0) to (-1,0);
                \draw[very thick, red] ({3/2*cos(150)},{3/2*sin(150)}) to (-1,0);
                \draw[very thick, red] ({3/2*cos(210)},{3/2*sin(210)}) to (-1,0);
                \draw[very thick] (0,0) ellipse (3/2 and 3/2);
                \draw[very thick, out=-50, in=230] (-3/2,0) to (3/2,0);
                \node[left] at ({3/2*cos(150)},{3/2*sin(150)}) {$P_5$};
                \node[left] at ({3/2*cos(210)},{3/2*sin(210)}) {$P_6$};
                \node[right] at ({3/2*cos(30)},{3/2*sin(30)}) {$P_2$};
                \node[right] at ({3/2*cos(-30)},{3/2*sin(-30)}) {$P_1$};
                \node[above] at (3/4,-.1) {$P_b$};
                \node[above] at (-3/4,-.1) {$P_a$};
                \node[above] at (0,1/2+.1) {$P_4$};
                \node[below] at (0,-1/2-.1) {$P_3$};
                \draw[draw=red,fill=red] (-1,0) circle (.07);
                \draw[draw=red,fill=red] (1,0) circle (.07);
                \draw[draw=red,fill=red] (-1/2,0) circle (.07);
                \draw[draw=red,fill=red] (1/2,0) circle (.07);
            \end{scope}
        \end{tikzpicture}.  \label{eq:three boundary sphere four point partition function}
\end{multline}
We have temporarily restored the sphere boundaries in representing the conformal blocks in order to emphasize that the last conformal block should be interpreted as a state in the Hilbert space of the outer sphere boundary with a loop of Wilson lines in the interior, \emph{not} as a higher-genus conformal block. In particular we can remove the loop by recalling the TQFT identity \cite{Collier:2023fwi}
\begin{equation}\label{eq:Wilson line loop identity}
    \begin{tikzpicture}[baseline={([yshift=-.5ex]current bounding box.center)}]
        \draw[very thick,red] (-9/8,0) to (-3/8,0);
        \draw[very thick,red] (0,0) circle (3/8);
        \draw[very thick,red] (3/8,0) to (9/8,0);
        \node[above] at (-3/4,0) {$P_a$};
        \node[above] at (3/4,0) {$P_b$};
        \node[above] at (0,3/8) {$P_4$};
        \node[below] at (0,-3/8) {$P_3$};
        \draw[draw=red,fill=red] (-3/8,0) circle (.07);
        \draw[draw=red,fill=red] (3/8,0) circle (.07);
    \end{tikzpicture}
    = \frac{\delta(P_a-P_b)}{\rho_0(P_a)C_0(P_3,P_4,P_a)}
    \begin{tikzpicture}[baseline={([yshift=-2.3ex]current bounding box.center)}]
        \draw[very thick,red] (-9/8,0) to (9/8,0);
        \node[above] at (0,0) {$P_a$};
    \end{tikzpicture}\  ,
\end{equation}
which leads us to
\begin{multline}
    Z_{\text{Vir}}(M_3) = \int_0^\infty \d P_a \ \rho_0(P_a) C_0(P_1,P_2,P_a)C_0(P_3,P_4,P_a)C_0(P_5,P_6,P_a)\\
    \times 
    \begin{tikzpicture}[baseline={([yshift=-.5ex]current bounding box.center)}]
            \draw[very thick,red] (-3/4,1/2) to (-3/8,0);
            \draw[very thick,red] (-3/4,-1/2) to (-3/8,0);
            \draw[very thick,red] (-3/8,0) to (3/8,0);
            \draw[very thick,red] (3/8,0) to (3/4,1/2);
            \draw[very thick,red] (3/8,0) to (3/4,-1/2);
            \node[left] at (-3/4,1/2) {$P_1$}; 
            \node[left] at (-3/4,-1/2) {$P_2$};
            \node[right] at (3/4,1/2) {$P_4$};
            \node[right] at (3/4,-1/2) {$P_3$};
            \node[above] at (0,0) {$P_a$};
            \draw[draw=red,fill=red] (-3/8,0) circle (.07);
            \draw[draw=red,fill=red] (3/8,0) circle (.07);
            \begin{scope}[shift={(3,0)}]
                \draw[very thick,red] (-3/4,1/2) to (-3/8,0);
                \draw[very thick,red] (-3/4,-1/2) to (-3/8,0);
                \draw[very thick,red] (-3/8,0) to (3/8,0);
                \draw[very thick,red] (3/8,0) to (3/4,1/2);
                \draw[very thick,red] (3/8,0) to (3/4,-1/2);
                \node[left] at (-3/4,1/2) {$P_4$}; 
                \node[left] at (-3/4,-1/2) {$P_3$};
                \node[right] at (3/4,1/2) {$P_5$};
                \node[right] at (3/4,-1/2) {$P_6$};
                \node[above] at (0,0) {$P_a$};
                \draw[draw=red,fill=red] (-3/8,0) circle (.07);
                \draw[draw=red,fill=red] (3/8,0) circle (.07);
            \end{scope}
            \begin{scope}[shift={(6,0)}]
                \draw[very thick,red] (-3/4,1/2) to (-3/8,0);
                \draw[very thick,red] (-3/4,-1/2) to (-3/8,0);
                \draw[very thick,red] (-3/8,0) to (3/8,0);
                \draw[very thick,red] (3/8,0) to (3/4,1/2);
                \draw[very thick,red] (3/8,0) to (3/4,-1/2);
                \node[left] at (-3/4,1/2) {$P_5$}; 
                \node[left] at (-3/4,-1/2) {$P_6$};
                \node[right] at (3/4,1/2) {$P_2$};
                \node[right] at (3/4,-1/2) {$P_1$};
                \node[above] at (0,0) {$P_a$};
                \draw[draw=red,fill=red] (-3/8,0) circle (.07);
                \draw[draw=red,fill=red] (3/8,0) circle (.07);
            \end{scope}
        \end{tikzpicture}.  
\label{eq:three boundary sphere four point partition function 2}
\end{multline}
The generalization to the case of $k$ four-punctured sphere boundaries follows immediately by viewing $M_k$ as a compression body with $k-1$ inner boundaries and repeated application of the identity (\ref{eq:Wilson line loop identity}). Upon squaring the TQFT partition function to obtain the 3d gravity partition function, we hence verify (\ref{eq:k-boundary four-point wormhole prediction}), the prediction from the averaged product of $k$ sphere four-point functions in the Gaussian ensemble. Much like the case of the two-boundary Euclidean wormhole revisited in \cite{Collier:2023fwi}, the correspondence between the averaged CFT quantities and the gravity partition function on a fixed topology persists beyond the semiclassical limit.

\subsection{Four-boundary non-Gaussianity wormhole} \label{subsec:four-boundary wormhole}

Consider a wormhole with four three-punctured spheres as asymptotic boundaries, with defects threading the bulk of the wormhole in the following tetrahedral configuration 
\begin{equation}
    M = \begin{tikzpicture}[baseline={([yshift=-.5ex]current bounding box.center)},scale=1]
        \draw[thick, out = 30, in = 150] ({3*cos(210)+cos(-60)},{3*sin(210)+sin(-60)}) to ({{3*cos(330)+cos(240)}},{3*sin(330)+sin(240)});
            \draw[thick, out = 150, in = -90] ({{3*cos(330)+cos(60)}},{3*sin(330)+sin(60)}) to ({{3*cos(90)+cos(0)}},{3*sin(90)+sin(0)});
            \draw[thick, out = -90, in = 30] ({{3*cos(90)+cos(180)}},{3*sin(90)+sin(180)}) to ({{3*cos(210)+cos(120)}},{3*sin(210)+sin(120)});
            \begin{scope}[shift={({3*cos(210)},3*sin(210))}]
                    \draw[very thick] (0,0) ellipse (1 and 1);
                    \draw[very thick, out=-50, in=230] (-1,0) to (1,0);
                    \draw[very thick, densely dashed, out=50, in=130] (-1,0) to (1,0);
            \end{scope}
            \begin{scope}[shift={({3*cos(330)},{+3*sin(330)})}]
                    \draw[very thick] (0,0) ellipse (1 and 1);
                    \draw[very thick, out=-50, in=230] (-1,0) to (1,0);
                    \draw[very thick, densely dashed, out=50, in=130] (-1,0) to (1,0);
            \end{scope}
            \begin{scope}[shift={({0+3*cos(90)},{+3*sin(90)})}]
                    \draw[very thick] (0,0) ellipse (1 and 1);
                    \draw[very thick, out=-50, in=230] (-1,0) to (1,0);
                    \draw[very thick, densely dashed, out=50, in=130] (-1,0) to (1,0);
            \end{scope}

            \draw[very thick, red, out=30, in = 150] ({3*cos(210)+4/5*cos(-60)},{3*sin(210)+4/5*sin(-60)}) to ({{3*cos(330)+4/5*cos(240)}},{3*sin(330)+4/5*sin(240)});
            \draw[very thick, red, out=150, in = -90] ({{3*cos(330)+4/5*cos(60)}},{3*sin(330)+4/5*sin(60)}) to ({{3*cos(90)+4/5*cos(0)}},{3*sin(90)+4/5*sin(0)});
            \draw[very thick, red, out=-90, in = 30] ({{3*cos(90)+4/5*cos(180)}},{3*sin(90)+4/5*sin(180)}) to ({{3*cos(210)+4/5*cos(120)}},{3*sin(210)+4/5*sin(120)});

            \draw[very thick, red] ({3*cos(210)+4/5*cos(30)},{3*sin(210)+4/5*sin(30)}) to ({4/5*cos(210)},{4/5*sin(210)});
            \draw[very thick, red] ({{3*cos(330)+4/5*cos(150)}},{3*sin(330)+4/5*sin(150)}) to ({4/5*cos(330)},{4/5*sin(330)}); 
            \draw[very thick, red] ({{3*cos(90)+4/5*cos(-90)}},{3*sin(90)+4/5*sin(-90)}) to ({4/5*cos(90)},{4/5*sin(90)});

            \draw[fill=gray,opacity=.5] ({3*cos(210)+cos(-60)},{3*sin(210)+sin(-60)}) to[out = 30, in = 150] ({{3*cos(330)+cos(240)}},{3*sin(330)+sin(240)}) to[out=150, in=150, looseness=1.7] ({{3*cos(330)+cos(60)}},{3*sin(330)+sin(60)}) to [out=150, in = -90] ({{3*cos(90)+cos(0)}},{3*sin(90)+sin(0)}) to[out = -90, in = -90, looseness=1.7] ({{3*cos(90)+cos(180)}},{3*sin(90)+sin(180)}) to[out=-90, in = 30] ({{3*cos(210)+cos(120)}},{3*sin(210)+sin(120)}) to[out=30, in = 30, looseness=1.7] ({{3*cos(210)+cos(-60)}},{3*sin(210)+sin(-60)});

            \begin{scope}
                    \draw[fill=white, draw=white] (0,0) ellipse (1 and 1);
                    \draw[very thick, red] ({cos(210)},{sin(210)}) to ({4/5*cos(210)},{4/5*sin(210)});
                    \draw[very thick, red] ({cos(330)},{sin(330)}) to ({4/5*cos(330)},{4/5*sin(330)}); 
                    \draw[very thick, red] (({cos(90)},{sin(90)}) to ({4/5*cos(90)},{4/5*sin(90)});
                    \draw[very thick] (0,0) ellipse (1 and 1);
                    \draw[very thick, out=-50, in=230] (-1,0) to (1,0);
                    \draw[very thick, densely dashed, out=50, in=130] (-1,0) to (1,0);
            \end{scope}

            \node at ({{3*cos(330)+4/5*cos(150)+1/5}},{3*sin(330)+4/5*sin(150)-1/5}) {$3$};
            \node at ({4/5*cos(-30)-1/5},{4/5*sin(-30)+1/5}) {$3$};

            \node at ({3*cos(210)+4/5*cos(30)-1/5},{3*sin(210)+4/5*sin(30)-1/5}) {$2$};
            \node at ({4/5*cos(210)+1/5},{4/5*sin(210)+1/5}) {$2$};

            \node[above] at ({{3*cos(90)+4/5*cos(-90)}},{3*sin(90)+4/5*sin(-90)}) {$t$};
            \node[below] at ({4/5*cos(90)},{4/5*sin(90)}) {$t$};

            \node[right] at ({{3*cos(330)+4/5*cos(240)}},{3*sin(330)+4/5*sin(240)}) {$s$};
            \node[left] at ({3*cos(210)+4/5*cos(-60)},{3*sin(210)+4/5*sin(-60)}) {$s$};

            \node at ({{3*cos(330)+4/5*cos(60)+1/10}},{3*sin(330)+4/5*sin(60)-1/10}) {$4$};
            \node at ({{3*cos(90)+4/5*cos(0)-1/10}},{3*sin(90)+4/5*sin(0)+1/10}) {$4$};

            \node at ({{3*cos(210)+4/5*cos(120)-1/10}},{3*sin(210)+4/5*sin(120)-1/10}) {$1$};
            \node at ({{3*cos(90)+4/5*cos(180)+1/10}},{3*sin(90)+4/5*sin(180)+1/10}) {$1$};
    \end{tikzpicture} \label{eq:four-boundary wormhole}
\end{equation}
The gravity path integral on this wormhole should compute the following connected part of the fourth moment of structure constants in the dual description of 3d gravity in terms of an ensemble of CFT data
\begin{equation}\label{eq:fourth moment}
    |Z_{\rm Vir}(M)|^2 \leftrightarrow \overline{c_{12s}c_{s34}c_{14t}c_{t32}}.
\end{equation}

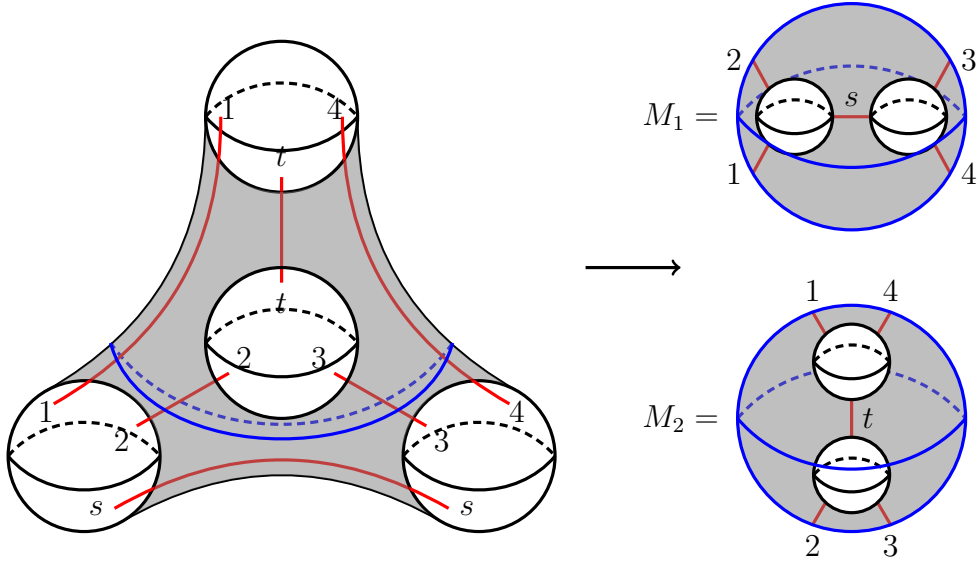
\begin{figure}[ht]
    \centering
    \begin{tikzpicture}
        \begin{scope}[shift={(-1,-1)}]
            \draw[very thick, densely dashed, blue, out = 150, in = 30,bend left = 55] (2.25,0) to (-2.25,0);

            \draw[thick, out = 30, in = 150] ({3*cos(210)+cos(-60)},{3*sin(210)+sin(-60)}) to ({{3*cos(330)+cos(240)}},{3*sin(330)+sin(240)});
            \draw[thick, out = 150, in = -90] ({{3*cos(330)+cos(60)}},{3*sin(330)+sin(60)}) to ({{3*cos(90)+cos(0)}},{3*sin(90)+sin(0)});
            \draw[thick, out = -90, in = 30] ({{3*cos(90)+cos(180)}},{3*sin(90)+sin(180)}) to ({{3*cos(210)+cos(120)}},{3*sin(210)+sin(120)});
            \begin{scope}[shift={({3*cos(210)},3*sin(210))}]
                    \draw[very thick] (0,0) ellipse (1 and 1);
                    \draw[very thick, out=-50, in=230] (-1,0) to (1,0);
                    \draw[very thick, densely dashed, out=50, in=130] (-1,0) to (1,0);
            \end{scope}
            \begin{scope}[shift={({3*cos(330)},{+3*sin(330)})}]
                    \draw[very thick] (0,0) ellipse (1 and 1);
                    \draw[very thick, out=-50, in=230] (-1,0) to (1,0);
                    \draw[very thick, densely dashed, out=50, in=130] (-1,0) to (1,0);
            \end{scope}
            \begin{scope}[shift={({0+3*cos(90)},{+3*sin(90)})}]
                    \draw[very thick] (0,0) ellipse (1 and 1);
                    \draw[very thick, out=-50, in=230] (-1,0) to (1,0);
                    \draw[very thick, densely dashed, out=50, in=130] (-1,0) to (1,0);
            \end{scope}

            \draw[very thick, red, out=30, in = 150] ({3*cos(210)+4/5*cos(-60)},{3*sin(210)+4/5*sin(-60)}) to ({{3*cos(330)+4/5*cos(240)}},{3*sin(330)+4/5*sin(240)});
            \draw[very thick, red, out=150, in = -90] ({{3*cos(330)+4/5*cos(60)}},{3*sin(330)+4/5*sin(60)}) to ({{3*cos(90)+4/5*cos(0)}},{3*sin(90)+4/5*sin(0)});
            \draw[very thick, red, out=-90, in = 30] ({{3*cos(90)+4/5*cos(180)}},{3*sin(90)+4/5*sin(180)}) to ({{3*cos(210)+4/5*cos(120)}},{3*sin(210)+4/5*sin(120)});

            \draw[very thick, red] ({3*cos(210)+4/5*cos(30)},{3*sin(210)+4/5*sin(30)}) to ({4/5*cos(210)},{4/5*sin(210)});
            \draw[very thick, red] ({{3*cos(330)+4/5*cos(150)}},{3*sin(330)+4/5*sin(150)}) to ({4/5*cos(330)},{4/5*sin(330)}); 
            \draw[very thick, red] ({{3*cos(90)+4/5*cos(-90)}},{3*sin(90)+4/5*sin(-90)}) to ({4/5*cos(90)},{4/5*sin(90)});

            \draw[fill=gray,opacity=.5] ({3*cos(210)+cos(-60)},{3*sin(210)+sin(-60)}) to[out = 30, in = 150] ({{3*cos(330)+cos(240)}},{3*sin(330)+sin(240)}) to[out=150, in=150, looseness=1.7] ({{3*cos(330)+cos(60)}},{3*sin(330)+sin(60)}) to [out=150, in = -90] ({{3*cos(90)+cos(0)}},{3*sin(90)+sin(0)}) to[out = -90, in = -90, looseness=1.7] ({{3*cos(90)+cos(180)}},{3*sin(90)+sin(180)}) to[out=-90, in = 30] ({{3*cos(210)+cos(120)}},{3*sin(210)+sin(120)}) to[out=30, in = 30, looseness=1.7] ({{3*cos(210)+cos(-60)}},{3*sin(210)+sin(-60)});

            \begin{scope}
                    \draw[fill=white, draw=white] (0,0) ellipse (1 and 1);
                    \draw[very thick, red] ({cos(210)},{sin(210)}) to ({4/5*cos(210)},{4/5*sin(210)});
                    \draw[very thick, red] ({cos(330)},{sin(330)}) to ({4/5*cos(330)},{4/5*sin(330)}); 
                    \draw[very thick, red] (({cos(90)},{sin(90)}) to ({4/5*cos(90)},{4/5*sin(90)});
                    \draw[very thick] (0,0) ellipse (1 and 1);
                    \draw[very thick, out=-50, in=230] (-1,0) to (1,0);
                    \draw[very thick, densely dashed, out=50, in=130] (-1,0) to (1,0);
            \end{scope}

            \node at ({{3*cos(330)+4/5*cos(150)+1/5}},{3*sin(330)+4/5*sin(150)-1/5}) {$3$};
            \node at ({4/5*cos(-30)-1/5},{4/5*sin(-30)+1/5}) {$3$};

            \node at ({3*cos(210)+4/5*cos(30)-1/5},{3*sin(210)+4/5*sin(30)-1/5}) {$2$};
            \node at ({4/5*cos(210)+1/5},{4/5*sin(210)+1/5}) {$2$};

            \node[above] at ({{3*cos(90)+4/5*cos(-90)}},{3*sin(90)+4/5*sin(-90)}) {$t$};
            \node[below] at ({4/5*cos(90)},{4/5*sin(90)}) {$t$};

            \node[right] at ({{3*cos(330)+4/5*cos(240)}},{3*sin(330)+4/5*sin(240)}) {$s$};
            \node[left] at ({3*cos(210)+4/5*cos(-60)},{3*sin(210)+4/5*sin(-60)}) {$s$};

            \node at ({{3*cos(330)+4/5*cos(60)+1/10}},{3*sin(330)+4/5*sin(60)-1/10}) {$4$};
            \node at ({{3*cos(90)+4/5*cos(0)-1/10}},{3*sin(90)+4/5*sin(0)+1/10}) {$4$};

            \node at ({{3*cos(210)+4/5*cos(120)-1/10}},{3*sin(210)+4/5*sin(120)-1/10}) {$1$};
            \node at ({{3*cos(90)+4/5*cos(180)+1/10}},{3*sin(90)+4/5*sin(180)+1/10}) {$1$};

            \draw[very thick, blue, out = 210, in = -30,bend left = 75] (2.25,0) to (-2.25,0);
        \end{scope}

        \draw[very thick,->] (3,0) to (4.25,0);

        \begin{scope}[shift={(6.5,2)}]
        \node at (-2.25,0) {$M_1=$};
        \draw[very thick, blue, densely dashed,out=50, in=130] (-3/2,0) to (3/2,0);

        \draw[very thick, red] (-3/4+1/2,0) to (3/4-1/2,0);
        \draw[very thick, red] ({3/4+7/16*cos(45)},{7/16*sin(45)}) to ({3/2*cos(30)},{3/2*sin(30)});
        \draw[very thick, red] ({3/4+7/16*cos(45)},{-7/16*sin(45)}) to ({3/2*cos(30)},{-3/2*sin(30)});
        \draw[very thick, red] ({-3/4-7/16*cos(45)},{7/16*sin(45)}) to ({-3/2*cos(30)},{3/2*sin(30)});
        \draw[very thick, red] ({-3/4-7/16*cos(45)},{-7/16*sin(45)}) to ({-3/2*cos(30)},{-3/2*sin(30)});

        \draw[fill=gray, opacity=.5, draw=gray] (0,0) ellipse (3/2 and 3/2);
        \draw[fill=white, draw=white] (-3/4,0) ellipse (1/2 and 1/2);
        \draw[fill=white, draw=white] (3/4,0) ellipse (1/2 and 1/2);

        \draw[very thick, fill=white] (-3/4,0) ellipse (1/2 and 1/2);
        \draw[very thick, out=-50, in=230] (-3/4-1/2,0) to (-3/4+1/2,0);
        \draw[very thick, densely dashed, out=50, in=130] (-3/4-1/2,0) to (-3/4+1/2,0);

        \draw[very thick, fill=white] (3/4,0) ellipse (1/2 and 1/2);
        \draw[very thick, out=-50, in=230] (3/4-1/2,0) to (3/4+1/2,0);
        \draw[very thick, densely dashed, out=50, in=130] (3/4-1/2,0) to (3/4+1/2,0);

        \draw[very thick, blue] (0,0) ellipse (3/2 and 3/2);
        \draw[very thick, blue, out=-50, in=230] (-3/2,0) to (3/2,0);

        \node[right] at ({3/2*cos(30)},{3/2*sin(30)}) {$3$};
        \node[right] at ({3/2*cos(30)},{-3/2*sin(30)}) {$4$};
        \node[left] at ({-3/2*cos(30)},{3/2*sin(30)}) {$2$};
        \node[left] at ({-3/2*cos(30)},{-3/2*sin(30)}) {$1$};
        \node[above] at (0,0) {$s$};

        \end{scope}

        \begin{scope}[shift={(6.5,-2)}]
        \node at (-2.25,0) {$M_2=$};
        \draw[very thick, blue, densely dashed,out=50, in=130] (-3/2,0) to (3/2,0);

        \draw[very thick, red] (0,3/4-1/2) to (0,-3/4+1/2);
        \draw[very thick, red] ({0+7/16*cos(135)},{3/4+7/16*sin(135)}) to ({3/2*cos(110)},{3/2*sin(110)});
        \draw[very thick, red] ({0+7/16*cos(45)},{3/4+7/16*sin(45)}) to ({3/2*cos(70)},{3/2*sin(70)});
        \draw[very thick, red] ({0+7/16*cos(225)},{-3/4+7/16*sin(225)}) to ({3/2*cos(250)},{3/2*sin(250)});
        \draw[very thick, red] ({0+7/16*cos(-45)},{-3/4+7/16*sin(-45)}) to ({3/2*cos(290)},{3/2*sin(290)});

        \draw[fill=gray, opacity=.5, draw=gray] (0,0) ellipse (3/2 and 3/2);
        \draw[fill=white, draw=white] (0,-3/4) ellipse (1/2 and 1/2);
        \draw[fill=white, draw=white] (0,3/4) ellipse (1/2 and 1/2);

        \draw[very thick,fill=white] (0,3/4) ellipse (1/2 and 1/2);
        \draw[very thick, out=-50, in=230] (-1/2,3/4) to (1/2,3/4);
        \draw[very thick, densely dashed, out=50, in=130] (-1/2,3/4) to (1/2, 3/4);

        \draw[very thick,fill=white] (0,-3/4) ellipse (1/2 and 1/2);
        \draw[very thick, out=-50, in=230] (-1/2,-3/4) to (1/2,-3/4);
        \draw[very thick, densely dashed, out=50, in=130] (-1/2,-3/4) to (1/2, -3/4);

        \draw[very thick, blue] (0,0) ellipse (3/2 and 3/2);
        \draw[very thick, blue, out=-50, in=230] (-3/2,0) to (3/2,0);

        \node[above] at ({3/2*cos(110)},{3/2*sin(110)}) {$1$};
        \node[above] at ({3/2*cos(70)},{3/2*sin(70)}) {$4$};
        \node[below] at ({3/2*cos(250)},{3/2*sin(250)}) {$2$};
        \node[below] at ({3/2*cos(290)},{3/2*sin(290)}) {$3$};
        \node[right] at (0,0) {$t$};
        \end{scope}
    \end{tikzpicture}
    \caption{The Heegaard splitting of the four-boundary wormhole $M$ into two generalized compression bodies $M_1$ and $M_2$, each with the topology of a three-ball with two three-balls drilled out in its interior and with Wilson lines connecting the two-sphere boundaries as shown in the figure. The TQFT path integral on each of $M_1$ and $M_2$ prepares a state in the Hilbert space of the four-punctured sphere.}\label{fig:four boundary wormhole heegaard splitting}
\end{figure}

\paragraph{Computation via Heegaard splitting.} It straightforward to apply the Heegaard splitting technique described in detail in \cite{Collier:2023fwi} to compute the Virasoro TQFT partition function on the four-boundary wormhole. For instance, we can cut $M$ along a four-punctured sphere through the bulk of the wormhole as pictured in figure \ref{fig:four boundary wormhole heegaard splitting}. This cuts the four-boundary wormhole into two generalized compression bodies $M_1$ and $M_2$. Each compression body has an outer boundary given by a four-punctured sphere and two three-punctured sphere inner boundaries. The Virasoro TQFT path integral on each compression body prepares a state in the Hilbert space of the four-punctured sphere, and the inner product of these states computes the TQFT partition function on the four-boundary wormhole. Using (\ref{eq:definition juncture}) to write the three-punctured sphere boundaries in terms of trivalent Wilson line junctions, the TQFT partition functions on the compression bodies are given by
\begin{subequations}
    \begin{align}
        \bra{Z_{\rm Vir}(M_1)} &= C_0(P_1,P_2,P_s)C_0(P_3,P_4,P_s)\Bigg\langle
        \begin{tikzpicture}[baseline={([yshift=-.5ex]current bounding box.center)}]
            \draw[very thick, red] (-3/4,1/2) to (-3/8,0);
            \draw[very thick, red] (-3/4,-1/2) to (-3/8,0);
            \draw[very thick, red] (-3/8,0) to (3/8,0);
            \draw[very thick, red] (3/8,0) to (3/4,1/2);
            \draw[very thick, red] (3/8,0) to (3/4,-1/2);
            \node[left] at (-3/4,1/2) {$P_1$}; 
            \node[left] at (-3/4,-1/2) {$P_2$};
            \node[right] at (3/4,1/2) {$P_4$};
            \node[right] at (3/4,-1/2) {$P_3$};
            \node[above] at (0,0) {$P_s$};
            \draw[draw=red,fill=red] (-3/8,0) circle (.07);
            \draw[draw=red,fill=red] (3/8,0) circle (.07);
        \end{tikzpicture}
        \Bigg|\ ,\\
        \ket{Z_{\rm Vir}(M_2)} &= C_0(P_1,P_4,P_t)C_0(P_2,P_3,P_t)
        \Bigg|
        \begin{tikzpicture}[baseline={([yshift=-.5ex]current bounding box.center)}]
            \draw[very thick, red] (-1/2,3/4) to (0,3/8);
            \draw[very thick, red] (1/2,3/4) to (0,3/8);
            \draw[very thick, red] (0,3/8) to (0,-3/8);
            \draw[very thick, red] (0,-3/8) to (-1/2,-3/4);
            \draw[very thick, red] (0,-3/8) to (1/2,-3/4);
            \node[left] at (-1/2,3/4) {$P_1$};
            \node[left] at (-1/2,-3/4) {$P_2$};
            \node[right] at (1/2,3/4) {$P_4$};
            \node[right] at (1/2,-3/4) {$P_3$};
            \node[left] at (0,0) {$P_t$};
            \draw[draw=red,fill=red] (0,-3/8) circle (.07);
            \draw[draw=red,fill=red] (0,3/8) circle (.07);
        \end{tikzpicture}
        \Bigg\rangle\ .
    \end{align} \label{eq:sphere four point compression body}%
\end{subequations}
Up to the $C_0$ factors, the compression body partition functions are given by individual sphere four-point conformal blocks in the $s$- and  the $t$-channel. The inner product of these states is proportional to the Virasoro fusion kernel essentially by definition: 
\begin{align}
    &\Bigg\langle
        \begin{tikzpicture}[baseline={([yshift=-.5ex]current bounding box.center)}]
            \draw[very thick, red] (-3/4,1/2) to (-3/8,0);
            \draw[very thick, red] (-3/4,-1/2) to (-3/8,0);
            \draw[very thick, red] (-3/8,0) to (3/8,0);
            \draw[very thick, red] (3/8,0) to (3/4,1/2);
            \draw[very thick, red] (3/8,0) to (3/4,-1/2);
            \node[left] at (-3/4,1/2) {$P_1$}; 
            \node[left] at (-3/4,-1/2) {$P_2$};
            \node[right] at (3/4,1/2) {$P_4$};
            \node[right] at (3/4,-1/2) {$P_3$};
            \node[above] at (0,0) {$P_s$};
            \draw[draw=red,fill=red] (-3/8,0) circle (.07);
            \draw[draw=red,fill=red] (3/8,0) circle (.07);
        \end{tikzpicture}
        \Bigg|
        \begin{tikzpicture}[baseline={([yshift=-.5ex]current bounding box.center)}]
            \draw[very thick, red] (-1/2,3/4) to (0,3/8);
            \draw[very thick, red] (1/2,3/4) to (0,3/8);
            \draw[very thick, red] (0,3/8) to (0,-3/8);
            \draw[very thick, red] (0,-3/8) to (-1/2,-3/4);
            \draw[very thick, red] (0,-3/8) to (1/2,-3/4);
            \node[left] at (-1/2,3/4) {$P_1$};
            \node[left] at (-1/2,-3/4) {$P_2$};
            \node[right] at (1/2,3/4) {$P_4$};
            \node[right] at (1/2,-3/4) {$P_3$};
            \node[left] at (0,0) {$P_t$};
            \draw[draw=red,fill=red] (0,-3/8) circle (.07);
            \draw[draw=red,fill=red] (0,3/8) circle (.07);
        \end{tikzpicture}
        \Bigg\rangle\nonumber\\
        &\qquad\qquad= \,  \int \d P_s' \ \mathbb{F}_{P_t P_s'}\begin{bmatrix}P_1 & P_2 \\ P_4 & P_3 \end{bmatrix}
        \Bigg\langle
        \begin{tikzpicture}[baseline={([yshift=-.5ex]current bounding box.center)}]
            \draw[very thick, red] (-3/4,1/2) to (-3/8,0);
            \draw[very thick, red] (-3/4,-1/2) to (-3/8,0);
            \draw[very thick, red] (-3/8,0) to (3/8,0);
            \draw[very thick, red] (3/8,0) to (3/4,1/2);
            \draw[very thick, red] (3/8,0) to (3/4,-1/2);
            \node[left] at (-3/4,1/2) {$P_1$}; 
            \node[left] at (-3/4,-1/2) {$P_2$};
            \node[right] at (3/4,1/2) {$P_4$};
            \node[right] at (3/4,-1/2) {$P_3$};
            \node[above] at (0,0) {$P_s$};
            \draw[draw=red,fill=red] (-3/8,0) circle (.07);
            \draw[draw=red,fill=red] (3/8,0) circle (.07);
        \end{tikzpicture}
        \Bigg|
        \begin{tikzpicture}[baseline={([yshift=-.5ex]current bounding box.center)}]
            \draw[very thick, red] (-3/4,1/2) to (-3/8,0);
            \draw[very thick, red] (-3/4,-1/2) to (-3/8,0);
            \draw[very thick, red] (-3/8,0) to (3/8,0);
            \draw[very thick, red] (3/8,0) to (3/4,1/2);
            \draw[very thick, red] (3/8,0) to (3/4,-1/2);
            \node[left] at (-3/4,1/2) {$P_1$}; 
            \node[left] at (-3/4,-1/2) {$P_2$};
            \node[right] at (3/4,1/2) {$P_4$};
            \node[right] at (3/4,-1/2) {$P_3$};
            \node[above] at (0,0) {$P_s'$};
            \draw[draw=red,fill=red] (-3/8,0) circle (.07);
            \draw[draw=red,fill=red] (3/8,0) circle (.07);
        \end{tikzpicture}
        \Bigg\rangle \\
        &\qquad\qquad=  \frac{\mathbb{F}_{P_t P_s}\begin{bmatrix} P_1 & P_2 \\ P_4 & P_3 \end{bmatrix}}{\rho_0(P_s)C_0(P_1,P_2,P_s)C_0(P_3,P_4,P_s)}\\ 
        &\qquad\qquad=  \frac{\mathbb{F}_{P_s  P_t}\begin{bmatrix} P_1 & P_4 \\ P_2 & P_3 \end{bmatrix}}{\rho_0(P_t)C_0(P_1,P_4,P_t)C_0(P_2,P_3,P_t)}\ .
\end{align}
In the penultimate line we computed the inner product by expanding the $t$-channel block in a complete basis of $s$-channel blocks using the Ponsot-Teschner fusion kernel \cite{Ponsot:1999uf,Ponsot:2000mt}, and in the last line we did the reverse. The equivalence of these two expressions is not a priori obvious without appealing to consistency of the conformal block inner product, but it is guaranteed by for example a special case of the pentagon identity, which is one of the Moore-Seiberg consistency conditions satisfied by the fusion kernel. In fact, this combination has a tetrahedral symmetry inherited from the bulk Wilson line configuration that is obscured by this presentation. Indeed, it can be rewritten in a manifestly tetrahedrally symmetric form in terms of the Virasoro $6j$ symbol in the Racah-Wigner normalization \cite{Teschner:2012em} as follows
\vspace{2.5ex}
\setlength{\jot}{-2ex}
\begin{multline}
    \Bigg\langle
        \begin{tikzpicture}[baseline={([yshift=-.5ex]current bounding box.center)}]
            \draw[very thick, red] (-3/4,1/2) to (-3/8,0);
            \draw[very thick, red] (-3/4,-1/2) to (-3/8,0);
            \draw[very thick, red] (-3/8,0) to (3/8,0);
            \draw[very thick, red] (3/8,0) to (3/4,1/2);
            \draw[very thick, red] (3/8,0) to (3/4,-1/2);
            \node[left] at (-3/4,1/2) {$P_1$}; 
            \node[left] at (-3/4,-1/2) {$P_2$};
            \node[right] at (3/4,1/2) {$P_4$};
            \node[right] at (3/4,-1/2) {$P_3$};
            \node[above] at (0,0) {$P_s$};
            \draw[draw=red,fill=red] (-3/8,0) circle (.07);
            \draw[draw=red,fill=red] (3/8,0) circle (.07);
        \end{tikzpicture}
        \Bigg|
        \begin{tikzpicture}[baseline={([yshift=-.5ex]current bounding box.center)}]
            \draw[very thick, red] (-1/2,3/4) to (0,3/8);
            \draw[very thick, red] (1/2,3/4) to (0,3/8);
            \draw[very thick, red] (0,3/8) to (0,-3/8);
            \draw[very thick, red] (0,-3/8) to (-1/2,-3/4);
            \draw[very thick, red] (0,-3/8) to (1/2,-3/4);
            \node[left] at (-1/2,3/4) {$P_1$};
            \node[left] at (-1/2,-3/4) {$P_2$};
            \node[right] at (1/2,3/4) {$P_4$};
            \node[right] at (1/2,-3/4) {$P_3$};
            \node[left] at (0,0) {$P_t$};
            \draw[draw=red,fill=red] (0,-3/8) circle (.07);
            \draw[draw=red,fill=red] (0,3/8) circle (.07);
        \end{tikzpicture}
        \Bigg\rangle\\
=\frac{\begin{Bmatrix} P_1 & P_2 & P_s \\ P_3 & P_4 & P_t \end{Bmatrix}}{\sqrt{C_0(P_1,P_2,P_s)C_0(P_3,P_4,P_s)C_0(P_1,P_4,P_t)C_0(P_2,P_3,P_t)}}\ .
\end{multline}
\setlength{\jot}{.5ex}
The upshot is that the Virasoro TQFT partition function on the four-boundary wormhole can be expressed in terms of the Virasoro $6j$ symbol via the following inner product in the Hilbert space of the four-punctured sphere
\begin{align}
        Z_{\rm Vir}(M) &=  \Braket{Z_{\rm Vir}(M_1)|Z_{\rm Vir}(M_2)}\\
        &= C_0(P_1,P_2,P_s)C_0(P_3,P_4,P_s)C_0(P_1,P_4,P_t)C_0(P_2,P_3,P_t) \nonumber\\
        & \qquad \times \Bigg\langle
        \begin{tikzpicture}[baseline={([yshift=-.5ex]current bounding box.center)}]
            \draw[very thick, red] (-3/4,1/2) to (-3/8,0);
            \draw[very thick, red] (-3/4,-1/2) to (-3/8,0);
            \draw[very thick, red] (-3/8,0) to (3/8,0);
            \draw[very thick, red] (3/8,0) to (3/4,1/2);
            \draw[very thick, red] (3/8,0) to (3/4,-1/2);
            \node[left] at (-3/4,1/2) {$P_1$}; 
            \node[left] at (-3/4,-1/2) {$P_2$};
            \node[right] at (3/4,1/2) {$P_4$};
            \node[right] at (3/4,-1/2) {$P_3$};
            \node[above] at (0,0) {$P_s$};
            \draw[draw=red,fill=red] (-3/8,0) circle (.07);
            \draw[draw=red,fill=red] (3/8,0) circle (.07);
        \end{tikzpicture}
        \Bigg|
        \begin{tikzpicture}[baseline={([yshift=-.5ex]current bounding box.center)}]
            \draw[very thick, red] (-1/2,3/4) to (0,3/8);
            \draw[very thick, red] (1/2,3/4) to (0,3/8);
            \draw[very thick, red] (0,3/8) to (0,-3/8);
            \draw[very thick, red] (0,-3/8) to (-1/2,-3/4);
            \draw[very thick, red] (0,-3/8) to (1/2,-3/4);
            \node[left] at (-1/2,3/4) {$P_1$};
            \node[left] at (-1/2,-3/4) {$P_2$};
            \node[right] at (1/2,3/4) {$P_4$};
            \node[right] at (1/2,-3/4) {$P_3$};
            \node[left] at (0,0) {$P_t$};
            \draw[draw=red,fill=red] (0,-3/8) circle (.07);
            \draw[draw=red,fill=red] (0,3/8) circle (.07);
        \end{tikzpicture}
        \Bigg\rangle\\
        &= \sqrt{C_0(P_1,P_2,P_s)C_0(P_3,P_4,P_s)C_0(P_1,P_4,P_t)C_0(P_2,P_3,P_t)} \begin{Bmatrix} P_1 & P_2 & P_s \\ P_3 & P_4 & P_t \end{Bmatrix}\ . \label{eq:four boundary wormhole partition function}
\end{align}

\paragraph{Consistency with boundary ensemble description.} The gravity partition function $Z_{\rm grav}(M) = |Z_{\rm Vir}(M)|^2$ on the four-boundary wormhole (\ref{eq:four boundary wormhole partition function}) makes a concrete prediction for the connected contribution to the fourth moment of structure constants in the description of 3d gravity in terms of an ensemble of CFT data:\footnote{This fourth moment has previously appeared in \cite{Belin:2023efa} where it was argued for by requiring that the variance of the crossing equation vanish, and in \cite{Belin:2021ryy} where it followed from genus-three modular invariance (using similar logic as that which shows that the variance should be given by the $C_0$ formula).}
\begin{equation}\label{eq:fourth moment 6j symbol}
    \overline{c_{12s}c_{s34}c_{14t}c_{t32}} \supset |Z_{\rm Vir}(M)|^2 = 
\sqrt{\overline{c^2_{12s}}\,\overline{c^2_{s34}}\,\overline{c^2_{14t}}\,\overline{c^2_{t32}}}
    \left|\begin{Bmatrix} P_1 & P_2 & P_s \\ P_3 & P_4 & P_t \end{Bmatrix}\right|^2
\end{equation}
This represents the leading correction to the Gaussian ensemble elucidated in \cite{Chandra:2022bqq}. We say that the fourth moment \emph{contains} this contribution (rather than being literally equal to it) because there may be corrections to (\ref{eq:fourth moment 6j symbol}) associated with wormholes with the same boundaries but with  higher topology in the bulk. It is expected that in the semiclassical limit such contributions are parametrically suppressed and hence that (\ref{eq:fourth moment 6j symbol}) represents the leading contribution to the fourth moment.

Here we will see that this non-Gaussian correction in fact exactly ensures the internal consistency of the results predicted by the Gaussian ensemble. 

To illustrate the point, consider the two-boundary Euclidean wormhole with the topology of a (possibly punctured) Riemann surface $\Sigma$ times an interval. The gravity path integral on the Euclidean wormhole is given by the square of (\ref{eq:VTQFT Euclidean wormhole}), the corresponding observable in Liouville CFT with the moduli on the two sides paired. This agrees with the averaged product of CFT observables in the Gaussian ensemble (\ref{eq:ensemble definition}). However the computation in the Gaussian ensemble often relies on the choice of a specific channel in the conformal block decomposition; this is obviously inconsistent with crossing symmetry of the ensemble. Associativity of the OPE then requires non-Gaussian statistics in order to reproduce this result in other channels. Relatedly, while the Gaussian ensemble is crossing symmetric on average, higher moments of the crossing equation do not vanish; this has recently been emphasized in \cite{Belin:2023efa}.

For concreteness, consider in particular the averaged product of four-point functions of local operators $\mathcal{O}_i$. In the Gaussian ensemble, we have
\begin{align}
        &\overline{\langle \mathcal{O}_1(0)\mathcal{O}_2(z,\bar z)\mathcal{O}_3(1)\mathcal{O}_4(\infty)\rangle\langle \mathcal{O}_1(0)\mathcal{O}_2(z',\bar z')\mathcal{O}_3(1)\mathcal{O}_4(\infty)\rangle^*}\nonumber\\
        &\quad= \sum_{s,s'}\overline{c_{12s}c_{34s}c_{12s'}^*c_{34s'}^*}\Bigg|\begin{tikzpicture}[baseline={([yshift=-.5ex]current bounding box.center)}]
            \draw[very thick, red] (-3/4,1/2) to (-3/8,0);
            \draw[very thick, red] (-3/4,-1/2) to (-3/8,0);
            \draw[very thick, red] (-3/8,0) to (3/8,0);
            \draw[very thick, red] (3/8,0) to (3/4,1/2);
            \draw[very thick, red] (3/8,0) to (3/4,-1/2);
            \node[left] at (-3/4,1/2) {$P_1$}; 
            \node[left] at (-3/4,-1/2) {$P_2$};
            \node[right] at (3/4,1/2) {$P_4$};
            \node[right] at (3/4,-1/2) {$P_3$};
            \node[above] at (0,0) {$P_s$};
            \draw[draw=red,fill=red] (-3/8,0) circle (.07);
            \draw[draw=red,fill=red] (3/8,0) circle (.07);
        \end{tikzpicture}\!\!
        (z)\begin{tikzpicture}[baseline={([yshift=-.5ex]current bounding box.center)}]
            \draw[very thick, red] (-3/4,1/2) to (-3/8,0);
            \draw[very thick, red] (-3/4,-1/2) to (-3/8,0);
            \draw[very thick, red] (-3/8,0) to (3/8,0);
            \draw[very thick, red] (3/8,0) to (3/4,1/2);
            \draw[very thick, red] (3/8,0) to (3/4,-1/2);
            \node[left] at (-3/4,1/2) {$P_1$}; 
            \node[left] at (-3/4,-1/2) {$P_2$};
            \node[right] at (3/4,1/2) {$P_4$};
            \node[right] at (3/4,-1/2) {$P_3$};
            \node[above] at (0,0) {$P_s'$};
            \draw[draw=red,fill=red] (-3/8,0) circle (.07);
            \draw[draw=red,fill=red] (3/8,0) circle (.07);
        \end{tikzpicture}\!\!
        (z')\Bigg|^2\\
        &\quad=  \left|\int \d P_s \, \rho_0(P_s)C_0(P_1,P_2,P_s)C_0(P_3,P_4,P_s)\!\!
        \begin{tikzpicture}[baseline={([yshift=-.5ex]current bounding box.center)}]
            \draw[very thick, red] (-3/4,1/2) to (-3/8,0);
            \draw[very thick, red] (-3/4,-1/2) to (-3/8,0);
            \draw[very thick, red] (-3/8,0) to (3/8,0);
            \draw[very thick, red] (3/8,0) to (3/4,1/2);
            \draw[very thick, red] (3/8,0) to (3/4,-1/2);
            \node[left] at (-3/4,1/2) {$P_1$}; 
            \node[left] at (-3/4,-1/2) {$P_2$};
            \node[right] at (3/4,1/2) {$P_4$};
            \node[right] at (3/4,-1/2) {$P_3$};
            \node[above] at (0,0) {$P_s$};
            \draw[draw=red,fill=red] (-3/8,0) circle (.07);
            \draw[draw=red,fill=red] (3/8,0) circle (.07);
        \end{tikzpicture}\!\!
        (z)\!\!
        \begin{tikzpicture}[baseline={([yshift=-.5ex]current bounding box.center)}]
            \draw[very thick, red] (-3/4,1/2) to (-3/8,0);
            \draw[very thick, red] (-3/4,-1/2) to (-3/8,0);
            \draw[very thick, red] (-3/8,0) to (3/8,0);
            \draw[very thick, red] (3/8,0) to (3/4,1/2);
            \draw[very thick, red] (3/8,0) to (3/4,-1/2);
            \node[left] at (-3/4,1/2) {$P_1$}; 
            \node[left] at (-3/4,-1/2) {$P_2$};
            \node[right] at (3/4,1/2) {$P_4$};
            \node[right] at (3/4,-1/2) {$P_3$};
            \node[above] at (0,0) {$P_s$};
            \draw[draw=red,fill=red] (-3/8,0) circle (.07);
            \draw[draw=red,fill=red] (3/8,0) circle (.07);
        \end{tikzpicture}\!\!
        (z')
        \right|^2\\
        &\quad=  \left|Z_{\text{Liouville}}(P_1,P_2,P_3,P_4|z,z')\right|^2\ . \label{eq:averaged product of four pt functions}
\end{align}
Here we expanded the four-point functions in the same OPE channel, and performed the Gaussian contractions in the third line using eq.~\eqref{eq:ensemble definition}. If we had instead expanded one four-point function in the S-channel and the other in the T-channel, we would have gotten zero in the Gaussian ensemble since
\begin{equation}
    \left.\overline{c_{12s}c_{s34}c_{14t}c_{t32}}\right|_{\rm Gaussian} = 0
\end{equation}
for distinct external operators.\footnote{We thank Vladimir Narovlansky for asking a question that raised this point.} This is obviously inconsistent with basic principles of conformal field theory. The result for the averaged product of four-point functions in terms of the four-point function in Liouville CFT is equal to the partition function of 3d gravity coupled to point particles on the Euclidean wormhole with the topology of a four-punctured sphere times an interval, so we seek a correction to the Gaussian ensemble that preserves (\ref{eq:averaged product of four pt functions}). If we supplement the Gaussian ensemble with the fourth moment (\ref{eq:fourth moment 6j symbol}) as computed by the four-boundary wormhole, we instead have
\begin{align}
         &\overline{\langle \mathcal{O}_1(0)\mathcal{O}_2(z,\bar z)\mathcal{O}_3(1)\mathcal{O}_4(\infty)\rangle\langle \mathcal{O}_1(0)\mathcal{O}_2(z',\bar z')\mathcal{O}_3(1)\mathcal{O}_4(\infty)\rangle^*}\nonumber\\
        &\quad= \sum_{s,t}\overline{c_{12s}c_{s34}c_{41t}^*c_{t23}^*}\Bigg|
        \!\begin{tikzpicture}[baseline={([yshift=-.5ex]current bounding box.center)}]
            \draw[very thick, red] (-3/4,1/2) to (-3/8,0);
            \draw[very thick, red] (-3/4,-1/2) to (-3/8,0);
            \draw[very thick, red] (-3/8,0) to (3/8,0);
            \draw[very thick, red] (3/8,0) to (3/4,1/2);
            \draw[very thick, red] (3/8,0) to (3/4,-1/2);
            \node[left] at (-3/4,1/2) {$P_1$}; 
            \node[left] at (-3/4,-1/2) {$P_2$};
            \node[right] at (3/4,1/2) {$P_4$};
            \node[right] at (3/4,-1/2) {$P_3$};
            \node[above] at (0,0) {$P_s$};
            \draw[draw=red,fill=red] (-3/8,0) circle (.07);
            \draw[draw=red,fill=red] (3/8,0) circle (.07);
        \end{tikzpicture}\!\!
        (z)
        \begin{tikzpicture}[baseline={([yshift=-.5ex]current bounding box.center)}]
            \draw[very thick, red] (-1/2,3/4) to (0,3/8);
            \draw[very thick, red] (1/2,3/4) to (0,3/8);
            \draw[very thick, red] (0,3/8) to (0,-3/8);
            \draw[very thick, red] (0,-3/8) to (-1/2,-3/4);
            \draw[very thick, red] (0,-3/8) to (1/2,-3/4);
            \node[left] at (-1/2,3/4) {$P_1$};
            \node[left] at (-1/2,-3/4) {$P_2$};
            \node[right] at (1/2,3/4) {$P_4$};
            \node[right] at (1/2,-3/4) {$P_3$};
            \node[left] at (0,0) {$P_t$};
            \draw[draw=red,fill=red] (0,-3/8) circle (.07);
            \draw[draw=red,fill=red] (0,3/8) circle (.07);
        \end{tikzpicture}\!\!
        (z')\Bigg|^2\\
        &\quad= \Bigg|\int \d P_s \, \d P_t\, \rho_0(P_s)C_0(P_1,P_2,P_s)C_0(P_3,P_4,P_s)\mathbb{F}_{P_s\, P_t}\begin{bmatrix} P_1 & P_4 \\ P_2 & P_3 \end{bmatrix}\nonumber\\
        & \qquad \times\!
        \begin{tikzpicture}[baseline={([yshift=-.5ex]current bounding box.center)}]
            \draw[very thick, red] (-3/4,1/2) to (-3/8,0);
            \draw[very thick, red] (-3/4,-1/2) to (-3/8,0);
            \draw[very thick, red] (-3/8,0) to (3/8,0);
            \draw[very thick, red] (3/8,0) to (3/4,1/2);
            \draw[very thick, red] (3/8,0) to (3/4,-1/2);
            \node[left] at (-3/4,1/2) {$P_1$}; 
            \node[left] at (-3/4,-1/2) {$P_2$};
            \node[right] at (3/4,1/2) {$P_4$};
            \node[right] at (3/4,-1/2) {$P_3$};
            \node[above] at (0,0) {$P_s$};
            \draw[draw=red,fill=red] (-3/8,0) circle (.07);
            \draw[draw=red,fill=red] (3/8,0) circle (.07);
        \end{tikzpicture}\!\!
        (z)\!
        \begin{tikzpicture}[baseline={([yshift=-.5ex]current bounding box.center)}]
            \draw[very thick, red] (-1/2,3/4) to (0,3/8);
            \draw[very thick, red] (1/2,3/4) to (0,3/8);
            \draw[very thick, red] (0,3/8) to (0,-3/8);
            \draw[very thick, red] (0,-3/8) to (-1/2,-3/4);
            \draw[very thick, red] (0,-3/8) to (1/2,-3/4);
            \node[left] at (-1/2,3/4) {$P_1$};
            \node[left] at (-1/2,-3/4) {$P_2$};
            \node[right] at (1/2,3/4) {$P_4$};
            \node[right] at (1/2,-3/4) {$P_3$};
            \node[left] at (0,0) {$P_t$};
            \draw[draw=red,fill=red] (0,-3/8) circle (.07);
            \draw[draw=red,fill=red] (0,3/8) circle (.07);
        \end{tikzpicture}\!\!
        (z')
        \Bigg|^2\\
        &\quad= \left|Z_{\text{Liouville}}(P_1,P_2,P_3,P_4|z,z')\right|^2,
\end{align}
in agreement with the previous computation and with the wormhole partition function. 

\paragraph{On braiding and the $u$-channel.} In the discussion so far we have suppressed an important subtlety. 
In 2d CFT, the structure constants are not strictly invariant under permutations of the three operators. For example, swapping a pair of operators leads to a sign that depends on the sum of the spins of the three operators
\begin{equation}\label{eq:exchanging operators}
    c_{ikj} = (-1)^{\ell_i+\ell_j+\ell_k}c_{ijk}.
\end{equation}
This is inherited from reality properties of the structure constants: they are real if the sum of spins is even and imaginary if the sum of spins is odd, and the swap complex-conjugates the structure constants, $c_{ikj} = c^*_{ijk}$.
Similarly, in the computation of wormhole partition functions with bulk Wilson lines via Heegaard splitting, there may be crossings of lines that need to be undone via braiding operations. These braidings introduce phases that depend on the conformal weights.

In general, we can read off the ordering of the structure constants from a bulk manifold by fixing a cyclic ordering and reading the labels around three-punctured boundaries cyclically. The same applies in CFT computations, where we read off the labels of the structure constants cyclically around every vertex in the conformal blocks.\footnote{The overall cyclic direction does not matter since every label appears twice and thus cancels if we reverse the overall cyclic direction.}

As a simple example of a wormhole computation for which such braidings are essential, consider the following four-boundary wormhole:
\begin{equation}\label{eq:four boundary wormhole u channel}
    M = \begin{tikzpicture}[baseline={([yshift=-.5ex]current bounding box.center)},scale=1]
        \draw[very thick] (-2,2) ellipse (1 and 1);
        \draw[very thick, out=-50, in=230] (-3,2) to (-1,2);
        \draw[very thick, densely dashed, out=50, in=130] (-3,2) to (-1,2);

        \draw[very thick] (2,2) ellipse (1 and 1);
        \draw[very thick, out=-50, in=230] (1,2) to (3,2);
        \draw[very thick, densely dashed, out=50, in=130] (1,2) to (3,2);

        \draw[very thick] (-2,-2) ellipse (1 and 1);
        \draw[very thick, out=-50, in=230] (-3,-2) to (-1,-2);
        \draw[very thick, densely dashed, out=50, in=130] (-3,-2) to (-1,-2);

        \draw[very thick] (2,-2) ellipse (1 and 1);
        \draw[very thick, out=-50, in=230] (1,-2) to (3,-2);
        \draw[very thick, densely dashed, out=50, in=130] (1,-2) to (3,-2);

        \draw[thick, out = 30, in = -30] ({-2-cos(60)},{-2+sin(60)}) to ({{-2-cos(60)}},{2-sin(60)}); 
        \draw[thick, out = 60, in = 120] ({-2+cos(30)},{-2-sin(30)}) to ({{2-cos(30)}},{-2-sin(30)});
        \draw[thick, out = 150, in = -150] ({2+cos(60)},{-2+sin(60)}) to ({{2+cos(60)}},{2-sin(60)}); 
        \draw[thick, out=-60, in=240] ({-2+cos(30)},{2+sin(30)}) to ({{2-cos(30)}},{2+sin(30)});

        \draw[very thick, red, out =-60, in = 240] ({-2+4/5*cos(30)},{2+4/5*sin(30)}) to ({{2-4/5*cos(30)}},{2+4/5*sin(30)});
        \draw[very thick, red, out = 30, in = -30] ({-2-4/5*cos(60)},{-2+4/5*sin(60)}) to ({{-2-4/5*cos(60)}},{2-4/5*sin(60)}); 
        \draw[very thick, red, out = 60, in = 120] ({-2+4/5*cos(30)},{-2-4/5*sin(30)}) to ({{2-4/5*cos(30)}},{-2-4/5*sin(30)});
        \draw[very thick, red, out = 150, in = -150] ({2+4/5*cos(60)},{-2+4/5*sin(60)}) to ({{2+4/5*cos(60)}},{2-4/5*sin(60)});
        \draw[very thick, red] (-2+.565685,2-.565685) to (2-.565685,-2+.565685);
        \draw[very thick, red] (-2+.565685,-2+.565685) to (0-1/10,0-1/10);
        \draw[very thick, red] (1/10,1/10) to (2-.565685,2-.565685); 

        \node[left] at ({-2+4/5*cos(30)},{2+4/5*sin(30)+3/20}) {$1$};
        \node[left] at (-2+.565685,2-.565685) {$2$};
        \node[left] at ({{-2-4/5*cos(60)+1/20}},{2-4/5*sin(60)+1/20}) {$s$};

        \node[left] at ({-2-4/5*cos(60)+1/20},{-2+4/5*sin(60)-1/20}) {$s$};
        \node[left] at (-2+.565685,-2+.565685) {$3$};
        \node[left] at ({-2+4/5*cos(30)},{-2-4/5*sin(30)-3/20}) {$4$};

        \node[right] at ({{2-4/5*cos(30)}},{-2-4/5*sin(30)-3/20}) {$4$};
        \node[right] at (2-.565685,-2+.565685) {$2$}; 
        \node[right] at ({2+4/5*cos(60)-1/20},{-2+4/5*sin(60)-1/20}) {$u$};

        \node[right] at ({{2+4/5*cos(60)-1/20}},{2-4/5*sin(60)+1/20}) {$u$};
        \node[right] at (2-.565685,2-.565685) {$3$};
        \node[right] at ({{2-4/5*cos(30)}},{2+4/5*sin(30)+3/20}) {$1$};

        \draw[fill=gray, opacity=0.5] ({-2-cos(60)},{-2+sin(60)}) to[out=30, in = -30] ({{-2-cos(60)}},{2-sin(60)}) to[out = -30, in=-60,looseness=1.35] ({{-2+cos(30)}},{2+sin(30)}) to[out=-60,in = 240] ({{2-cos(30)}},{2+sin(30)}) to[out = 240, in = -150,looseness=1.35] ({{2+cos(60)}},{2-sin(60)}) to[out=-150, in = 150] ({{2+cos(60)}},{-2+sin(60)}) to[out=150, in = 120,looseness=1.35] ({{2-cos(30)}},{-2-sin(30)}) to[out=120, in = 60] ({{-2+cos(30)}},{-2-sin(30)}) to[out=60, in = 30,looseness=1.35] ({{-2-cos(60)}},{-2+sin(60)});
    \end{tikzpicture}\ ,
\end{equation}
which is essentially the same as \eqref{eq:four-boundary wormhole}.
The boundaries of this wormhole are three-punctured spheres corresponding to the structure constants that appear in the $s$- and $u$-channel conformal block decompositions of the sphere four-point function $\langle \mathcal{O}_1\mathcal{O}_2\mathcal{O}_3\mathcal{O}_4\rangle$. 

We compute the TQFT partition function as before by splitting along a four-punctured sphere in the bulk. Undoing the crossing of the Wilson lines and computing the inner product in the Hilbert space of the splitting surface leads to the following result for the TQFT partition function on this four-boundary wormhole
\begin{equation}\label{eq:s to u four boundary wormhole}
    Z_{\text{Vir}}(M) =
    \sqrt{\mathsf{C}_{12s}\mathsf{C}_{s34}\mathsf{C}_{31u}\mathsf{C}_{u42}}
    \begin{Bmatrix}
        P_1 & P_2 & P_s \\ P_4 & P_3 & P_u
    \end{Bmatrix}
    \mathrm{e}^{\pi i (P_1^2+P_4^2-P_s^2-P_u^2)}\, .
\end{equation}
Here we have introduced the shorthand
\begin{equation}
    \mathsf{C}_{ijk} \equiv C_0(P_i,P_j,P_k)\, ,
\end{equation}
We notice the presence of an additional phase compared to (\ref{eq:four boundary wormhole partition function}). This result follows from taking the inner product between an $s$- and a $u$-channel Virasoro conformal block, and hence this phase may be understood in terms of the crossing transformation that relates $s$- and $u$-channel blocks. 
This crossing transformation is given by
\begin{equation}
    \begin{tikzpicture}[scale=.75, baseline={([yshift=-.5ex]current bounding box.center)}]
        \draw[very thick, red] (0,0) to (3,0);
        \draw[very thick, red] (1,0) to (1,1);
        \draw[very thick, red] (2,0) to (2,1);
        \node[above] at (0,0) {$1$};
        \node[above] at (3,0) {$4$};
        \node[above] at (1,1) {$2$};
        \node[above] at (2,1) {$3$};
        \node[below] at (3/2,0) {$s$};
        \draw[draw=red,fill=red] (1,0) circle (.07);
        \draw[draw=red,fill=red] (2,0) circle (.07);
    \end{tikzpicture}
    = \int_0^\infty \!\!\d P_u \, \mathrm{e}^{\pi i(P_1^2 + P_4^2 - P_s^2 - P_u^2)}\, \mathbb{F}_{P_s P_u}\begin{bmatrix}
            P_1 & P_3 \\ P_2 & P_4
        \end{bmatrix}
    \begin{tikzpicture}[scale=.75, baseline={([yshift=-.5ex]current bounding box.center)}]
        \draw[very thick, red] (0,0) to (3,0);
        \draw[very thick, red] (2,0) to (1,1);
        \fill[white] (1.5,.5) circle (.1);
        \draw[very thick, red] (1,0) to (2,1);
        \node[above] at (0,0) {$1$};
        \node[above] at (3,0) {$4$};
        \node[above] at (1,1) {$2$};
        \node[above] at (2,1) {$3$};
        \node[below] at (3/2,0) {$u$};
        \draw[draw=red,fill=red] (1,0) circle (.07);
        \draw[draw=red,fill=red] (2,0) circle (.07);
    \end{tikzpicture}\ . \label{eq:conformal block braiding}
\end{equation}
The combination that appears on the right-hand side is sometimes referred to as the ``R-matrix.'' The semiclassical near-extremal limit of the R-matrix governs the out-of-time-order four-point function in the Schwarzian theory \cite{Mertens:2017mtv}.

Hence for the following fourth moment of CFT structure constants we find
\begin{equation}
    \overline{c_{12s}c_{s34}c_{31u}c_{u42}} \supset |Z_{\text{Vir}}(M)|^2 = (-1)^{\ell_1+\ell_4+\ell_s+\ell_u}\sqrt{\overline{c^2_{12s}}\, \overline{c^2_{s34}}\, \overline{c^2_{13u}}\, \overline{c^2_{u24}}}
    \left|\begin{Bmatrix}
        P_1 & P_2 & P_s \\ P_4 & P_3 & P_u
    \end{Bmatrix}\right|^2.
\end{equation}
This is exactly consistent with the previous result (\ref{eq:fourth moment 6j symbol}) upon relabeling $t\to u$, $4\leftrightarrow 3$ and making use of the exchange property (\ref{eq:exchanging operators}). It is also consistent with the averaged product of sphere four-point functions in the Gaussian ensemble, where we expand one four-point function in the $s$-channel and the other in the $u$-channel.

\subsection{Many-boundary wormholes and higher non-Gaussianities}

\subsubsection{A simple six-boundary example} 
\begin{figure}[ht]
    \centering
    \begin{tikzpicture}[baseline={([yshift=-.5ex]current bounding box.center)},scale=.9]
            \draw[very thick, densely dashed, blue, out = 150, in = 30,looseness=.25] (2.15,0) to (-2.15,0);

            \draw[thick, out = 120, in = 240] ({2+cos(30)},{-3+sin(30)}) to ({{2+cos(-30)}},{3+sin(-30)});
            \draw[thick, out = 240, in = -60] ({2+cos(150)},{3+sin(150)}) to ({{-2+cos(30)}},{3+sin(30)});
            \draw[thick, out = -60, in = 60] ({-2+cos(210)},{3+sin(210)}) to ({{-2+cos(150)}},{-3+sin(150)});
            \draw[thick, out = 60, in = 120] ({-2+cos(-30)},{-3+sin(-30)}) to ({{2+cos(210)}},{-3+sin(210)});

            \begin{scope}[shift={(-2,-3)}]
                    \draw[very thick] (0,0) ellipse (1 and 1);
                    \draw[very thick, out=-50, in=230] (-1,0) to (1,0);
                    \draw[very thick, densely dashed, out=50, in=130] (-1,0) to (1,0);
            \end{scope}
            \begin{scope}[shift={(2,-3)}]
                    \draw[very thick] (0,0) ellipse (1 and 1);
                    \draw[very thick, out=-50, in=230] (-1,0) to (1,0);
                    \draw[very thick, densely dashed, out=50, in=130] (-1,0) to (1,0);
            \end{scope}
            \begin{scope}[shift={(-2,3)}]
                    \draw[very thick] (0,0) ellipse (1 and 1);
                    \draw[very thick, out=-50, in=230] (-1,0) to (1,0);
                    \draw[very thick, densely dashed, out=50, in=130] (-1,0) to (1,0);
            \end{scope}
            \begin{scope}[shift={(2,3)}]
                    \draw[very thick] (0,0) ellipse (1 and 1);
                    \draw[very thick, out=-50, in=230] (-1,0) to (1,0);
                    \draw[very thick, densely dashed, out=50, in=130] (-1,0) to (1,0);
            \end{scope}

            \draw[very thick, red, out = 120, in = 240] ({2+4/5*cos(30)},{-3+4/5*sin(30)}) to ({{2+4/5*cos(-30)}},{3+4/5*sin(-30)});
            \draw[very thick, red, out = 240, in = -60] ({2+4/5*cos(150)},{3+4/5*sin(150)}) to ({{-2+4/5*cos(30)}},{3+4/5*sin(30)});
            \draw[very thick, red, out = -60, in = 60] ({-2+4/5*cos(210)},{3+4/5*sin(210)}) to ({{-2+4/5*cos(150)}},{-3+4/5*sin(150)});
            \draw[very thick, red, out = 60, in = 120] ({-2+4/5*cos(-30)},{-3+4/5*sin(-30)}) to ({{2+4/5*cos(210)}},{-3+4/5*sin(210)});

            \draw[very thick, red] ({-2+4/5*cos(60)},{-3+4/5*sin(60)}) to ({4/5*cos(210)},{-1.25+4/5*sin(210)});
            \draw[very thick, red]  ({2+4/5*cos(120)},{-3+4/5*sin(120)}) to ({4/5*cos(-30)},{-1.25+4/5*sin(-30)});
            \draw[very thick, red] ({0},{-1.25+4/5}) to ({0},{1.25-4/5});
            \draw[very thick, red] ({{-2+4/5*cos(-60)}},{3+4/5*sin(-60)}) to ({4/5*cos(150)},{1.25+4/5*sin(150)});
            \draw[very thick, red] ({2+4/5*cos(240)},{3+4/5*sin(240)}) to ({4/5*cos(30)},{1.25+4/5*sin(30)});

            \draw[fill=gray, opacity=.5] ({2+cos(30)},{-3+sin(30)}) to[out = 120, in = 240] ({{2+cos(-30)}},{3+sin(-30)}) to[out = 240, in = 240, looseness=1.7] ({{2+cos(150)}},{3+sin(150)}) to[out = 240, in = -60] ({{-2+cos(30)}},{3+sin(30)}) to[out = -60, in = -60, looseness=1.7] ({{-2+cos(210)}},{3+sin(210)}) to[out = -60, in = 60] ({{-2+cos(150)}},{-3+sin(150)}) to[out= 60, in = 60, looseness=1.7] ({{-2+cos(-30)}},{-3+sin(-30)}) to[out = 60, in = 120] ({{2+cos(210)}},{-3+sin(210)}) to[out= 120, in = 120, looseness=1.7] ({{2+cos(30)}},{-3+sin(30)});

            \begin{scope}[shift={(0,1.25)}]
                    \draw[very thick, fill = white, draw = white] (0,0) ellipse (1 and 1);
                    \draw[very thick, red] ({cos(31)},{sin(31)}) to ({4/5*cos(30)},{4/5*sin(30)});
                    \draw[very thick, red] ({cos(149)},{sin(149)}) to ({4/5*cos(150)},{4/5*sin(150)}); 
                    \draw[very thick, red] (({cos(-90)},{sin(-90)}) to ({4/5*cos(-90)},{4/5*sin(-90)});
                    \draw[very thick] (0,0) ellipse (1 and 1);
                    \draw[very thick, out=-50, in=230] (-1,0) to (1,0);
                    \draw[very thick, densely dashed, out=50, in=130] (-1,0) to (1,0);
            \end{scope}
            \begin{scope}[shift={(0,-1.25)}]
                    \draw[very thick, fill = white, draw = white] (0,0) ellipse (1 and 1);
                    \draw[very thick, red] ({cos(-31)},{sin(-31)}) to ({4/5*cos(-30)},{4/5*sin(-30)});
                    \draw[very thick, red] ({cos(211)},{sin(211)}) to ({4/5*cos(210)},{4/5*sin(210)}); 
                    \draw[very thick, red] (({cos(90)},{sin(90)}) to ({4/5*cos(90)},{4/5*sin(90)});
                    \draw[very thick] (0,0) ellipse (1 and 1);
                    \draw[very thick, out=-50, in=230] (-1,0) to (1,0);
                    \draw[very thick, densely dashed, out=50, in=130] (-1,0) to (1,0);
            \end{scope}

            \node[above] at ({0},{1.25-4/5}) {$a$};
            \node[below] at ({0},{-1.25+4/5}) {$a$};

            \node at ({4/5*cos(150)+1/10},{1.25+4/5*sin(150)-1/10}) {$1$};
            \node at ({{-2+4/5*cos(-60)-1/10}},{3+4/5*sin(-60)+1/10}) {$1$};

            \node at ({4/5*cos(30)-1/10},{1.25+4/5*sin(30)-1/10}) {$2$};
            \node at ({2+4/5*cos(240)+1/10},{3+4/5*sin(240)+1/10}) {$2$};

            \node at ({4/5*cos(210)+1/10},{-1.25+4/5*sin(210)+1/10}) {$5$};
            \node at ({-2+4/5*cos(60)-1/10},{-3+4/5*sin(60)-1/10}) {$5$};

            \node at ({4/5*cos(-30)-1/10},{-1.25+4/5*sin(-30)+1/10}) {$4$};
            \node at ({2+4/5*cos(120)+1/10},{-3+4/5*sin(120)-1/10}) {$4$};

            \node[above] at ({{2+4/5*cos(-30)}},{3+4/5*sin(-30)}) {$3$};
            \node[below] at ({2+4/5*cos(30)},{-3+4/5*sin(30)}) {$3$};

            \node[right] at ({2+4/5*cos(150)},{3+4/5*sin(150)}) {$b$};
            \node[left] at ({{-2+4/5*cos(30)}},{3+4/5*sin(30)}) {$b$};

            \node[above] at ({-2+4/5*cos(210)},{3+4/5*sin(210)}) {$6$};
            \node[below] at ({{-2+4/5*cos(150)}},{-3+4/5*sin(150)}) {$6$};

            \node[left] at ({-2+4/5*cos(-30)},{-3+4/5*sin(-30)}) {$c$};
            \node[right] at ({{2+4/5*cos(210)}},{-3+4/5*sin(210)}) {$c$};

            \draw[very thick, blue, out = 210, in = -30, looseness=.25] (2.15,0) to (-2.15,0);

        \draw[very thick,->] (3,0) to (4.25,0);

        \begin{scope}[scale=.75,shift={(9.5,2.5)}]
                \draw[very thick, densely dashed, out = 50, in = 130, blue] (-2,0) to (2,0);

                \draw[fill=gray, opacity=.5, draw=gray] (0,0) ellipse (2 and 2);
                \draw[fill=white, draw=white] (-1,1/2) ellipse (1/2 and 1/2);
                \draw[fill=white, draw=white] (1,1/2) ellipse (1/2 and 1/2);
                \draw[fill=white, draw=white] (0,-5/4) ellipse (1/2 and 1/2);

                \draw[very thick,fill=white] (-1, 1/2) ellipse (.5 and .5);
                \draw[very thick, out=-50, in=230] (-3/2,1/2) to (-1/2,1/2);
                \draw[very thick, densely dashed, out = 50, in = 130] (-3/2,1/2) to (-1/2,1/2);

                \draw[very thick, fill=white] (1, 1/2) ellipse (.5 and .5);
                \draw[very thick, out=-50, in=230] (1/2,1/2) to (3/2,1/2);
                \draw[very thick, densely dashed, out = 50, in = 130] (1/2,1/2) to (3/2,1/2);

                \draw[very thick, fill=white] (0,-5/4) ellipse (.5 and .5);
                \draw[very thick, out=-50, in=230] (-1/2,-5/4) to (1/2,-5/4);
                \draw[very thick,densely dashed, out = 50, in = 130] (-1/2,-5/4) to (1/2,-5/4);

                \draw[very thick, red] ({2*cos(150)},{2*sin(150)}) to ({-1+3/8*cos(150)},{1/2+3/8*sin(150)});
                \draw[very thick, red] ({2*cos(30)},{2*sin(30)}) to ({1+3/8*cos(30)},{1/2+3/8*sin(30)});
                \draw[very thick, red] (-1/2,1/2) to (1/2,1/2);
                \draw[very thick, red] ({1+3/8*cos(240)},{1/2+3/8*sin(240)}) to ({0+3/8*cos(50)},{-5/4+3/8*sin(50)});
                \draw[very thick, red] ({-1+3/8*cos(-50)},{1/2+3/8*sin(-50)}) to ({0+3/8*cos(120)},{-5/4+3/8*sin(120)});
                \draw[very thick, red] (0,-5/4-3/8) to (0,-2);

                \draw[very thick, blue] (0,0) ellipse (2 and 2);
                \draw[very thick, out = -50, in = 230, blue] (-2,0) to (2,0); 

                \node[left] at ({2*cos(150)},{2*sin(150)}) {$6$};
                \node[right] at ({2*cos(30)},{2*sin(30)}) {$3$};
                \node[below] at (0,-2) {$a$};
                \node[below] at (0,1/2) {$b$};
                \node[left] at (-1/2,-1/2) {$1$};
                \node[right] at (1/2,-1/2) {$2$};
        \end{scope}

        \begin{scope}[scale=.75,shift={(9.5,-2.5)}]
                \draw[very thick, densely dashed, out = 50, in = 130, blue] (-2,0) to (2,0);

                \draw[fill=gray, opacity=.5, draw=gray] (0,0) ellipse (2 and 2);
                \draw[fill=white, draw=white] (-1,1/2) ellipse (1/2 and 1/2);
                \draw[fill=white, draw=white] (1,1/2) ellipse (1/2 and 1/2);
                \draw[fill=white, draw=white] (0,-5/4) ellipse (1/2 and 1/2);

                \draw[very thick,fill=white] (-1, 1/2) ellipse (.5 and .5);
                \draw[very thick, out=-50, in=230] (-3/2,1/2) to (-1/2,1/2);
                \draw[very thick, densely dashed, out = 50, in = 130] (-3/2,1/2) to (-1/2,1/2);

                \draw[very thick, fill=white] (1, 1/2) ellipse (.5 and .5);
                \draw[very thick, out=-50, in=230] (1/2,1/2) to (3/2,1/2);
                \draw[very thick, densely dashed, out = 50, in = 130] (1/2,1/2) to (3/2,1/2);

                \draw[very thick, fill=white] (0,-5/4) ellipse (.5 and .5);
                \draw[very thick, out=-50, in=230] (-1/2,-5/4) to (1/2,-5/4);
                \draw[very thick,densely dashed, out = 50, in = 130] (-1/2,-5/4) to (1/2,-5/4);

                \draw[very thick, red] ({2*cos(150)},{2*sin(150)}) to ({-1+3/8*cos(150)},{1/2+3/8*sin(150)});
                \draw[very thick, red] ({2*cos(30)},{2*sin(30)}) to ({1+3/8*cos(30)},{1/2+3/8*sin(30)});
                \draw[very thick, red] (-1/2,1/2) to (1/2,1/2);
                \draw[very thick, red] ({1+3/8*cos(240)},{1/2+3/8*sin(240)}) to ({0+3/8*cos(50)},{-5/4+3/8*sin(50)});
                \draw[very thick, red] ({-1+3/8*cos(-50)},{1/2+3/8*sin(-50)}) to ({0+3/8*cos(120)},{-5/4+3/8*sin(120)});
                \draw[very thick, red] (0,-5/4-3/8) to (0,-2);

                \draw[very thick, blue] (0,0) ellipse (2 and 2);
                \draw[very thick, out = -50, in = 230, blue] (-2,0) to (2,0); 

                \node[left] at ({2*cos(150)},{2*sin(150)}) {$3$};
                \node[right] at ({2*cos(30)},{2*sin(30)}) {$6$};
                \node[below] at (0,-2) {$a$};
                \node[below] at (0,1/2) {$c$};
                \node[left] at (-1/2,-1/2) {$4$};
                \node[right] at (1/2,-1/2) {$5$};
        \end{scope}

        \node at (-4.25,0) {$M=$};
        \node at (4.75,1.875) {$M_1=$};
        \node at (4.75,-1.875) {$M_2=$};
    \end{tikzpicture}
    \caption{A Heegaard splitting of the wormhole with six three-punctured sphere boundaries. Each constituent compression body is equivalent to the four-boundary wormhole studied in section \ref{subsec:four-boundary wormhole}.}\label{fig:six boundary wormhole heegaard splitting}
\end{figure}
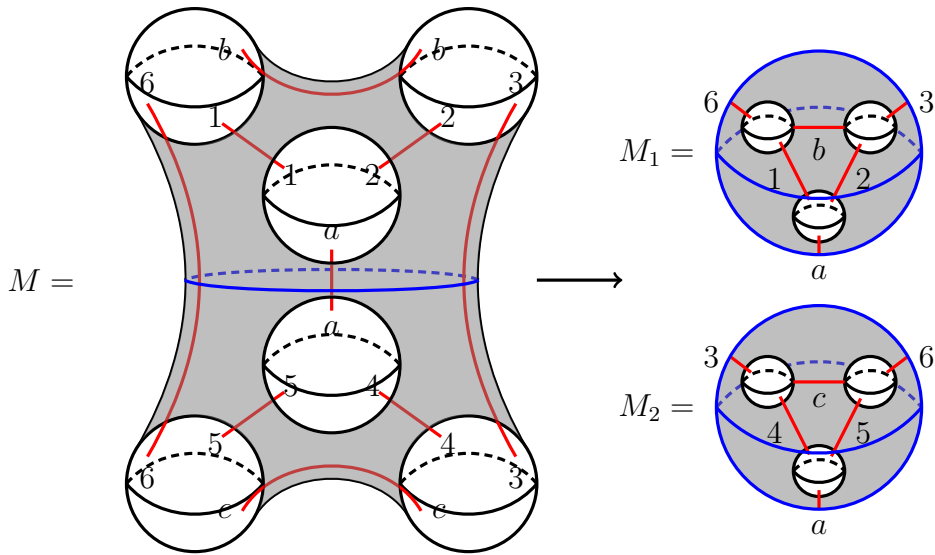

Consider the following wormhole with six three-punctured sphere boundaries 
\begin{equation}\label{eq:six booundary wormhole easy} 
    M \, = \, \begin{tikzpicture}[baseline={([yshift=-.5ex]current bounding box.center)},scale=.8]
            \draw[thick, out = 120, in = 240] ({2+cos(30)},{-3+sin(30)}) to ({{2+cos(-30)}},{3+sin(-30)});
            \draw[thick, out = 240, in = -60] ({2+cos(150)},{3+sin(150)}) to ({{-2+cos(30)}},{3+sin(30)});
            \draw[thick, out = -60, in = 60] ({-2+cos(210)},{3+sin(210)}) to ({{-2+cos(150)}},{-3+sin(150)});
            \draw[thick, out = 60, in = 120] ({-2+cos(-30)},{-3+sin(-30)}) to ({{2+cos(210)}},{-3+sin(210)});

            \begin{scope}[shift={(-2,-3)}]
                    \draw[very thick] (0,0) ellipse (1 and 1);
                    \draw[very thick, out=-50, in=230] (-1,0) to (1,0);
                    \draw[very thick, densely dashed, out=50, in=130] (-1,0) to (1,0);
            \end{scope}
            \begin{scope}[shift={(2,-3)}]
                    \draw[very thick] (0,0) ellipse (1 and 1);
                    \draw[very thick, out=-50, in=230] (-1,0) to (1,0);
                    \draw[very thick, densely dashed, out=50, in=130] (-1,0) to (1,0);
            \end{scope}
            \begin{scope}[shift={(-2,3)}]
                    \draw[very thick] (0,0) ellipse (1 and 1);
                    \draw[very thick, out=-50, in=230] (-1,0) to (1,0);
                    \draw[very thick, densely dashed, out=50, in=130] (-1,0) to (1,0);
            \end{scope}
            \begin{scope}[shift={(2,3)}]
                    \draw[very thick] (0,0) ellipse (1 and 1);
                    \draw[very thick, out=-50, in=230] (-1,0) to (1,0);
                    \draw[very thick, densely dashed, out=50, in=130] (-1,0) to (1,0);
            \end{scope}

            \draw[very thick, red, out = 120, in = 240] ({2+4/5*cos(30)},{-3+4/5*sin(30)}) to ({{2+4/5*cos(-30)}},{3+4/5*sin(-30)});
            \draw[very thick, red, out = 240, in = -60] ({2+4/5*cos(150)},{3+4/5*sin(150)}) to ({{-2+4/5*cos(30)}},{3+4/5*sin(30)});
            \draw[very thick, red, out = -60, in = 60] ({-2+4/5*cos(210)},{3+4/5*sin(210)}) to ({{-2+4/5*cos(150)}},{-3+4/5*sin(150)});
            \draw[very thick, red, out = 60, in = 120] ({-2+4/5*cos(-30)},{-3+4/5*sin(-30)}) to ({{2+4/5*cos(210)}},{-3+4/5*sin(210)});

            \draw[very thick, red] ({-2+4/5*cos(60)},{-3+4/5*sin(60)}) to ({4/5*cos(210)},{-1.25+4/5*sin(210)});
            \draw[very thick, red]  ({2+4/5*cos(120)},{-3+4/5*sin(120)}) to ({4/5*cos(-30)},{-1.25+4/5*sin(-30)});
            \draw[very thick, red] ({0},{-1.25+4/5}) to ({0},{1.25-4/5});
            \draw[very thick, red] ({{-2+4/5*cos(-60)}},{3+4/5*sin(-60)}) to ({4/5*cos(150)},{1.25+4/5*sin(150)});
            \draw[very thick, red] ({2+4/5*cos(240)},{3+4/5*sin(240)}) to ({4/5*cos(30)},{1.25+4/5*sin(30)});

            \draw[fill=gray, opacity=.5] ({2+cos(30)},{-3+sin(30)}) to[out = 120, in = 240] ({{2+cos(-30)}},{3+sin(-30)}) to[out = 240, in = 240, looseness=1.7] ({{2+cos(150)}},{3+sin(150)}) to[out = 240, in = -60] ({{-2+cos(30)}},{3+sin(30)}) to[out = -60, in = -60, looseness=1.7] ({{-2+cos(210)}},{3+sin(210)}) to[out = -60, in = 60] ({{-2+cos(150)}},{-3+sin(150)}) to[out= 60, in = 60, looseness=1.7] ({{-2+cos(-30)}},{-3+sin(-30)}) to[out = 60, in = 120] ({{2+cos(210)}},{-3+sin(210)}) to[out= 120, in = 120, looseness=1.7] ({{2+cos(30)}},{-3+sin(30)});

            \begin{scope}[shift={(0,1.25)}]
                    \draw[very thick, fill = white, draw = white] (0,0) ellipse (1 and 1);
                    \draw[very thick, red] ({cos(31)},{sin(31)}) to ({4/5*cos(30)},{4/5*sin(30)});
                    \draw[very thick, red] ({cos(149)},{sin(149)}) to ({4/5*cos(150)},{4/5*sin(150)}); 
                    \draw[very thick, red] (({cos(-90)},{sin(-90)}) to ({4/5*cos(-90)},{4/5*sin(-90)});
                    \draw[very thick] (0,0) ellipse (1 and 1);
                    \draw[very thick, out=-50, in=230] (-1,0) to (1,0);
                    \draw[very thick, densely dashed, out=50, in=130] (-1,0) to (1,0);
            \end{scope}
            \begin{scope}[shift={(0,-1.25)}]
                    \draw[very thick, fill = white, draw = white] (0,0) ellipse (1 and 1);
                    \draw[very thick, red] ({cos(-31)},{sin(-31)}) to ({4/5*cos(-30)},{4/5*sin(-30)});
                    \draw[very thick, red] ({cos(211)},{sin(211)}) to ({4/5*cos(210)},{4/5*sin(210)}); 
                    \draw[very thick, red] (({cos(90)},{sin(90)}) to ({4/5*cos(90)},{4/5*sin(90)});
                    \draw[very thick] (0,0) ellipse (1 and 1);
                    \draw[very thick, out=-50, in=230] (-1,0) to (1,0);
                    \draw[very thick, densely dashed, out=50, in=130] (-1,0) to (1,0);
            \end{scope}

            \node[above] at ({0},{1.25-4/5}) {$a$};
            \node[below] at ({0},{-1.25+4/5}) {$a$};

            \node at ({4/5*cos(150)+1/10},{1.25+4/5*sin(150)-1/10}) {$1$};
            \node at ({{-2+4/5*cos(-60)-1/10}},{3+4/5*sin(-60)+1/10}) {$1$};

            \node at ({4/5*cos(30)-1/10},{1.25+4/5*sin(30)-1/10}) {$2$};
            \node at ({2+4/5*cos(240)+1/10},{3+4/5*sin(240)+1/10}) {$2$};

            \node at ({4/5*cos(210)+1/10},{-1.25+4/5*sin(210)+1/10}) {$5$};
            \node at ({-2+4/5*cos(60)-1/10},{-3+4/5*sin(60)-1/10}) {$5$};

            \node at ({4/5*cos(-30)-1/10},{-1.25+4/5*sin(-30)+1/10}) {$4$};
            \node at ({2+4/5*cos(120)+1/10},{-3+4/5*sin(120)-1/10}) {$4$};

            \node[above] at ({{2+4/5*cos(-30)}},{3+4/5*sin(-30)}) {$3$};
            \node[below] at ({2+4/5*cos(30)},{-3+4/5*sin(30)}) {$3$};

            \node[right] at ({2+4/5*cos(150)},{3+4/5*sin(150)}) {$b$};
            \node[left] at ({{-2+4/5*cos(30)}},{3+4/5*sin(30)}) {$b$};

            \node[above] at ({-2+4/5*cos(210)},{3+4/5*sin(210)}) {$6$};
            \node[below] at ({{-2+4/5*cos(150)}},{-3+4/5*sin(150)}) {$6$};

            \node[left] at ({-2+4/5*cos(-30)},{-3+4/5*sin(-30)}) {$c$};
            \node[right] at ({{2+4/5*cos(210)}},{-3+4/5*sin(210)}) {$c$};

    \end{tikzpicture}\ .
\end{equation}
As indicated by the diagram, it contributes to the following sixth moment of CFT structure constants
\begin{equation}
    |Z_{\rm Vir}(M)|^2 \leftrightarrow \overline{c_{12a}c_{2b3}c_{3c4}c_{45a}c_{5c6}c_{6b1}}.
\end{equation}
There are several Heegaard splittings that one could employ to compute the Virasoro TQFT partition function on this wormhole, but the simplest is indicated in figure \ref{fig:six boundary wormhole heegaard splitting}: we cut the wormhole through the bulk along a three-punctured sphere. This divides $M$ into two generalized compression bodies $M_1$ and $M_2$, each of which is itself a four-boundary wormhole of the type described in the previous subsection.
The partition function of Virasoro TQFT on the generalized compression bodies was computed in (\ref{eq:four boundary wormhole partition function}) as
\begin{subequations}
    \begin{align}
        \ket{Z_{\rm Vir}(M_1)} &= 
\sqrt{\mathsf{C}_{12a}\mathsf{C}_{2b3}\mathsf{C}_{6b1}\mathsf{C}_{a36}}
        \begin{Bmatrix}
            P_1 & P_2 & P_a \\
            P_3 & P_6 & P_b
        \end{Bmatrix}\\
        \ket{Z_{\rm Vir}(M_2)} &= \sqrt{\mathsf{C}_{3c4}\mathsf{C}_{45a}\mathsf{C}_{5c6}\mathsf{C}_{a36}}
        \begin{Bmatrix}
            P_3 & P_4 & P_c \\
            P_5 & P_6 & P_a
        \end{Bmatrix}
    \end{align}
\end{subequations}
Then the Virasoro TQFT partition function on the six-boundary wormhole is given by the following inner product between these states in the Hilbert space of the shared three-punctured sphere boundary
\begin{align}
        Z_{\rm Vir}(M) &= \braket{Z_{\rm Vir}(M_1)|Z_{\rm Vir}(M_2)}\\
        &=\sqrt{\mathsf{C}_{12a}\mathsf{C}_{2b3}\mathsf{C}_{3c4}\mathsf{C}_{45a}\mathsf{C}_{5c6}\mathsf{C}_{6b1}}
        \begin{Bmatrix}
            P_1 & P_2 & P_a \\
            P_3 & P_6 & P_b
        \end{Bmatrix}
        \begin{Bmatrix}
            P_3 & P_4 & P_c \\
            P_5 & P_6 & P_a
        \end{Bmatrix}\ .\label{eq:six boundary wormhole easy partition function}
\end{align}
Notice that here the only effect of the inner product is to divide by the extra factor of $C_0(P_3,P_a,P_6)$. 

This particular Heegaard splitting of the six-boundary wormhole is far from unique: for example, we could have cut it through a five-punctured sphere, or along three four-punctured spheres. In all cases, the corresponding splittings yield the same result (\ref{eq:six boundary wormhole easy partition function}) for the TQFT partition function.

This wormhole partition function implies that the corresponding sixth moment of CFT structure constants is given by
\begin{equation}\label{eq:sixth moment easy}
    \overline{c_{12a}c_{2b3}c_{3c4}c_{45a}c_{5c6}c_{6b1}} \supset \sqrt{\overline{c^2_{12a}}\,\overline{c^2_{2b3}}\,\overline{c^2_{3c4}}\,\overline{c^2_{45a}}\,\overline{c^2_{5c6}}\, \overline{c^2_{6b1}}}\left|
    \begin{Bmatrix}
            P_1 & P_2 & P_a \\
            P_3 & P_6 & P_b
        \end{Bmatrix}
        \begin{Bmatrix}
            P_3 & P_4 & P_c \\
            P_5 & P_6 & P_a
        \end{Bmatrix}
    \right|^2.
\end{equation}

\paragraph{Consistency with boundary ensemble description.}
Much like the fourth moment of the structure constants inferred from the four-boundary wormhole of section \ref{subsec:four-boundary wormhole}, the sixth moment (\ref{eq:sixth moment easy}) is needed for consistency of the description of the boundary theory in terms of an ensemble of CFT data. There are a variety of ways to see this. Roughly, for each Heegaard splitting of the wormhole, there is a corresponding product of CFT observables for which consistency of the ensemble description requires that the appropriate moment of CFT data is correctly computed by the wormhole. 

For concreteness, consider the average of the following product of five-point functions
\begin{equation}\label{eq:average of five point functions}
    \overline{\langle \mathcal{O}_b\mathcal{O}_1\mathcal{O}_a\mathcal{O}_4\mathcal{O}_c\rangle\langle \mathcal{O}_b\mathcal{O}_1\mathcal{O}_a\mathcal{O}_4\mathcal{O}_c\rangle^*}.
\end{equation}
This is associated with splitting the wormhole (\ref{eq:six booundary wormhole easy}) along a five-punctured sphere in the bulk.
The average (\ref{eq:average of five point functions}), which corresponds to the two-boundary sphere five-point function wormhole, is given by the corresponding five-point function in Liouville CFT as in (\ref{eq:averaged product of four pt functions}).
In the Gaussian ensemble this however requires that we expand the two five-point functions in aligned channels when taking the ensemble average. Of course we are free to expand the five-point functions in different channels, in which case we need to invoke the non-Gaussian statistics. The combination of OPE channels that is associated to the particular Heegaard splitting is determined by the combination of sphere three-point boundaries that appear in each compression body of the Heegaard splitting. For example, if we compute the averaged product of sphere five-point functions by expanding in the following channel where there is not a Gaussian contraction
\begin{align}
        &\overline{\langle \mathcal{O}_b\mathcal{O}_1\mathcal{O}_a\mathcal{O}_4\mathcal{O}_c\rangle\langle \mathcal{O}_b\mathcal{O}_1\mathcal{O}_a\mathcal{O}_4\mathcal{O}_c\rangle^*}\nonumber\\
        &\quad= \sum_{\mathcal{O}_{2,3,5,6}}\overline{c_{12a}c_{2b3}c_{3c4}c_{4a5}^*c_{56c}^*c_{61b}^*}
        \bigg|\!
        \begin{tikzpicture}[scale=.75, baseline={([yshift=-.5ex]current bounding box.center)}]
            \draw[very thick, red] (0,0) to (3,0);
            \draw[very thick, red] (1,0) to (1,1) to (1-.707,1+.707);
            \draw[very thick, red] (1,1) to (1+.707,1+.707);
            \draw[very thick, red] (2,0) to (2,1);
            \node[left] at (0,0) {$b$};
            \node[left] at (1-.707,1+.707) {$1$};
            \node[right] at (1+.707,1+.707) {$a$};
            \node[right] at (2,1-1/10) {$4$};
            \node[right] at (3,0) {$c$};
            \node[below] at (3/2,0) {$3$};
            \node[right] at (1,1/2) {$2$};
            \draw[draw=red,fill=red] (1,0) circle (.07);
            \draw[draw=red,fill=red] (2,0) circle (.07);
            \draw[draw=red,fill=red] (1,1) circle (.07);
        \end{tikzpicture}
        \!\!\!
        (m_1)
        \begin{tikzpicture}[scale=.75, baseline={([yshift=-.5ex]current bounding box.center)}]
            \draw[very thick, red] (0,0) to (3,0);
            \draw[very thick, red] (1,0) to (1,1);
            \draw[very thick, red] (2,0) to (2,1) to (2-.707,1+.707);
            \draw[very thick, red] (2,1) to (2+.707, 1+.707);
            \node[left] at (0,0) {$b$};
            \node[left] at (1,1) {$1$};
            \node[left] at (2-.707,1+.707) {$a$};
            \node[right] at (2+.707,1+.707) {$4$};
            \node[right] at (3,0) {$c$};
            \node[below] at (3/2,0) {$6$};
            \node[right] at (2,1/2) {$5$};
            \draw[draw=red,fill=red] (1,0) circle (.07);
            \draw[draw=red,fill=red] (2,0) circle (.07);
            \draw[draw=red,fill=red] (2,1) circle (.07);
        \end{tikzpicture}
        \!\!\!(m_2)
        \bigg|^2\\
        &\quad= \bigg| \int_0^\infty \d P_3\, \d P_6 \, \rho_0(P_3)\rho_0(P_6) C_0(P_1,P_b,P_6)C_0(P_6,P_a,P_3)C_0(P_3,P_c,P_4)\nonumber\\
        &\qquad\times 
        \!\!\begin{tikzpicture}[baseline={([yshift=-.5ex]current bounding box.center)}]
            \draw[very thick, red] (0,0) to (4*3/4,0);
            \draw[very thick, red] (1*3/4,0) to (1*3/4,1*3/4);
            \draw[very thick, red] (2*3/4,0) to (2*3/4,1*3/4);
            \draw[very thick, red] (3*3/4,0) to (3*3/4,1*3/4);
            \node[above] at (0,0) {$b$};
            \node[below] at (3/4+3/8,0) {$6$};
            \node[below] at (2*3/4+3/8,0) {$3$};
            \node[above] at (3/4,3/4) {$1$};
            \node[above] at (2*3/4,3/4) {$a$};
            \node[above] at (3*3/4,3/4) {$4$};
            \node[above] at (4*3/4,0) {$c$};
            \draw[draw=red,fill=red] (3/4,0) circle (.07);
            \draw[draw=red,fill=red] (3/2,0) circle (.07);
            \draw[draw=red,fill=red] (9/4,0) circle (.07);
        \end{tikzpicture}
        \!
        (m_1)
        \begin{tikzpicture}[baseline={([yshift=-.5ex]current bounding box.center)}]
            \draw[very thick, red] (0,0) to (4*3/4,0);
            \draw[very thick, red] (1*3/4,0) to (1*3/4,1*3/4);
            \draw[very thick, red] (2*3/4,0) to (2*3/4,1*3/4);
            \draw[very thick, red] (3*3/4,0) to (3*3/4,1*3/4);
            \node[above] at (0,0) {$b$};
            \node[below] at (3/4+3/8,0) {$6$};
            \node[below] at (2*3/4+3/8,0) {$3$};
            \node[above] at (3/4,3/4) {$1$};
            \node[above] at (2*3/4,3/4) {$a$};
            \node[above] at (3*3/4,3/4) {$4$};
            \node[above] at (4*3/4,0) {$c$};
            \draw[draw=red,fill=red] (3/4,0) circle (.07);
            \draw[draw=red,fill=red] (3/2,0) circle (.07);
            \draw[draw=red,fill=red] (9/4,0) circle (.07);
        \end{tikzpicture}
        \!(m_2)\bigg|^2\\
        &\quad= |Z_{\rm Liouville}(P_b,P_1,P_a,P_4,P_c|m_1,m_2)|^2\, ,
    \end{align}
then making use of the sixth moment (\ref{eq:sixth moment easy}) and the fact that the $6j$ symbols implement crossing transformations on the conformal blocks, we reproduce exactly the result from the Gaussian contraction, the sphere five-point function in Liouville theory. Here $m_i$ collectively denote the moduli of each five-point function.

We could have considered other Heegaard splittings, corresponding to averaged CFT observables that receive contributions from this combination of structure constants in a particular OPE channel. For example, the following averaged product of three four-point functions
\begin{equation}
    \overline{\langle \mathcal{O}_6\mathcal{O}_1\mathcal{O}_2\mathcal{O}_3\rangle \langle \mathcal{O}_1\mathcal{O}_2\mathcal{O}_4\mathcal{O}_5\rangle\langle \mathcal{O}_6\mathcal{O}_5\mathcal{O}_4\mathcal{O}_3\rangle}
\end{equation}
receives contributions from the sixth moment (\ref{eq:sixth moment easy}) in a specific OPE channel that precisely reproduce the result (\ref{eq:three boundary sphere four point partition function 2}) for the averaged product in the Gaussian ensemble.

\subsubsection{A more nontrivial six-boundary example}\label{subsubsec:six boundary wormhole hard}
Here we consider another wormhole with six three-punctured sphere boundaries, but with the defects arranged slightly differently between the boundaries

\begin{equation}\label{eq:six boundary wormhole borromean rings}
    M = 
    \begin{tikzpicture}[baseline={([yshift=-.5ex]current bounding box.center)},scale=.8]
        \draw[very thick] (0,3) ellipse (1 and 1);
        \draw[very thick, out=-50, in=230] (-1,3) to (1,3);
        \draw[very thick, densely dashed, out=50, in=130] (-1,3) to (1,3);

        \draw[very thick] ({3*cos(30)},{3*sin(30)}) ellipse (1 and 1);
        \draw[very thick, out=-50, in=230] ({{3*cos(30)-1}},{3*sin(30)}) to ({{3*cos(30)+1}},{3*sin(30)});
        \draw[very thick, densely dashed, out=50, in=130] ({{3*cos(30)-1}},{3*sin(30)}) to ({{3*cos(30)+1}},{3*sin(30)});

        \draw[very thick] ({3*cos(-30)},{3*sin(-30)}) ellipse (1 and 1);
        \draw[very thick, out=-50, in=230] ({{3*cos(-30)-1}},{3*sin(-30)}) to ({{3*cos(-30)+1}},{3*sin(-30)});
        \draw[very thick, densely dashed, out=50, in=130] ({{3*cos(-30)-1}},{3*sin(-30)}) to ({{3*cos(-30)+1}},{3*sin(-30)});

        \draw[very thick] (0,-3) ellipse (1 and 1);
        \draw[very thick, out=-50, in=230] (-1,-3) to (1,-3);
        \draw[very thick, densely dashed, out=50, in=130] (-1,-3) to (1,-3);

        \draw[very thick] ({3*cos(150)},{3*sin(150)}) ellipse (1 and 1);
        \draw[very thick, out=-50, in=230] ({{3*cos(150)-1}},{3*sin(150)}) to ({{3*cos(150)+1}},{3*sin(150)});
        \draw[very thick, densely dashed, out=50, in=130] ({{3*cos(150)-1}},{3*sin(150)}) to ({{3*cos(150)+1}},{3*sin(150)});

        \draw[very thick] ({3*cos(210)},{3*sin(210)}) ellipse (1 and 1);
        \draw[very thick, out=-50, in=230] ({{3*cos(210)-1}},{3*sin(210)}) to ({{3*cos(210)+1}},{3*sin(210)});
        \draw[very thick, densely dashed, out=50, in=130] ({{3*cos(210)-1}},{3*sin(210)}) to ({{3*cos(210)+1}},{3*sin(210)});

        \draw[very thick, out=-90, in = -30] (0-1,3) to ({{3*cos(150)+cos(60)}},{3*sin(150)+sin(60)});
        \draw[very thick, out=-30, in = 30] ({{3*cos(150)+cos(240)}},{3*sin(150)+sin(240)}) to ({{3*cos(210)+cos(120)}},{3*sin(210)+sin(120)});
        \draw[very thick, out=30, in=90] ({{3*cos(210)+cos(-60)}},{3*sin(210)+sin(-60)}) to ({{0-1}},{-3});
        \draw[very thick, out=90, in = 150] (1,-3) to ({{3*cos(-30)+cos(240)}},{3*sin(-30)+sin(240)});
        \draw[very thick, out=150, in=210] ({{3*cos(-30)+cos(60)}},{3*sin(-30)+sin(60)}) to ({{3*cos(30)+cos(-60)}},{3*sin(30)+sin(-60)});
        \draw[very thick, out=210, in=-90] ({{3*cos(30)+cos(120)}},{3*sin(30)+sin(120)}) to ({0+1},{3});

        \draw[very thick, red, out=-90, in = -30] (0-4/5,3) to ({{3*cos(150)+4/5*cos(60)}},{3*sin(150)+4/5*sin(60)});
        \draw[very thick, red, out=-30, in = 30] ({{3*cos(150)+4/5*cos(240)}},{3*sin(150)+4/5*sin(240)}) to ({{3*cos(210)+4/5*cos(120)}},{3*sin(210)+4/5*sin(120)});
        \draw[very thick, red, out=30, in=90] ({{3*cos(210)+4/5*cos(-60)}},{3*sin(210)+4/5*sin(-60)}) to ({{0-4/5}},{-3});
        \draw[very thick, red, out=90, in = 150] (4/5,-3) to ({{3*cos(-30)+4/5*cos(240)}},{3*sin(-30)+4/5*sin(240)});
        \draw[very thick, red, out=150, in=210] ({{3*cos(-30)+4/5*cos(60)}},{3*sin(-30)+4/5*sin(60)}) to ({{3*cos(30)+4/5*cos(-60)}},{3*sin(30)+4/5*sin(-60)});
        \draw[very thick, red, out=210, in=-90] ({{3*cos(30)+4/5*cos(120)}},{3*sin(30)+4/5*sin(120)}) to ({0+4/5},{3});

        \draw[very thick, red] ({{3*cos(210)+4/5*cos(30)}},{3*sin(210)+4/5*sin(30)}) to ({{3*cos(30)+4/5*cos(240)}},{3*sin(30)+4/5*sin(240)});
        \draw[very thick, red] ({{3*cos(150)+4/5*cos(-30)}},{3*sin(150)+4/5*sin(-30)}) to ({{3*cos(-30)+4/5*cos(120)}},{3*sin(-30)+4/5*sin(120)});
        \draw[fill=white,draw=white] (0,-1/4) circle (1/10);
        \draw[fill=white,draw=white] (0,1/4) circle (1/10);
        \draw[very thick, red] (0,{3-4/5}) to (0,{-3+4/5});

        \node[above] at (0,{3-4/5}) {$a$};
        \node[below] at (0,{-3+4/5}) {$a$};
        \node[right] at (0-4/5,3) {$1$};
        \node[left] at ({{3*cos(150)+4/5*cos(60)}},{3*sin(150)+4/5*sin(60)}) {$1$};
        \node[left] at ({{3*cos(150)+4/5*cos(240)}},{3*sin(150)+4/5*sin(240)+1/10}) {$6$};
        \node[left] at ({{3*cos(210)+4/5*cos(120)}},{3*sin(210)+4/5*sin(120)-1/10}) {$6$};
        \node[left] at ({{3*cos(210)+4/5*cos(-60)}},{3*sin(210)+4/5*sin(-60)}) {$5$};
        \node[right] at ({{0-4/5}},{-3}) {$5$};
        \node[left] at (4/5,-3) {$4$};
        \node[right] at ({{3*cos(-30)+4/5*cos(240)}},{3*sin(-30)+4/5*sin(240)}) {$4$};
        \node[right] at ({{3*cos(-30)+4/5*cos(60)}},{3*sin(-30)+4/5*sin(60)-1/10}) {$3$};
        \node[right] at ({{3*cos(30)+4/5*cos(-60)}},{3*sin(30)+4/5*sin(-60)+1/10}) {$3$};
        \node[right] at ({{3*cos(30)+4/5*cos(120)}},{3*sin(30)+4/5*sin(120)}) {$2$};
        \node[left] at ({0+4/5},{3}) {$2$};
        \node[above] at ({{3*cos(150)+4/5*cos(-30)-3/20}},{3*sin(150)+4/5*sin(-30)}) {$c$};
        \node[right] at ({{3*cos(30)+4/5*cos(240)}},{3*sin(30)+4/5*sin(240)+1/10}) {$b$};
        \node[below] at ({{3*cos(210)+4/5*cos(30)-3/20-1/10}},{3*sin(210)+4/5*sin(30)}) {$b$};
        \node[right] at ({{3*cos(-30)+4/5*cos(120)}},{3*sin(-30)+4/5*sin(120)-1/10}) {$c$};

        \draw[fill=white,draw=white] (0,1/4-.05) circle (1/10);
        \draw[very thick, red] ({.12*cos(158.5)},{1/4-.027+.12*sin(158.5)}) to ({.12*cos(-28.5)},{1/4-.027+.12*sin(-28.5)});
        \draw[fill=white,draw=white] (.433,0) circle (1/10);
        \draw[very thick, red] ({.433+.12*cos(207)},{-.008+.12*sin(207)}) to ({.433+.12*cos(23)},{-.008+.12*sin(23)});

        \draw[fill=gray,opacity=.5] (0-1,3) to[out=-90,in=-30] ({{3*cos(150)+cos(60)}},{3*sin(150)+sin(60)}) to[out=-30,in=-30,looseness=1.7] ({{3*cos(150)+cos(240)}},{3*sin(150)+sin(240)}) to[out=-30, in = 30] ({{3*cos(210)+cos(120)}},{3*sin(210)+sin(120)}) to[out=30,in=30,looseness=1.7] ({{3*cos(210)+cos(-60)}},{3*sin(210)+sin(-60)}) to[out=30, in = 90] ({{0-1}},{-3}) to[out=90,in=90,looseness=1.7] (1,-3) to[out=90, in = 150] ({{3*cos(-30)+cos(240)}},{3*sin(-30)+sin(240)}) to[out=150, in=150,looseness=1.7] ({{3*cos(-30)+cos(60)}},{3*sin(-30)+sin(60)}) to[out=150, in = 210] ({{3*cos(30)+cos(-60)}},{3*sin(30)+sin(-60)}) to[out=210, in= 210,looseness=1.7] ({{3*cos(30)+cos(120)}},{3*sin(30)+sin(120)}) to[out=210, in = -90] ({0+1},{3}) to[out=-90,in=-90,looseness=1.7] (0-1,3);
    \end{tikzpicture}
\end{equation}
This contributes to a different sixth moment of the structure constants
\begin{equation}
    |Z_{\text{Vir}}(M)|^2 \leftrightarrow \overline{c_{1a2}c_{2b3}c_{3c4}c_{4a5}c_{5b6}c_{6c1}}.
\end{equation}

We could of course compute the TQFT partition function on this wormhole by a straightforward Heegaard splitting, for example along three four-punctured spheres. In this case it turns out to be most convenient to replace the three-punctured sphere boundaries with trivalent junctions as in (\ref{eq:definition juncture}) and hence regard the wormhole as a network of Wilson lines embedded in $\mathrm{S}^3$:
\begin{align}
    Z_{\text{Vir}}(M) &= 
\mathsf{C}_{12a}\mathsf{C}_{23b}\mathsf{C}_{34c}\mathsf{C}_{45a}\mathsf{C}_{56b}\mathsf{C}_{61c}\,
    Z_{\text{Vir}}\Bigg(
    \vcenter{\hbox{
    \begin{tikzpicture}[scale=.6, baseline={([yshift=-.5ex]current bounding box.center)}]
        \path[use as bounding box] (0,0) ellipse (2.4 and 2.4);
        \draw[very thick, red] ({2*cos(19)},{2*sin(19)}) to ({2*cos(210)},{2*sin(210)});
        \draw[fill=white,draw=white] (.433,0) circle (1/10);
        \draw[very thick, red] ({2*cos(150)},{2*sin(150)}) to ({2*cos(-20)},{2*sin(-20)}); 
        \draw[fill=white,draw=white] (0,-1/4) circle (1/10);
        \draw[very thick, red] (0,{2}) to (0,{-2});
        \node[above] at (0,{2}) {$a$};
        \node[below] at (0,{-2}) {$a$};
        \node at ({{2*cos(150)-1/5}},{2*sin(150)+1/5}) {$c$};
        \node at ({{2*cos(20)+1/5}},{2*sin(20)+1/5}) {$b$};
        \node at ({{2*cos(210)-1/5}},{2*sin(210)-1/5}) {$b$};
        \node at ({{2*cos(-20)+1/5}},{2*sin(-20)-1/5}) {$c$};
        \node at ({2*cos(120)-1/5},{2*sin(120)+1/5}) {$1$};
        \node[left] at (-2,0) {$6$};
        \node at ({2*cos(240)-1/5},{2*sin(240)-1/5}) {$5$};
        \node at ({2*cos(-60)+1/5},{2*sin(-60)-1/5}) {$4$};
        \node[right] at (2,0) {$3$};
        \draw[fill=white,draw=white] (0,1/4-.05) circle (1/10);
        \draw[very thick, red] ({.11*cos(158.5)},{1/4-.05+.11*sin(158.5)}) to ({.11*cos(-28.5)},{1/4-.05+.110*sin(-28.5)});
        \draw[fill=white,draw=white] (.433,0) circle (1/10);
        \draw[very thick, red] ({.433+.12*cos(207)},{-.008+.12*sin(207)}) to ({.433+.12*cos(22.2)},{-.008+.12*sin(22.2)});
        \draw[very thick, red] (0,0) circle (2);
        \draw[very thick, red, fill=red] (0,2) circle (.07);
        \draw[very thick, red, fill=red] ({2*cos(20)},{2*sin(20)}) circle (.07);
        \draw[very thick, red, fill=red] ({2*cos(-20)},{2*sin(-20)}) circle (.07);
        \draw[very thick, red, fill=red] (0,-2) circle (.07);
        \draw[very thick, red, fill=red] ({2*cos(210)},{2*sin(210)}) circle (.07);
        \draw[very thick, red, fill=red] ({2*cos(150)},{2*sin(150)}) circle (.07);
        \node at ({2*cos(60)+1/5},{2*sin(60)+1/5}) {$2$};
    \end{tikzpicture}
    }}
    \Bigg) \label{eq:6 bdy wormhole as Wilson line network}\\
    &\equiv \mathsf{C}_{12a}\mathsf{C}_{23b}\mathsf{C}_{34c}\mathsf{C}_{45a}\mathsf{C}_{56b}\mathsf{C}_{61c} Z_{\text{Vir}}(M')\  . 
\end{align}
Here $M'$ is the network of Wilson lines depicted on the right-hand side of (\ref{eq:6 bdy wormhole as Wilson line network}) embedded in $S^3$. 
Braiding the Wilson lines and applying a fusion transformation, the TQFT partition function may then be simplified as follows\footnote{Here
\begin{equation}
    \mathbb{B}_{P_i}^{P_jP_k} = e^{\pi i(P_i^2-P_j^2-P_k^2-\frac{Q^2}{4})}
\end{equation}
is the braiding phase.}
\begin{align}
    Z_{\text{Vir}}(M') &=
    (\mathbb{B}_{P_4}^{P_a P_5}\mathbb{B}_{P_6}^{P_cP_1}\mathbb{B}_{P_4}^{P_3P_c})^{-1}
    \,
    Z_{\text{Vir}}\Bigg(
    \begin{tikzpicture}[scale=.55, baseline={([yshift=-.5ex]current bounding box.center)}]
        \path[use as bounding box] (.1,0) ellipse (2.9 and 2.8);
        \draw[very thick, red] (0,0) circle (2);
        \draw[very thick, red] ({2*cos(20)},{2*sin(20)}) to ({2*cos(210)},{2*sin(210)});
        \draw[fill=white,draw=white] ({2*cos(170)},{2*sin(170)}) circle (1/10);
        \draw[fill=white,draw=white] ({2*cos(60)},{2*sin(60)}) circle (1/10);
        \node[left] at ({2*cos(120)-1/10},{2*sin(120)}) {$1$};
        \node[left] at (-2,0) {$6$};
        \node[left] at ({2*cos(240)},{2*sin(240)-1/10}) {$5$};
        \node[right] at ({2*cos(-60)},{2*sin(-60)-1/10}) {$4$};
        \node[right] at (2,0) {$3$};
        \node[above] at (0,-1/5) {$b$};
        \node[below] at (0,-2-1/10) {$a$};
        \node[right] at (5/2+1/10,0) {$c$};
        \draw[very thick, red, looseness=1.75] (0,-2) to[out=240, in = 170] ({{2*cos(170)}},{2*sin(170)})  to [out=-10, in = -90] (0,2);
        \draw[very thick, red, looseness=1.75] ({2*cos(150)},{2*sin(150)}) to[out = 150, in=60] ({{2*cos(60)}},{2*sin(60)}) to[out=240, in = 210] ({{2*cos(45)}},{2*sin(45)}) to [out=30, in = -30] ({{2*cos(-20)}},{2*sin(-20)});
        \draw[fill=white,draw=white] ({2*cos(45)},{2*sin(45)}) circle (1/10);
        \draw[very thick, red] ({2*cos(45)+.11*cos(136.5)},{2*sin(45)+.11*sin(136.5)}) to ({2*cos(45)+.11*cos(-46.5)},{2*sin(45)+.11*sin(-46.5)});
        \draw[very thick, red, fill=red] (0,2) circle (.07);
        \draw[very thick, red, fill=red] ({2*cos(20)},{2*sin(20)}) circle (.07);
        \draw[very thick, red, fill=red] ({2*cos(-20)},{2*sin(-20)}) circle (.07);
        \draw[very thick, red, fill=red] (0,-2) circle (.07);
        \draw[very thick, red, fill=red] ({2*cos(210)},{2*sin(210)}) circle (.07);
        \draw[very thick, red, fill=red] ({2*cos(150)},{2*sin(150)}) circle (.07);
        \node[right] at ({2*cos(60)+1/5},{2*sin(60)}) {$2$};
    \end{tikzpicture}
    \Bigg)\\
    &= (\mathbb{B}_{P_4}^{P_a P_5}\mathbb{B}_{P_6}^{P_cP_1}\mathbb{B}_{P_4}^{P_3P_c})^{-1} \int \d P_d \, \mathbb{F}_{P_3 P_d}
    \begin{bmatrix}
        P_2 & P_c \\ P_b & P_4
    \end{bmatrix}
    Z_{\text{Vir}}\Bigg(
    \begin{tikzpicture}[scale=.55, baseline={([yshift=-.5ex]current bounding box.center)}]
        \path[use as bounding box] (-.1,0) ellipse (2.75 and 2.7);
        \draw[very thick, red] (0,0) circle (2);
        \draw[very thick, red] ({2*cos(-30)},{2*sin(-30)}) to ({2*cos(210)},{2*sin(210)});
        \draw[fill=white,draw=white] ({2*cos(170)},{2*sin(170)}) circle (1/10);
        \draw[fill=white,draw=white] ({2*cos(60)},{2*sin(60)}) circle (1/10);
        \node[left] at ({2*cos(120)},{2*sin(120)+1/10}) {$1$};
        \node[left] at (-2,0) {$6$};
        \node[left] at ({2*cos(240)},{2*sin(240)-1/10}) {$5$};
        \node at ({2*cos(-60)+1/5},{2*sin(-60)-1/5}) {$4$};
        \node[right] at (2,0) {$d$};
        \node[above] at (0,-1) {$b$};
        \node[below] at (0,-2-1/10) {$a$};
        \node[right] at (2+1/10,1) {$c$};
        \draw[very thick, red, looseness=1.75] (0,-2) to[out=240, in = 170] ({{2*cos(170)}},{2*sin(170)})  to [out=-10, in = -90] (0,2);
        \draw[very thick, red, looseness=1.75] ({2*cos(150)},{2*sin(150)}) to[out = 150, in=60] ({{2*cos(60)}},{2*sin(60)}) to[out=240, in = 210] ({{2*cos(45)}},{2*sin(45)}) to [out=30, in = 20] ({{2*cos(20)}},{2*sin(20)});
        \draw[fill=white,draw=white] ({2*cos(45)},{2*sin(45)}) circle (1/10);
        \draw[very thick, red] ({2*cos(45)+.11*cos(136.5)},{2*sin(45)+.11*sin(136.5)}) to ({2*cos(45)+.11*cos(-46.5)},{2*sin(45)+.11*sin(-46.5)});
        \draw[very thick, red, fill=red] (0,2) circle (.07);
        \draw[very thick, red, fill=red] ({2*cos(20)},{2*sin(20)}) circle (.07);
        \draw[very thick, red, fill=red] ({2*cos(-30)},{2*sin(-30)}) circle (.07);
        \draw[very thick, red, fill=red] (0,-2) circle (.07);
        \draw[very thick, red, fill=red] ({2*cos(210)},{2*sin(210)}) circle (.07);
        \draw[very thick, red, fill=red] ({2*cos(150)},{2*sin(150)}) circle (.07);
        \node[right] at ({2*cos(60)+1/5},{2*sin(60)}) {$2$};
    \end{tikzpicture}
    \Bigg)\ .
\end{align}
We then recognize the following Wilson line identity (see \cite[eq.~(3.44)]{Collier:2023fwi}) 
\begin{equation}
    \begin{tikzpicture}[baseline={([yshift=-2.5ex]current bounding box.center)},scale=1]
        \draw[very thick, red] (-.433,-1/4) to (.433,-1/4) to (0,1/2) to (-.433,-1/4);
        \draw[very thick, red] (0,1/2) to (0,1);
        \draw[very thick, red] (-.433,-1/4) to (-.866,-1/2);
        \draw[very thick, red] (.433,-1/4) to (.866,-1/2);
        \node[left] at (-.866,-1/2) {$2$};
        \node[right] at (.866,-1/2) {$3$};
        \node[above] at (0,1) {$t$};
        \node at (-.433,1/4) {$1$};
        \node at (.433,1/4) {$4$};
        \node[below] at (0,-1/4) {$s$};
        \draw[very thick, fill=red, draw=red] (0,1/2) circle (.07);
        \draw[very thick, fill=red, draw=red] (-.433,-1/4) circle (.07);
        \draw[very thick, fill=red, draw=red] (.433,-1/4) circle (.07);
    \end{tikzpicture}
    = \sqrt{\frac{\mathsf{C}_{23t}}{\mathsf{C}_{12s}\mathsf{C}_{34s}\mathsf{C}_{14t}}}
    \begin{Bmatrix}
        P_1 & P_2 & P_s \\ P_3 & P_4 & P_t
    \end{Bmatrix}
    \begin{tikzpicture}[baseline={([yshift=-2.5ex]current bounding box.center)},scale=1]
        \draw[very thick, red] (0,0) to (0,1);
        \draw[very thick, red] (0,0) to (.866,-1/2);
        \draw[very thick, red] (0,0) to (-.866, -1/2);
        \node[left] at (-.866,-1/2) {$2$};
        \node[right] at (.866,-1/2) {$3$};
        \node[above] at (0,1) {$t$};
        \draw[very thick, fill=red, draw=red] (0,0) circle (.07);
    \end{tikzpicture}\, ,
\end{equation}
which allows us to recast the TQFT partition function as
\begin{multline}
        Z_{\text{Vir}}(M') =  (\mathbb{B}_{P_4}^{P_a P_5}\mathbb{B}_{P_6}^{P_cP_1}\mathbb{B}_{P_4}^{P_3P_c})^{-1} \int \d P_d \, \rho_0(P_d) \sqrt{\frac{\mathsf{C}_{2cd}{\mathsf{C}_{a6d}}}{\mathsf{C}_{2b3}\mathsf{C}_{3c4}\mathsf{C}_{4a5}\mathsf{C}_{5b6}}} \\
        \times \begin{Bmatrix}
            P_2 & P_3 & P_b \\
            P_4 & P_d & P_c
        \end{Bmatrix}
        \begin{Bmatrix}
            P_4 & P_5 & P_a \\ 
            P_6 & P_d & P_b
        \end{Bmatrix}
        Z_{\text{Vir}}\Bigg(
        \begin{tikzpicture}[scale=.55, baseline={([yshift=-.5ex]current bounding box.center)}]
            \path[use as bounding box] (-.15,0) ellipse (2.65 and 2.7);
            \draw[very thick, red] (0,0) circle (2);
            \draw[fill=white,draw=white] ({2*cos(170)},{2*sin(170)}) circle (1/10);
            \draw[fill=white,draw=white] ({2*cos(60)},{2*sin(60)}) circle (1/10);
            \node[left] at ({2*cos(120)},{2*sin(120)+1/10}) {$1$};
            \node[left] at ({2*cos(225)},{2*sin(225)}) {$6$};
            \node[right] at ({2*cos(-45)},{2*sin(-45)}) {$d$};
            \node[below] at (0,-2-1/10) {$a$};
            \node[right] at (2+1/10,1) {$c$};
            \draw[very thick, red, looseness=1.75] (0,-2) to[out=240, in = 170] ({{2*cos(170)}},{2*sin(170)})  to [out=-10, in = -90] (0,2);
            \draw[very thick, red, looseness=1.75] ({2*cos(150)},{2*sin(150)}) to[out = 150, in=60] ({{2*cos(60)}},{2*sin(60)}) to[out=240, in = 210] ({{2*cos(45)}},{2*sin(45)}) to [out=30, in = 20] ({{2*cos(20)}},{2*sin(20)});
            \draw[fill=white,draw=white] ({2*cos(45)},{2*sin(45)}) circle (1/10);
            \draw[very thick, red] ({2*cos(45)+.11*cos(136.5)},{2*sin(45)+.11*sin(136.5)}) to ({2*cos(45)+.11*cos(-46.5)},{2*sin(45)+.11*sin(-46.5)});
            \draw[very thick, red, fill=red] (0,2) circle (.07);
            \draw[very thick, red, fill=red] ({2*cos(20)},{2*sin(20)}) circle (.07);
            \draw[very thick, red, fill=red] (0,-2) circle (.07);
            \draw[very thick, red, fill=red] ({2*cos(150)},{2*sin(150)}) circle (.07);
            \node[right] at ({2*cos(60)+1/5},{2*sin(60)}) {$2$};
        \end{tikzpicture}
        \Bigg)\ .
\end{multline}
Finally, we undo the crossings by braiding the Wilson lines and recognize the remaining configuration as the four-boundary wormhole studied in section \ref{subsec:four-boundary wormhole} to arrive at
\begin{multline}
    Z_{\text{Vir}}(M) = \sqrt{\mathsf{C}_{12a}\mathsf{C}_{23b}\mathsf{C}_{34c}\mathsf{C}_{45a}\mathsf{C}_{56b}\mathsf{C}_{61c}}\mathrm{e}^{\pi i(P_1^2+P_3^2+P_5^2-2P_2^2-2P_4^2-2P_6^2)}\\
     \times \int \d P_d \, \rho_0(P_d) \, \mathrm{e}^{3\pi i P_d^2} 
    \begin{Bmatrix}
        P_6 & P_1 & P_c \\
        P_2 & P_d & P_a
    \end{Bmatrix}
    \begin{Bmatrix}
            P_2 & P_3 & P_b \\
            P_4 & P_d & P_c
        \end{Bmatrix}
        \begin{Bmatrix}
            P_4 & P_5 & P_a \\ 
            P_6 & P_d & P_b
        \end{Bmatrix}\ .
        \label{eq:Zvir six boundary wormhole hard}
\end{multline}
Once the dust has settled, as in previous examples the wormhole partition function is given by factors of $\sqrt{C_0}$ for each sphere three-point boundary together with a suitable combination of Virasoro $6j$ symbols associated with the Wilson line crossings.

This wormhole partition function implies that the corresponding sixth moment for the CFT structure constants receives the following contribution
\begin{multline}
    \overline{c_{12a}c_{23b}c_{34c}c_{45a}c_{56b}c_{61c}} \supset \sqrt{\overline{c^2_{12a}}\,\overline{c^2_{23b}}\,\overline{c^2_{34c}}\,\overline{c^2_{45a}}\,\overline{c^2_{56b}} \, \overline{c^2_{61c}}} (-1)^{\ell_1+\ell_3+\ell_5} \\
    \times \left|\int \d P_d\,\rho_0(P_d) \mathrm{e}^{3\pi i P_d^2}\begin{Bmatrix}
        P_6 & P_1 & P_c \\
        P_2 & P_d & P_a
    \end{Bmatrix}
    \begin{Bmatrix}
            P_2 & P_3 & P_b \\
            P_4 & P_d & P_c
        \end{Bmatrix}
        \begin{Bmatrix}
            P_4 & P_5 & P_a \\ 
            P_6 & P_d & P_b
        \end{Bmatrix}\right|^2\, .\label{eq:sixth moment borromean}
\end{multline}
As in previous examples, this sixth moment precisely affirms the  internal consistency of the description in terms of an ensemble of CFT data. Indeed, if one expands for example the product of two sphere five-point functions or three sphere four-point functions in certain OPE channels where there is not a Gaussian contraction, this leads to a result consistent with the computation in the Gaussian ensemble. For concreteness, consider the following averaged product of three four-point functions, all expanded in the $u$-channel 
\begin{align}
        &\qquad \overline{\langle\mathcal{O}_2\mathcal{O}_c\mathcal{O}_a\mathcal{O}_6\rangle\langle\mathcal{O}_6\mathcal{O}_a\mathcal{O}_b\mathcal{O}_4\rangle
    \langle\mathcal{O}_4\mathcal{O}_b\mathcal{O}_c\mathcal{O}_2\rangle}\nonumber\\
    &= \sum_{\mathcal{O}_{1,3,5}} \overline{c_{1a2}c_{2b3}c_{3c4}c_{4a5}c_{5b6}c_{6c1}}\, 
    \Bigg|
    \begin{tikzpicture}[scale=.75, baseline={([yshift=-.32ex]current bounding box.center)}]
        \draw[very thick, red] (0,0) to (3,0);
        \draw[very thick, red] (1,0) to (2,1);
        \fill[white] (1.5,.5) circle (.1);
        \draw[very thick, red] (2,0) to (1,1);
        \node[above] at (0,0) {$2$};
        \node[above] at (3,0) {$6$};
        \node[above] at (1,1) {$c$};
        \node[above] at (2,1) {$a$};
        \node[below] at (3/2,0) {$1$};
        \draw[draw=red,fill=red] (1,0) circle (.07);
        \draw[draw=red,fill=red] (2,0) circle (.07);
    \end{tikzpicture} \!\!
    \begin{tikzpicture}[scale=.7, baseline={([yshift=-.5ex]current bounding box.center)}]
        \draw[very thick, red] (0,0) to (3,0);
        \draw[very thick, red] (1,0) to (2,1);
        \fill[white] (1.5,.5) circle (.1);
        \draw[very thick, red] (2,0) to (1,1);
        \node[above] at (0,0) {$6$};
        \node[above] at (3,0) {$4$};
        \node[above] at (1,1) {$a$};
        \node[above] at (2,1) {$b$};
        \node[below] at (3/2,0) {$5$};
        \draw[draw=red,fill=red] (1,0) circle (.07);
        \draw[draw=red,fill=red] (2,0) circle (.07);
    \end{tikzpicture} \!\!
    \begin{tikzpicture}[scale=.7,      baseline={([yshift=-.5ex]current bounding box.center)}]
        \draw[very thick, red] (0,0) to (3,0);
        \draw[very thick, red] (1,0) to (2,1);
        \fill[white] (1.5,.5) circle (.1);
        \draw[very thick, red] (2,0) to (1,1);
        \node[above] at (0,0) {$4$};
        \node[above] at (3,0) {$2$};
        \node[above] at (1,1) {$b$};
        \node[above] at (2,1) {$c$};
        \node[below] at (3/2,0) {$3$};
        \draw[draw=red,fill=red] (1,0) circle (.07);
        \draw[draw=red,fill=red] (2,0) circle (.07);
    \end{tikzpicture}
    \Bigg|^2 \!\! \\
    &= \Bigg|
        \int \d P_1\, \rho_0(P_1)\, \d P_3\, \rho_0(P_3)\, \d P_5\, \rho_0(P_5) \sqrt{\mathsf{C}_{1a2}\mathsf{C}_{2b3}\mathsf{C}_{3c4}\mathsf{C}_{4a5}\mathsf{C}_{5b6}\mathsf{C}_{6c1}}\, \mathrm{e}^{\pi i (P_1^2+P_3^2+P_5^2)}\nonumber\\
        &\qquad \times \mathrm{e}^{-2\pi i(P_2^2+P_4^2+P_6^2)}\int \d P_d\, \rho_0(P_d) \, \mathrm{e}^{3\pi i P_d^2}
        \begin{Bmatrix}
        P_6 & P_1 & P_c \\
        P_2 & P_d & P_a
        \end{Bmatrix}
        \begin{Bmatrix}
            P_2 & P_3 & P_b \\
            P_4 & P_d & P_c
        \end{Bmatrix}
        \begin{Bmatrix}
            P_4 & P_5 & P_a \\ 
            P_6 & P_d & P_b
        \end{Bmatrix}\nonumber\\
        &\qquad\times \begin{tikzpicture}[scale=.7, baseline={([yshift=-.15ex]current bounding box.center)}]
        \draw[very thick, red] (0,0) to (3,0);
        \draw[very thick, red] (1,0) to (2,1);
        \fill[white] (1.5,.5) circle (.1);
        \draw[very thick, red] (2,0) to (1,1);
        \node[above] at (0,0) {$2$};
        \node[above] at (3,0) {$6$};
        \node[above] at (1,1) {$c$};
        \node[above] at (2,1) {$a$};
        \node[below] at (3/2,0) {$1$};
        \draw[draw=red,fill=red] (1,0) circle (.07);
        \draw[draw=red,fill=red] (2,0) circle (.07);
    \end{tikzpicture}
    \begin{tikzpicture}[scale=.7, baseline={([yshift=-.5ex]current bounding box.center)}]
        \draw[very thick, red] (0,0) to (3,0);
        \draw[very thick, red] (1,0) to (2,1);
        \fill[white] (1.5,.5) circle (.1);
        \draw[very thick, red] (2,0) to (1,1);
        \node[above] at (0,0) {$6$};
        \node[above] at (3,0) {$4$};
        \node[above] at (1,1) {$a$};
        \node[above] at (2,1) {$b$};
        \node[below] at (3/2,0) {$5$};
        \draw[draw=red,fill=red] (1,0) circle (.07);
        \draw[draw=red,fill=red] (2,0) circle (.07);
    \end{tikzpicture}
    \begin{tikzpicture}[scale=.7, baseline={([yshift=-.5ex]current bounding box.center)}]
        \draw[very thick, red] (0,0) to (3,0);
        \draw[very thick, red] (1,0) to (2,1);
        \fill[white] (1.5,.5) circle (.1);
        \draw[very thick, red] (2,0) to (1,1);
        \node[above] at (0,0) {$4$};
        \node[above] at (3,0) {$2$};
        \node[above] at (1,1) {$b$};
        \node[above] at (2,1) {$c$};
        \node[below] at (3/2,0) {$3$};
        \draw[draw=red,fill=red] (1,0) circle (.07);
        \draw[draw=red,fill=red] (2,0) circle (.07);
    \end{tikzpicture}
    \Bigg|^2\\
    &= \Bigg|\int \d P_d \,\rho_0(P_d) \mathsf{C}_{2cd}\mathsf{C}_{4bd}\mathsf{C}_{6ad}
    \begin{tikzpicture}[scale=.7, baseline={([yshift=-.15ex]current bounding box.center)}]
        \draw[very thick, red] (0,0) to (3,0);
        \draw[very thick, red] (1,0) to (1,1);
        \draw[very thick, red] (2,0) to (2,1);
        \node[above] at (0,0) {$2$};
        \node[above] at (3,0) {$6$};
        \node[above] at (1,1) {$c$};
        \node[above] at (2,1) {$a$};
        \node[below] at (3/2,0) {$d$};
        \draw[draw=red,fill=red] (1,0) circle (.07);
        \draw[draw=red,fill=red] (2,0) circle (.07);
    \end{tikzpicture} \!\!
    \begin{tikzpicture}[scale=.7, baseline={([yshift=-.5ex]current bounding box.center)}]
        \draw[very thick, red] (0,0) to (3,0);
        \draw[very thick, red] (1,0) to (1,1);
        \draw[very thick, red] (2,0) to (2,1);
        \node[above] at (0,0) {$6$};
        \node[above] at (3,0) {$4$};
        \node[above] at (1,1) {$a$};
        \node[above] at (2,1) {$b$};
        \node[below] at (3/2,0) {$d$};
        \draw[draw=red,fill=red] (1,0) circle (.07);
        \draw[draw=red,fill=red] (2,0) circle (.07);
    \end{tikzpicture} \!\!
    \begin{tikzpicture}[scale=.7, baseline={([yshift=-.5ex]current bounding box.center)}]
        \draw[very thick, red] (0,0) to (3,0);
        \draw[very thick, red] (1,0) to (1,1);
        \draw[very thick, red] (2,0) to (2,1);
        \node[above] at (0,0) {$4$};
        \node[above] at (3,0) {$2$};
        \node[above] at (1,1) {$b$};
        \node[above] at (2,1) {$c$};
        \node[below] at (3/2,0) {$d$};
        \draw[draw=red,fill=red] (1,0) circle (.07);
        \draw[draw=red,fill=red] (2,0) circle (.07);
    \end{tikzpicture}
    \Bigg|^2\ .
\end{align}
So we see that applying the statistics (\ref{eq:sixth moment borromean}) precisely reproduces the result (\ref{eq:three boundary sphere four point partition function 2}) anticipated from the Gaussian ensemble.

Notice that in this case the corresponding sixth moment receives contributions from configurations in which the Wilson lines have a different pattern of over- and under-crossings in the bulk, in addition to those with higher topology in the bulk. In principle, we could consider contributions from the manifolds formed by cutting $M$ along a six-punctured sphere in the bulk and gluing in another six-punctured sphere with any tangle formed by three strands in the bulk.
As a simple example, we could have considered the following six-boundary wormhole
\begin{equation}
    M = 
    \begin{tikzpicture}[baseline={([yshift=-.5ex]current bounding box.center)},scale=.8]
        \draw[very thick] (0,3) ellipse (1 and 1);
        \draw[very thick, out=-50, in=230] (-1,3) to (1,3);
        \draw[very thick, densely dashed, out=50, in=130] (-1,3) to (1,3);

        \draw[very thick] ({3*cos(30)},{3*sin(30)}) ellipse (1 and 1);
        \draw[very thick, out=-50, in=230] ({{3*cos(30)-1}},{3*sin(30)}) to ({{3*cos(30)+1}},{3*sin(30)});
        \draw[very thick, densely dashed, out=50, in=130] ({{3*cos(30)-1}},{3*sin(30)}) to ({{3*cos(30)+1}},{3*sin(30)});

        \draw[very thick] ({3*cos(-30)},{3*sin(-30)}) ellipse (1 and 1);
        \draw[very thick, out=-50, in=230] ({{3*cos(-30)-1}},{3*sin(-30)}) to ({{3*cos(-30)+1}},{3*sin(-30)});
        \draw[very thick, densely dashed, out=50, in=130] ({{3*cos(-30)-1}},{3*sin(-30)}) to ({{3*cos(-30)+1}},{3*sin(-30)});

        \draw[very thick] (0,-3) ellipse (1 and 1);
        \draw[very thick, out=-50, in=230] (-1,-3) to (1,-3);
        \draw[very thick, densely dashed, out=50, in=130] (-1,-3) to (1,-3);

        \draw[very thick] ({3*cos(150)},{3*sin(150)}) ellipse (1 and 1);
        \draw[very thick, out=-50, in=230] ({{3*cos(150)-1}},{3*sin(150)}) to ({{3*cos(150)+1}},{3*sin(150)});
        \draw[very thick, densely dashed, out=50, in=130] ({{3*cos(150)-1}},{3*sin(150)}) to ({{3*cos(150)+1}},{3*sin(150)});

        \draw[very thick] ({3*cos(210)},{3*sin(210)}) ellipse (1 and 1);
        \draw[very thick, out=-50, in=230] ({{3*cos(210)-1}},{3*sin(210)}) to ({{3*cos(210)+1}},{3*sin(210)});
        \draw[very thick, densely dashed, out=50, in=130] ({{3*cos(210)-1}},{3*sin(210)}) to ({{3*cos(210)+1}},{3*sin(210)});

        \draw[very thick, out=-90, in = -30] (0-1,3) to ({{3*cos(150)+cos(60)}},{3*sin(150)+sin(60)});
        \draw[very thick, out=-30, in = 30] ({{3*cos(150)+cos(240)}},{3*sin(150)+sin(240)}) to ({{3*cos(210)+cos(120)}},{3*sin(210)+sin(120)});
        \draw[very thick, out=30, in=90] ({{3*cos(210)+cos(-60)}},{3*sin(210)+sin(-60)}) to ({{0-1}},{-3});
        \draw[very thick, out=90, in = 150] (1,-3) to ({{3*cos(-30)+cos(240)}},{3*sin(-30)+sin(240)});
        \draw[very thick, out=150, in=210] ({{3*cos(-30)+cos(60)}},{3*sin(-30)+sin(60)}) to ({{3*cos(30)+cos(-60)}},{3*sin(30)+sin(-60)});
        \draw[very thick, out=210, in=-90] ({{3*cos(30)+cos(120)}},{3*sin(30)+sin(120)}) to ({0+1},{3});

        \draw[very thick, red, out=-90, in = -30] (0-4/5,3) to ({{3*cos(150)+4/5*cos(60)}},{3*sin(150)+4/5*sin(60)});
        \draw[very thick, red, out=-30, in = 30] ({{3*cos(150)+4/5*cos(240)}},{3*sin(150)+4/5*sin(240)}) to ({{3*cos(210)+4/5*cos(120)}},{3*sin(210)+4/5*sin(120)});
        \draw[very thick, red, out=30, in=90] ({{3*cos(210)+4/5*cos(-60)}},{3*sin(210)+4/5*sin(-60)}) to ({{0-4/5}},{-3});
        \draw[very thick, red, out=90, in = 150] (4/5,-3) to ({{3*cos(-30)+4/5*cos(240)}},{3*sin(-30)+4/5*sin(240)});
        \draw[very thick, red, out=150, in=210] ({{3*cos(-30)+4/5*cos(60)}},{3*sin(-30)+4/5*sin(60)}) to ({{3*cos(30)+4/5*cos(-60)}},{3*sin(30)+4/5*sin(-60)});
        \draw[very thick, red, out=210, in=-90] ({{3*cos(30)+4/5*cos(120)}},{3*sin(30)+4/5*sin(120)}) to ({0+4/5},{3});

        \draw[very thick, red] ({{3*cos(210)+4/5*cos(30)}},{3*sin(210)+4/5*sin(30)}) to ({{3*cos(30)+4/5*cos(240)}},{3*sin(30)+4/5*sin(240)});
        \draw[very thick, red] ({{3*cos(150)+4/5*cos(-30)}},{3*sin(150)+4/5*sin(-30)}) to ({{3*cos(-30)+4/5*cos(120)}},{3*sin(-30)+4/5*sin(120)});
        \draw[fill=white,draw=white] (0,-1/4) circle (1/10);
        \draw[fill=white,draw=white] (0,1/4) circle (1/10);

        \node[above] at (0,{3-4/5}) {$a$};
        \node[below] at (0,{-3+4/5}) {$a$};
        \node[right] at (0-4/5,3) {$1$};
        \node[left] at ({{3*cos(150)+4/5*cos(60)}},{3*sin(150)+4/5*sin(60)}) {$1$};
        \node[left] at ({{3*cos(150)+4/5*cos(240)}},{3*sin(150)+4/5*sin(240)+1/10}) {$6$};
        \node[left] at ({{3*cos(210)+4/5*cos(120)}},{3*sin(210)+4/5*sin(120)-1/10}) {$6$};
        \node[left] at ({{3*cos(210)+4/5*cos(-60)}},{3*sin(210)+4/5*sin(-60)}) {$5$};
        \node[right] at ({{0-4/5}},{-3}) {$5$};
        \node[left] at (4/5,-3) {$4$};
        \node[right] at ({{3*cos(-30)+4/5*cos(240)}},{3*sin(-30)+4/5*sin(240)}) {$4$};
        \node[right] at ({{3*cos(-30)+4/5*cos(60)}},{3*sin(-30)+4/5*sin(60)-1/10}) {$3$};
        \node[right] at ({{3*cos(30)+4/5*cos(-60)}},{3*sin(30)+4/5*sin(-60)+1/10}) {$3$};
        \node[right] at ({{3*cos(30)+4/5*cos(120)}},{3*sin(30)+4/5*sin(120)}) {$2$};
        \node[left] at ({0+4/5},{3}) {$2$};
        \node[above] at ({{3*cos(150)+4/5*cos(-30)-3/20}},{3*sin(150)+4/5*sin(-30)}) {$c$};
        \node[right] at ({{3*cos(30)+4/5*cos(240)}},{3*sin(30)+4/5*sin(240)+1/10}) {$b$};
        \node[below] at ({{3*cos(210)+4/5*cos(30)-3/20-1/10}},{3*sin(210)+4/5*sin(30)}) {$b$};
        \node[right] at ({{3*cos(-30)+4/5*cos(120)}},{3*sin(-30)+4/5*sin(120)-1/10}) {$c$};

        \draw[very thick, red] ({.12*cos(158.5)},{1/4-.027+.12*sin(158.5)}) to ({.12*cos(-28.5)},{1/4-.027+.12*sin(-28.5)});
        \draw[fill=white,draw=white] (0,1/4-.05) circle (1/10);
        \draw[very thick, red] (0,{3-4/5}) to (0,{-3+4/5});
        
        \draw[fill=white,draw=white] (.433,0) circle (1/10);
        \draw[very thick, red] ({.433+.12*cos(207)},{-.008+.12*sin(207)}) to ({.433+.12*cos(23)},{-.008+.12*sin(23)});

        \draw[fill=gray,opacity=.5] (0-1,3) to[out=-90,in=-30] ({{3*cos(150)+cos(60)}},{3*sin(150)+sin(60)}) to[out=-30,in=-30,looseness=1.7] ({{3*cos(150)+cos(240)}},{3*sin(150)+sin(240)}) to[out=-30, in = 30] ({{3*cos(210)+cos(120)}},{3*sin(210)+sin(120)}) to[out=30,in=30,looseness=1.7] ({{3*cos(210)+cos(-60)}},{3*sin(210)+sin(-60)}) to[out=30, in = 90] ({{0-1}},{-3}) to[out=90,in=90,looseness=1.7] (1,-3) to[out=90, in = 150] ({{3*cos(-30)+cos(240)}},{3*sin(-30)+sin(240)}) to[out=150, in=150,looseness=1.7] ({{3*cos(-30)+cos(60)}},{3*sin(-30)+sin(60)}) to[out=150, in = 210] ({{3*cos(30)+cos(-60)}},{3*sin(30)+sin(-60)}) to[out=210, in= 210,looseness=1.7] ({{3*cos(30)+cos(120)}},{3*sin(30)+sin(120)}) to[out=210, in = -90] ({0+1},{3}) to[out=-90,in=-90,looseness=1.7] (0-1,3);
    \end{tikzpicture}
    \, .
\end{equation}
The TQFT partition function on this wormhole differs from (\ref{eq:Zvir six boundary wormhole hard}) in a subtle way
\begin{multline}
    Z_{\text{Vir}}(M) = \sqrt{\mathsf{C}_{1a2}\mathsf{C}_{2b3}\mathsf{C}_{3c4}\mathsf{C}_{4a5}\mathsf{C}_{5b6}\mathsf{C}_{6c1}}\mathrm{e}^{\pi i(-P_1^2+P_3^2+P_5^2-2P_4^2)}\\
     \times \int \d P_d \, \rho_0(P_d) \, \mathrm{e}^{\pi i P_d^2} 
    \begin{Bmatrix}
        P_6 & P_1 & P_c \\
        P_2 & P_d & P_a
    \end{Bmatrix}
    \begin{Bmatrix}
            P_2 & P_3 & P_b \\
            P_4 & P_d & P_c
        \end{Bmatrix}
        \begin{Bmatrix}
            P_4 & P_5 & P_a \\ 
            P_6 & P_d & P_b
        \end{Bmatrix}\ .
        \label{eq:Zvir six boundary wormhole hard 2}
\end{multline}
The only difference from (\ref{eq:Zvir six boundary wormhole hard}) are the phases, particularly that which appears in the integral over the intermediate Liouville momentum $P_d$. Although both contribute to the corresponding sixth moment of the structure constants,  between (\ref{eq:Zvir six boundary wormhole hard}) and (\ref{eq:Zvir six boundary wormhole hard 2}) is not a priori obvious which Wilson line configuration dominates in the semiclassical limit.

\subsubsection{Diagrammatic rules for multi-boundary wormholes and CFT statistics}

Although the intermediate details of the computations were nontrivial, there is an underlying simplicity to the previously discussed results for the Virasoro TQFT partition functions of wormholes with three-punctured sphere boundaries and trivial topology in the bulk, and hence for the leading contributions to the non-Gaussian statistics of CFT data in the boundary ensemble description of 3d gravity. 
In all cases, the wormhole partition function involves a factor of $\sqrt{C_0}$ for each three-punctured sphere boundary, together with a suitable combination of Virasoro $6j$ symbols. Here we describe diagrammatic rules that straightforwardly reproduce these results and that enable the computation of more nontrivial wormhole partition functions. These rules will turn out to be a slight generalization of the disk Feynman rules in JT gravity coupled to matter (see e.g. \cite{Jafferis:2022wez}).\footnote{SC is grateful to Baur Mukhametzhanov for discussions on this. Baur also independently observed that higher moments of CFT data required for internal consistency of the ensemble description of 3d gravity were reproduced by generalizations of the disk Feynman diagrams in JT gravity coupled to matter \cite{Mukhametzhanov:2023notes}.}

It is simplest to describe the situation in which the sphere boundaries are connected in a cyclic way, as in (\ref{eq:four boundary wormhole u channel}) and (\ref{eq:six boundary wormhole borromean rings}); the CFT statistics in other configurations may be obtained from the results in these cases by application of the swapping rule (\ref{eq:exchanging operators}). 

The idea is the following. Starting from a wormhole configuration with the boundaries connected in a cyclic way, replacing the punctured sphere boundaries with a trivalent vertex as follows
\begin{equation}
    \begin{tikzpicture}[baseline={([yshift=-2ex]current bounding box.center)},scale=1]
        \draw[very thick] (-2,2) ellipse (1 and 1);
        \draw[very thick, out=-50, in=230] (-3,2) to (-1,2);
        \draw[very thick, densely dashed, out=50, in=130] (-3,2) to (-1,2);

        \draw[thick] ({{-2-cos(60)}},{2-sin(60)}) to ({{-2-cos(60)+1/2}},{2-sin(60)-1/2}); 
        \draw[thick] ({-2+cos(30)},{2+sin(30)}) to ({{-2+cos(30)+1/2}},{2+sin(30)-1/2});

        \draw[very thick, red] ({-2+4/5*cos(30)},{2+4/5*sin(30)}) to ({-2+4/5*cos(30)+1/2},{2+4/5*sin(30)-1/2});
        \draw[very thick, red] ({{-2-4/5*cos(60)}},{2-4/5*sin(60)}) to ({{-2-4/5*cos(60)+1/2}},{2-4/5*sin(60)-1/2}); 

        \node[left] at ({-2+4/5*cos(30)},{2+4/5*sin(30)+3/20}) {$1$};
        \node[left] at (-2+.565685,2-.565685) {$2$};
        \node[left] at ({{-2-4/5*cos(60)+1/20}},{2-4/5*sin(60)+1/20}) {$3$};
        \draw[very thick, red] (-2+.565685,2-.565685) to ({-2+.565685+1/4},{2-.565685-1/4});
        \draw[fill=gray, opacity=0.5] ({{-2-cos(60)+1/2}},{2-sin(60)-1/2}) to[out=135, in = -30] ({{-2-cos(60)}},{2-sin(60)}) to[out = -30, in=-60,looseness=1.35] ({{-2+cos(30)}},{2+sin(30)}) to[out=-60,in = 135] ({{-2+cos(30)+1/2}},{2+sin(30)-1/2});
    \end{tikzpicture}
    \longrightarrow 
    \begin{tikzpicture}[baseline={([yshift=-0.5ex]current bounding box.center)}]
        \draw[very thick, red] (-1,0) to[out=90, in =180] (0,1);
        \draw[very thick, draw=red, fill=red] ({cos(135)},{sin(135)}) circle (.07);
        \draw[very thick, red] ({cos(135)},{sin(135)}) to ({.3*cos(135)},{.3*sin(135)});
        \node[below] at (-1,0) {$3$};
        \node[right] at (0,1) {$1$};
        \node at (0,0) {$2$};
    \end{tikzpicture}
\end{equation}
produces a disk diagram with lines that may cross in the interior of the disk, such as that drawn in (\ref{eq:6 bdy wormhole as Wilson line network}). It is important to keep track of the way that the lines over- and under-cross in the projection to a two-dimensional disk diagram. The TQFT partition function associated with this disk diagram is then computed according to the following simple Feynman rules:
\begin{itemize}
    \item Each trivalent vertex contributes a factor of $\sqrt{C_0}$:
\begin{equation}\label{eq:trivalent vertex}
    \begin{tikzpicture}[baseline={([yshift=-0.5ex]current bounding box.center)}]
        \draw[very thick, red] (-1,0) to[out=90, in =180] (0,1);
        \draw[very thick, draw=red, fill=red] ({cos(135)},{sin(135)}) circle (.07);
        \draw[very thick, red] ({cos(135)},{sin(135)}) to ({.3*cos(135)},{.3*sin(135)});
        \node[below] at (-1,0) {$3$};
        \node[right] at (0,1) {$1$};
        \node at (0,0) {$2$};
    \end{tikzpicture}
    = \sqrt{C_0(P_1,P_2,P_3)}\, .
\end{equation}
\item Each closed region in the interior of the disk is associated with a Liouville momentum $P$ that is integrated with the measure $\rho_0(P) \d P$.
\item Each crossing of a pair of lines in the interior of the disk contributes a Virasoro $6j$ symbol
\begin{equation}\label{eq:quartic vertex}
    \begin{tikzpicture}[baseline={([yshift=-2ex]current bounding box.center)}]
        \draw[very thick, red] (0,-1) to (0,1);
        \draw[fill=white, draw=white] (0,0) circle (1/10);
        \draw[very thick,red] (-1,0) to (1,0);
        \node at (-3/4,3/4) {$1$};
        \node at (3/4,3/4) {$2$};
        \node at (3/4,-3/4) {$3$};
        \node at (-3/4,-3/4) {$4$};
        \node[left] at (-1,0) {$t$};
        \node[above] at (0,1) {$s$};

        \draw[very thick, red, out = -90, in = 0, looseness=1.5] (-1/5,1) to (-1,1/5);
        \draw[very thick, red, out=-90, in = 180, looseness=1.5] (1/5,1) to (1,1/5);
        \draw[very thick, red, out = 90, in = 180] (1/5,-1) to (1,-1/5);
        \draw[very thick, red, out = 90, in = 0] (-1/5,-1) to (-1,-1/5);
    \end{tikzpicture}
    = \begin{Bmatrix} P_1 & P_2 & P_s \\ P_3 & P_4 & P_t
    \end{Bmatrix}
    \mathrm{e}^{\pi i(P_1^2 + P_3^2 - P_2^2 - P_4^2)}\, .
\end{equation}
Here the labels 1, 2, 3 and 4 are associated to the four faces delineated by the Wilson lines $s$ and $t$.
\end{itemize}
The Virasoro $6j$ symbol plays the role of a quartic vertex in these diagrammatic rules, dressed with a phase that keeps track of the way that the Wilson lines over- and under-cross. This reproduces the partition function on the four-boundary wormhole (\ref{eq:s to u four boundary wormhole}) essentially by design.

As a simple example, consider the six-boundary wormhole studied in section \ref{subsubsec:six boundary wormhole hard}. The two-dimensional projection of this configuration involves three crossings of Wilson lines and one closed region in the interior of the disk, so the TQFT partition function involves a single integral of three $6j$ symbols. Indeed, a straightforward application of these rules immediately reproduces the TQFT partition function (\ref{eq:Zvir six boundary wormhole hard}).

The Virasoro $6j$ symbol obeys many identities that facilitate the consistency of this description. For instance, it is often the case that there is an ambiguity of how to arrange the Wilson line crossings in the interior of the disk. The TQFT partition function as computed from these rules should be independent of such choices. For example, we should have
\begin{equation}
    \begin{tikzpicture}[baseline={([yshift=-0.5ex]current bounding box.center)}]
        \draw[very thick, red] (-1,1/2) to[out=-30, in = 180] (0,-1/4) to[out=0, in=210] (1,1/2);
        \draw[fill=white,draw=white] (-1/2,0) circle (1/10);
        \draw[fill=white,draw=white] (1/2,0) circle (1/10);
        \draw[very thick, red] (-1,-1/2) to[out=30, in = 180] (0,1/4) to[out=0, in=150] (1,-1/2);

        \draw[very thick, red] (-1,3/8) to[out = -30, in = 30] (-1,-3/8);
        \draw[very thick, red] (1,3/8) to[out=210, in = 150] (1,-3/8);
        \draw[very thick, red] (-3/4,1/2) to[out=-30, in = 210,looseness=.75] (3/4,1/2);
        \draw[very thick, red] (-3/4,-1/2) to[out=30, in = 150,looseness=.75] (3/4,-1/2);
        
        \node[left] at (-1,1/2) {$a$};
        \node[left] at (-1,-1/2) {$b$};
        \node[above] at (0,1/2) {$2$};
        \node[left] at (-1,0) {$1$};
        \node[right] at (1,0) {$3$};
        \node[below] at (0,-1/2) {$4$};
    \end{tikzpicture}
    = 
    \begin{tikzpicture}[baseline={([yshift=-0.5ex]current bounding box.center)}]
        \draw[very thick, red] (-1,1/2) to[out=-30, in = 210] (1,1/2);
        \draw[very thick, red] (-1,-1/2) to[out= 30, in = 150] (1,-1/2);
        
        \draw[very thick, red] (-1,3/8) to[out = -30, in = 30] (-1,-3/8);
        \draw[very thick, red] (1,3/8) to[out=210, in = 150] (1,-3/8);
        \draw[very thick, red] (-3/4,1/2) to[out=-30, in = 210,looseness=.75] (3/4,1/2);
        \draw[very thick, red] (-3/4,-1/2) to[out=30, in = 150,looseness=.75] (3/4,-1/2);

        \node[left] at (-1,1/2) {$a$};
        \node[left] at (-1,-1/2) {$b$};
        \node[above] at (0,1/2) {$2$};
        \node[left] at (-1,0) {$1$};
        \node[right] at (1,0) {$3$};
        \node[below] at (0,-1/2) {$4$};
    \end{tikzpicture}\, ,
\end{equation}
which is guaranteed by idempotency of the Virasoro $6j$ symbol
\begin{equation}
    \int \d P_s \, \rho_0(P_s) 
    \begin{Bmatrix}
        P_4 & P_1 & P_b \\
        P_2 & P_s & P_a
    \end{Bmatrix}
    \begin{Bmatrix}
        P_2 & P_3 & P_a \\
        P_4 & P_s & P_b
    \end{Bmatrix}
    = \frac{\delta(P_1-P_3)}{\rho_0(P_1)}.
\end{equation}
There is also a Yang-Baxter equation, which facilitates moving a line over a crossing as follows,
\begin{equation}
    \begin{tikzpicture}[baseline={([yshift=-0.5ex]current bounding box.center)}]
        \draw[very thick, red] (1,3/2) to (-1,-1/2);
        \draw[fill=white,draw=white] (0,1/2) circle (1/10);
        \draw[very thick, red] (-1,3/2) to (1,-1/2);
        \draw[fill=white,draw=white] (1/2,0) circle (1/10);
        \draw[fill=white,draw=white] (-1/2,0) circle (1/10);
        \draw[very thick, red] (-1,0) to (1,0);
        
        \draw[very thick, red] (-.9,3/2) to[out=-45, in = 225] (.9,3/2);
        \draw[very thick, red] (-1,1.4) to[out = -45, in = 0] (-1,.1);
        \draw[very thick, red] (-1,-.1) to[out=0, in =45] (-1,-.4);
        \draw[very thick, red] (-.9,-1/2) to[out=45, in = 135] (.9,-1/2);
        \draw[very thick, red] (1,-.4) to[out = 135, in = 180] (1,-.1);
        \draw[very thick, red] (1,.1) to[out = 180, in = 225] (1,1.4);

        \node at (-1-.15,3/2+.15) {$a$};
        \node at (1+.15,3/2+.15) {$b$};
        \node[left] at (-1,3/4) {$1$};
        \node[above] at (0,3/2) {$2$};
        \node[right] at (1,3/4) {$3$};
        \node[right] at (1, -.25) {$4$};
        \node[below] at (0,-1/2) {$5$};
        \node[left] at (-1,-.25) {$6$};
        \node[right] at (1,0+.05) {$c$};
    \end{tikzpicture}
    = 
    \begin{tikzpicture}[baseline={([yshift=-0.5ex]current bounding box.center)}]
        \draw[very thick, red] (1,3/2) to (-1,-1/2);
        \draw[fill=white,draw=white] (0,1/2) circle (1/10);
        \draw[very thick, red] (-1,3/2) to (1,-1/2);
        \draw[fill=white,draw=white] (1/2,1) circle (1/10);
        \draw[fill=white,draw=white] (-1/2,1) circle (1/10);
        \draw[very thick, red] (-1,1) to (1,1);

        \draw[very thick, red] (-.9,3/2) to[out=-45, in = 225] (.9,3/2);
        \draw[very thick, red] (-.9,-1/2) to[out=45, in = 135] (.9,-1/2);
        \draw[very thick, red] (-1,1.4) to[out=-45,in = 0] (-1,1.1);
        \draw[very thick, red] (1,1.4) to[out= 225, in = 180] (1,1.1);
        \draw[very thick, red] (1,.9) to [out= 180, in = 135] (1,-.4);
        \draw[very thick, red] (-1,.9) to [out= 0, in = 45] (-1,-.4);

        \node[left] at (-1, 1.25) {$1$};
        \node[above] at (0,3/2) {$2$};
        \node[right] at (1,1.25) {$3$};
        \node[right] at (1,.25) {$4$};
        \node[below] at (0,-1/2) {$5$};
        \node[left] at (-1,.25) {$6$};
        \node at (-1-.15,3/2+.15) {$a$};
        \node at (1+.15,3/2+.15) {$b$};
        \node[right] at (1,1-.05) {$c$};
        
    \end{tikzpicture}\ .
\end{equation}
In equations, this translates to
\begin{multline}
        \int \d P_d \, \rho_0(P_d)
        \begin{Bmatrix}
            P_1 & P_2 & P_a \\
            P_3 & P_d & P_b
        \end{Bmatrix}
        \begin{Bmatrix}
            P_5 & P_6 & P_b \\
            P_1 & P_d & P_c
        \end{Bmatrix}
        \begin{Bmatrix}
            P_3 & P_4 & P_c \\
            P_5 & P_d & P_a
        \end{Bmatrix}
        \mathrm{e}^{\pi i (P_d^2+P_2^2+P_4^2+P_6^2)}\\
    = \int \d P_e \, \rho_0(P_e) 
        \begin{Bmatrix} 
            P_6 & P_1 & P_c \\
            P_2 & P_e & P_a
        \end{Bmatrix}
        \begin{Bmatrix}
            P_2 & P_3 & P_b\\
            P_4 & P_e & P_c 
        \end{Bmatrix}
        \begin{Bmatrix}
            P_4 & P_5 & P_a\\
            P_6 & P_e & P_b
        \end{Bmatrix}
        \mathrm{e}^{\pi i (P_e^2+P_1^2+P_3^2 +P_5^2)}\ . \label{eq:Yang-Baxter equation}
\end{multline}
This identity follows from the consistency of braiding on the sphere. Indeed, the R-matrix also appears as the braiding matrix of conformal blocks as in eq.~\eqref{eq:conformal block braiding}. The Yang-Baxter equation then corresponds to the fundamental relation in the braid group as follows:
\begin{align}
    \begin{tikzpicture}[baseline={([yshift=-0.5ex]current bounding box.center)}]
        \draw[very thick, out=90, in=-90, red] (0,0) to (-1,1);
        \fill[white] (-.5,.5) circle (.1);
        \draw[very thick, out=90, in=-90, red] (-1,0) to (0,1); 
        \draw[very thick, red] (-1,1) to (-1,2);
        \draw[very thick, out=90, in=-90, red] (1,1) to (0,2);
        \fill[white] (.5,1.5) circle (.1);
        \draw[very thick, out=90, in=-90, red] (0,1) to (1,2);
        \draw[very thick, out=90, in=-90, red] (0,2) to (-1,3);
        \fill[white] (-.5,2.5) circle (.1);  
        \draw[very thick, out=90, in=-90, red] (-1,2) to (0,3);
        \draw[very thick, red] (1,0) to (1,1);
        \draw[very thick, red] (1,2) to (1,3);
        \draw[very thick, red] (-2,0) to (2,0);
        \fill[red] (-1,0) circle (.07);
        \fill[red] (0,0) circle (.07);
        \fill[red] (1,0) circle (.07);
        \node at (-2.3,0) {$1$};
        \node at (-.5,-.3) {$2$};
        \node at (.5,-.3) {$3$};
        \node at (2.3,0) {$4$};
        \node at (1,3.3) {$a$};
        \node at (0,3.3) {$b$};
        \node at (-1,3.3) {$c$};
    \end{tikzpicture} =
    \begin{tikzpicture}[baseline={([yshift=-0.5ex]current bounding box.center)}]
        \draw[very thick, out=90, in=-90, red] (1,0) to (0,1); 
        \fill[white] (.5,.5) circle (.1);
        \draw[very thick, out=90, in=-90, red] (0,0) to (1,1);
        \draw[very thick, red] (1,1) to (1,2);
        \draw[very thick, out=90, in=-90, red] (0,1) to (-1,2);
        \fill[white] (-.5,1.5) circle (.1);
        \draw[very thick, out=90, in=-90, red] (-1,1) to (0,2);
        \draw[very thick, out=90, in=-90, red] (1,2) to (0,3);
        \fill[white] (.5,2.5) circle (.1);  
        \draw[very thick, out=90, in=-90, red] (0,2) to (1,3);
        \draw[very thick, red] (-1,0) to (-1,1);
        \draw[very thick, red] (-1,2) to (-1,3);
        \draw[very thick, red] (-2,0) to (2,0);
        \fill[red] (-1,0) circle (.07);
        \fill[red] (0,0) circle (.07);
        \fill[red] (1,0) circle (.07);
        \node at (-2.3,0) {$1$};
        \node at (-.5,-.3) {$2$};
        \node at (.5,-.3) {$3$};
        \node at (2.3,0) {$4$};
        \node at (1,3.3) {$a$};
        \node at (0,3.3) {$b$};
        \node at (-1,3.3) {$c$};
    \end{tikzpicture}\ . 
\end{align}
Using \eqref{eq:conformal block braiding} to unbraid the left- and right-hand side and comparing the result leads to the Yang-Baxter equation \eqref{eq:Yang-Baxter equation}. 

These diagrammatic rules for wormhole partition functions are structurally identical to the disk Feynman rules for JT gravity coupled to matter, as described for example in \cite{Jafferis:2022wez}. The only differences are that here the trivalent vertex is given by $\sqrt{C_0}$, the quartic vertex is given by the Virasoro $6j$ symbol rather than the $\SL(2,\mathbb{R})$ $6j$ symbol, and one must keep track of the over- and under-crossings of the Wilson lines in the bulk, leading to extra phases in the quartic vertex. Indeed, these rules precisely reduce to the JT gravity + matter disk Feynman rules in the semiclassical near-extremal limit of \cite{Ghosh:2019rcj, Maxfield:2020ale}. In this limit one takes 
\begin{equation}
    c = 1+6(b+b^{-1})^2,\quad P_{\text{ext}} = bs_{\text{ext}},\quad P_{\text{int}} = \frac{i}{2}(b+b^{-1}-2b h_{\text{int}}),\quad b\to 0
\end{equation}
fixing $s_{\text{ext}}$ and $h_{\text{int}}$ in the semiclassical limit. Here $P_{\text{ext}}$ are the Liouville momenta of the Wilson lines forming the perimeter of the disk, $P_{\text{int}}$ are the Liouville momenta of those in the interior of the disk, and this limit corresponds to sending the external Wilson lines very near extremality while assigning the internal Wilson lines a fixed conformal weight $h_{\text{int}}$. With all external Wilson lines near extremality, the extra phase in the quartic vertex (\ref{eq:quartic vertex}) cancels and we no longer need to keep track of the over- and under-crossing of the Wilson lines in the semiclassical limit.

It is likely that these diagrammatic rules may be derived directly from the tensor model for AdS$_3$ gravity recently introduced in \cite{Belin:2023efa}, with tensor model diagrams corresponding to specific wormhole topologies. However we will not pursue this any further here. 

\subsection{Handle wormholes} 
In \cite{Chandra:2022bqq}, the on-shell action of a class of wormholes contributing to certain single-boundary observables was constructed. These wormholes admitted an elegant interpretation in terms of the Coleman-Giddings-Strominger mechanism \cite{Coleman:1988cy,Giddings:1987cg,Giddings:1988cx}, whereby the existence of Euclidean wormholes induce random bulk couplings in the low-energy effective theory. Here we demonstrate that the gravity partition function on these single-boundary ``handle wormholes'' is straightforward to compute using Virasoro TQFT.

For concreteness, consider the sphere four-point function of pairwise identical operators $\langle \mathcal{O}_1\mathcal{O}_2\mathcal{O}_2\mathcal{O}_1\rangle$. Suppose there is a third species of defect, dual to the operator $\mathcal{O}_3$. Naively, the trivalent coupling $\lambda_{123}$ in the bulk low-energy effective field theory vanishes since in the Gaussian ensemble the averaged structure constant vanishes $\overline{c_{123}} = 0$. However, there is a two-boundary wormhole that computes the variance $\overline{c^2_{123}}\ne 0$, so the conclusion that the defects are entirely non-interacting in the bulk cannot quite be correct. In particular, we expect a topology that corresponds to the exchange of $\mathcal{O}_3$ in the $\mathcal{O}_1\times\mathcal{O}_2$ OPE and hence contributes to the bulk-dual of the four-point function $\langle \mathcal{O}_1\mathcal{O}_2\mathcal{O}_2\mathcal{O}_1\rangle$.

Consider the following topology discussed in \cite{Chandra:2022bqq}
\begin{equation}\label{eq:handle wormhole}
    M = 
    \begin{tikzpicture}[baseline={([yshift=-0.5ex]current bounding box.center)}]
        \draw[very thick, densely dashed,out=50, in=130] (-3/2,0) to (3/2,0);

        \draw[very thick, red] (-3/4+1/2,0) to (3/4-1/2,0);
        \draw[very thick, red] ({3/4+7/16*cos(45)},{7/16*sin(45)}) to ({3/2*cos(30)},{3/2*sin(30)});
        \draw[very thick, red] ({3/4+7/16*cos(45)},{-7/16*sin(45)}) to ({3/2*cos(30)},{-3/2*sin(30)});
        \draw[very thick, red] ({-3/4-7/16*cos(45)},{7/16*sin(45)}) to ({-3/2*cos(30)},{3/2*sin(30)});
        \draw[very thick, red] ({-3/4-7/16*cos(45)},{-7/16*sin(45)}) to ({-3/2*cos(30)},{-3/2*sin(30)});

        \draw[fill=gray, opacity=.5, draw=gray] (0,0) ellipse (3/2 and 3/2);
        \draw[fill=white, draw=white] (-3/4,0) ellipse (1/2 and 1/2);
        \draw[fill=white, draw=white] (3/4,0) ellipse (1/2 and 1/2);

        \draw[very thick, fill=white] (-3/4,0) ellipse (1/2 and 1/2);
        \draw[very thick, out=-50, in=230] (-3/4-1/2,0) to (-3/4+1/2,0);
        \draw[very thick, densely dashed, out=50, in=130] (-3/4-1/2,0) to (-3/4+1/2,0);

        \draw[very thick, fill=white] (3/4,0) ellipse (1/2 and 1/2);
        \draw[very thick, out=-50, in=230] (3/4-1/2,0) to (3/4+1/2,0);
        \draw[very thick, densely dashed, out=50, in=130] (3/4-1/2,0) to (3/4+1/2,0);

        \draw[very thick] (0,0) ellipse (3/2 and 3/2);
        \draw[very thick, out=-50, in=230] (-3/2,0) to (3/2,0);

        \node[right] at ({3/2*cos(30)},{3/2*sin(30)}) {$1$};
        \node[right] at ({3/2*cos(30)},{-3/2*sin(30)}) {$2$};
        \node[left] at ({-3/2*cos(30)},{3/2*sin(30)}) {$1$};
        \node[left] at ({-3/2*cos(30)},{-3/2*sin(30)}) {$2$};
        \node[above] at (0,0) {$3$};

        \draw[very thick] ({-3/4-.05},{1/2-.1}) to (-3/4-.05, {1/2+.1});
        \draw[very thick] ({-3/4+.05},{1/2-.1}) to (-3/4+.05, {1/2+.1}); 

        \draw[very thick] ({3/4-.05},{1/2-.1}) to (3/4-.05, {1/2+.1});
        \draw[very thick] ({3/4+.05},{1/2-.1}) to (3/4+.05, {1/2+.1});
    \end{tikzpicture}\, .
\end{equation}
It is constructed by starting with a compression body whose outer boundary is a four-punctured sphere and two three-punctured sphere inner boundaries, and then identifying the two inner boundaries as shown in (\ref{eq:handle wormhole}). The Wilson lines corresponding to $\mathcal{O}_1$ and $\mathcal{O}_2$ traverse the resulting wormhole and that corresponding to $\mathcal{O}_3$ forms a closed loop in it. The TQFT partition function on the compression body (without the identification among the inner boundaries) is simply proportional to the corresponding sphere four-point conformal block
\begin{equation}
    Z_{\text{Vir}}
    \bigg(
    \!
    \begin{tikzpicture}[baseline={([yshift=-0.5ex]current bounding box.center)},scale=.6]
        \draw[very thick, densely dashed,out=50, in=130] (-3/2,0) to (3/2,0);

        \draw[very thick, red] (-3/4+1/2,0) to (3/4-1/2,0);
        \draw[very thick, red] ({3/4+7/16*cos(45)},{7/16*sin(45)}) to ({3/2*cos(30)},{3/2*sin(30)});
        \draw[very thick, red] ({3/4+7/16*cos(45)},{-7/16*sin(45)}) to ({3/2*cos(30)},{-3/2*sin(30)});
        \draw[very thick, red] ({-3/4-7/16*cos(45)},{7/16*sin(45)}) to ({-3/2*cos(30)},{3/2*sin(30)});
        \draw[very thick, red] ({-3/4-7/16*cos(45)},{-7/16*sin(45)}) to ({-3/2*cos(30)},{-3/2*sin(30)});

        \draw[fill=gray, opacity=.5, draw=gray] (0,0) ellipse (3/2 and 3/2);
        \draw[fill=white, draw=white] (-3/4,0) ellipse (1/2 and 1/2);
        \draw[fill=white, draw=white] (3/4,0) ellipse (1/2 and 1/2);

        \draw[very thick, fill=white] (-3/4,0) ellipse (1/2 and 1/2);
        \draw[very thick, out=-50, in=230] (-3/4-1/2,0) to (-3/4+1/2,0);
        \draw[very thick, densely dashed, out=50, in=130] (-3/4-1/2,0) to (-3/4+1/2,0);

        \draw[very thick, fill=white] (3/4,0) ellipse (1/2 and 1/2);
        \draw[very thick, out=-50, in=230] (3/4-1/2,0) to (3/4+1/2,0);
        \draw[very thick, densely dashed, out=50, in=130] (3/4-1/2,0) to (3/4+1/2,0);

        \draw[very thick] (0,0) ellipse (3/2 and 3/2);
        \draw[very thick, out=-50, in=230] (-3/2,0) to (3/2,0);

        \node[right] at ({3/2*cos(30)},{3/2*sin(30)}) {$1$};
        \node[right] at ({3/2*cos(30)},{-3/2*sin(30)}) {$2$};
        \node[left] at ({-3/2*cos(30)},{3/2*sin(30)}) {$1$};
        \node[left] at ({-3/2*cos(30)},{-3/2*sin(30)}) {$2$};
        \node[above] at (0,0) {$3$}; 
    \end{tikzpicture}
    \!
    \bigg)
    = C_0(P_1,P_2,P_3)^2 
    \begin{tikzpicture}[baseline={([yshift=-0.5ex]current bounding box.center)}]
        \draw[very thick, red] (-3/4,1/2) to (-3/8,0);
        \draw[very thick, red] (-3/4,-1/2) to (-3/8,0);
        \draw[very thick, red] (-3/8,0) to (3/8,0);
        \draw[very thick, red] (3/8,0) to (3/4,1/2);
        \draw[very thick, red] (3/8,0) to (3/4,-1/2);
        \node[left] at (-3/4,1/2) {$1$}; 
        \node[left] at (-3/4,-1/2) {$2$};
        \node[right] at (3/4,1/2) {$1$};
        \node[right] at (3/4,-1/2) {$2$};
        \node[above] at (0,0) {$3$};
        \fill[red] (-3/8,0) circle (0.07);
        \fill[red] (3/8,0) circle (0.07);
    \end{tikzpicture}\, .
\end{equation}
To implement the identification of the inner boundaries, we first view the partition function on the compression body as a state in the tensor product Hilbert space associated with the inner and outer boundaries $\mathcal{H}_{0,4}\otimes \mathcal{H}_{0,3}\otimes\mathcal{H}_{0,3}$. Taking the inner product between the states in the three-punctured sphere Hilbert spaces implements the identification between the inner boundaries and leaves us with the following state in $\mathcal{H}_{0,4}$:
\begin{equation}
    Z_{\text{Vir}}(M) = C_0(P_1,P_2,P_3) 
    \begin{tikzpicture}[baseline={([yshift=-0.5ex]current bounding box.center)}]
        \draw[very thick, red] (-3/4,1/2) to (-3/8,0);
        \draw[very thick, red] (-3/4,-1/2) to (-3/8,0);
        \draw[very thick, red] (-3/8,0) to (3/8,0);
        \draw[very thick, red] (3/8,0) to (3/4,1/2);
        \draw[very thick, red] (3/8,0) to (3/4,-1/2);
        \node[left] at (-3/4,1/2) {$1$}; 
        \node[left] at (-3/4,-1/2) {$2$};
        \node[right] at (3/4,1/2) {$1$};
        \node[right] at (3/4,-1/2) {$2$};
        \node[above] at (0,0) {$3$};
        \fill[red] (-3/8,0) circle (0.07);
        \fill[red] (3/8,0) circle (0.07);
    \end{tikzpicture}\, .
\end{equation}
Squaring the TQFT partition function leads to the expected contribution to the gravity path integral corresponding to the exchange of $\mathcal{O}_3$ in the $\mathcal{O}_1\times \mathcal{O}_2$ OPE, with squared OPE coefficient given by the corresponding variance in the Gaussian ensemble
\begin{equation}
    Z_{\text{grav}}(M) = \left|C_0(P_1,P_2,P_3)\right|^2\bigg|
    \begin{tikzpicture}[baseline={([yshift=-0.5ex]current bounding box.center)}]
        \draw[very thick, red] (-3/4,1/2) to (-3/8,0);
        \draw[very thick, red] (-3/4,-1/2) to (-3/8,0);
        \draw[very thick, red] (-3/8,0) to (3/8,0);
        \draw[very thick, red] (3/8,0) to (3/4,1/2);
        \draw[very thick, red] (3/8,0) to (3/4,-1/2);
        \node[left] at (-3/4,1/2) {$1$}; 
        \node[left] at (-3/4,-1/2) {$2$};
        \node[right] at (3/4,1/2) {$1$};
        \node[right] at (3/4,-1/2) {$2$};
        \node[above] at (0,0) {$3$};
        \fill[red] (-3/8,0) circle (0.07);
        \fill[red] (3/8,0) circle (0.07);
    \end{tikzpicture}
    \bigg|^2\, .
\end{equation}
This is precisely the result that was computed semiclassically in \cite{Chandra:2022bqq}.

\subsection{Twisted \texorpdfstring{$I$}{I}-bundles}
Let us discuss another interesting example which has appeared before in the literature on AdS$_3$ gravity known as a twisted $I$-bundle. It was studied in \cite{Yin:2007at} as a simple example of a non-handlebody saddle-point contribution to the 3d gravity path integral with a single higher-genus boundary. The name stems from the fact that these three-manifolds are constructed as a non-trivial $I$-bundle over a Riemann surface, where $I$ is an interval. Consider a hyperbolic Riemann surface $\Sigma$ together with an orientation-reversing (i.e.\ anti-holomorphic) fixed-point free involution $\Phi:\Sigma \to \Sigma$. We can then consider a quotient of the Euclidean wormhole $\Sigma \times [0,1]$ as follows:
\be 
M_\Phi=(\Sigma \times [0,1])/\{(z,x) \sim (\Phi(z),1-x)\}\ .
\ee
This identification is again orientation-preserving and thus we get an orientable hyperbolic manifold with a single boundary $\Sigma$, where the hyperbolic structure is inherited from the Euclidean wormhole.

$\Phi$ induces an involution on the boundary Teichm\"uller space which we also call $\Phi$ and hence the boundary moduli are constrained to lie on the fixed point set $\mathcal{T}^\Phi$. By the uniformization theorems of three-dimensional hyperbolic manifolds that we reviewed in the Appendix of \cite{Collier:2023fwi}, we are however guaranteed that the manifold with the same topology can also be defined away from the real locus in Teichm\"uller space. The construction then proceeds by taking a quotient of a quasi-Fuchsian wormhole, where the moduli of the left boundary are the image under $\Phi$ of the moduli of the right boundary. 

From the TQFT point of view, it is very simple to determine the Virasoro TQFT partition function on these manifolds. Indeed, we could squash the manifold to the surface $\Sigma \times \{\frac{1}{2}\}$ and the quotient by $\Phi$ simply produces $\widetilde{\Sigma} \times \{\frac{1}{2}\}$. Here,
\be 
\widetilde{\Sigma}=\Sigma/\{z \sim \Phi(z)\}
\ee
is the non-orientable surface obtained from quotienting $\Sigma$. Given that the Virasoro TQFT partition function on the Euclidean wormhole is simply the Liouville partition function, we see that $\Phi$ acts precisely by an orientifold projection. In other words, the partition function on the twisted $I$-bundle is simply the Liouville partition function on the non-orientable surface $\widetilde{\Sigma}$.

To see that this makes sense, recall that the conformal block expansion on a non-orientable surface involves a single conformal block on the doubled surface $\Sigma$ which hence defines a state in the boundary Hilbert space of the twisted $I$-bundle. Let us make this more concrete by recalling the precise construction of Liouville theory on a non-orientable surface. We can construct a non-orientable surface by including a number of cross-caps on an orientable surface.\footnote{Since two crosscaps are equivalent to a handle in the presence of another crosscap, one can restrict to one or two crosscaps.} E.g.\ on a torus with one puncture and a cross-cap, we have
\be 
\begin{tikzpicture}[baseline={([yshift=-.5ex]current bounding box.center)}]
    \begin{scope}
    \begin{scope}
        \draw[very thick] (-2.07,.43) to (-1.93,.57);
        \draw[very thick] (-2.07,.57) to (-1.93,.43);
        \end{scope}
        \draw[very thick, in=180, out=0] (-1.5,1) to (0,.5);
        \draw[very thick, in=180, out=0] (-1.5,-1) to (0,-.5);
        \draw[very thick, in=180, out=180, looseness=2] (-1.5,1) to (-1.5,-1);
        \draw[very thick, bend right=30] (-2,0.05) to (-1,0.05);
        \draw[very thick, bend left=30] (-1.9,0) to (-1.1,0);
        \begin{scope}[xscale=.5]
            \draw[very thick] (0,0) circle (.5);
            \draw[very thick] (-.354,-.354) to (.354,.354);
           \draw[very thick] (-.354,.354) to (.354,-.354);
        \end{scope}
        \end{scope}
    \end{tikzpicture}\ 
    =
\ \begin{tikzpicture}[baseline={([yshift=-.5ex]current bounding box.center)}]
       \begin{scope}
       \begin{scope}
        \draw[very thick] (-2.07,.43) to (-1.93,.57);
        \draw[very thick] (-2.07,.57) to (-1.93,.43);
        \end{scope}
       \begin{scope}
        \draw[very thick] (2.07,-.43) to (1.93,-.57);
        \draw[very thick] (2.07,-.57) to (1.93,-.43);
        \end{scope}
        \draw[densely dashed, thick] (-3.2,0) to (3.2,0);
        \draw[very thick, in=180, out=0] (-1.5,1) to (0,.5) to (1.5,1);
        \draw[very thick, in=180, out=0] (-1.5,-1) to (0,-.5) to (1.5,-1);
        \draw[very thick, in=180, out=180, looseness=2] (-1.5,1) to (-1.5,-1);
        \draw[very thick, in=0, out=0, looseness=2] (1.5,1) to (1.5,-1);
        \draw[very thick, bend right=30] (-2,0.05) to (-1,0.05);
        \draw[very thick, bend left=30] (-1.9,0) to (-1.1,0);
        \draw[very thick, bend left=30] (2,0.05) to (1,0.05);
        \draw[very thick, bend right=30] (1.9,0) to (1.1,0);
        \draw[very thick, bend right=60,->] (2.8,-.5) to (2.8,.5);
        \draw[thick, densely dashed] (0,0) circle (.25 and .5);
        \draw[very thick,<->] (-.5,1) to (.5,1);
        \end{scope}
    \end{tikzpicture}    
\ee
The orientifold acting reflects the right hand side of the picture to the left side and simultaneously rotates by $180$ degrees around the dashed horizontal line. This map has no fixed point and the quotient indeed leads to the crosscap state. On the level of the conformal blocks, this means that the conformal block of the Liouville partition function on this surface takes the form
\begin{multline} 
Z_\text{L}\Bigg(\!\!\!\!\!\begin{tikzpicture}[baseline={([yshift=-.5ex]current bounding box.center)}, scale=.8]
    \begin{scope}
    \begin{scope}
        \draw[very thick] (-2.07,.43) to (-1.93,.57);
        \draw[very thick] (-2.07,.57) to (-1.93,.43);
        \end{scope}
        \node at (-1.7,.5) {$P_0$};
        \draw[very thick, in=180, out=0] (-1.5,1) to (0,.5);
        \draw[very thick, in=180, out=0] (-1.5,-1) to (0,-.5);
        \draw[very thick, in=180, out=180, looseness=2] (-1.5,1) to (-1.5,-1);
        \draw[very thick, bend right=30] (-2,0.05) to (-1,0.05);
        \draw[very thick, bend left=30] (-1.9,0) to (-1.1,0);
        \begin{scope}[xscale=.5]
            \draw[very thick] (0,0) circle (.5);
            \draw[very thick] (-.354,-.354) to (.354,.354);
           \draw[very thick] (-.354,.354) to (.354,-.354);
        \end{scope}
        \end{scope}
    \end{tikzpicture}\Bigg)=\int_0^\infty \!\!\d P_1\, \d P_2\, \d P_3\ \rho_0(P_1) \rho_0(P_2)\Gamma(P_3)C_0(P_0,P_1,P_2)  \\
    \times C_0(P_1,P_2,P_3)\!\! \begin{tikzpicture}[baseline={([yshift=-.5ex]current bounding box.center)}]
       \begin{scope}
        \draw[red, very thick] (-2.4,0) to (-2.1,0);
        \draw[red, very thick] (-1.5,0) circle (.6 and .4);
        \draw[red, very thick] (2.4,0) to (2.1,0);
        \draw[red, very thick] (1.5,0) circle (.6 and .4);
        \draw[red, very thick] (-.9,0) to (.9,0);
        \fill[red] (-2.1,0) circle (.07);
        \fill[red] (-.9,0) circle (.07);
        \fill[red] (.9,0) circle (.07);
        \fill[red] (2.1,0) circle (.07);
       \begin{scope}
        \draw[very thick] (-2.47,-.07) to (-2.33,.07);
        \draw[very thick] (-2.47,.07) to (-2.33,-.07);
        \end{scope}
       \begin{scope}
        \draw[very thick] (2.47,-.07) to (2.33,.07);
        \draw[very thick] (2.47,.07) to (2.33,-.07);
        \end{scope}
        \draw[very thick, in=180, out=0] (-1.5,1) to (0,.5) to (1.5,1);
        \draw[very thick, in=180, out=0] (-1.5,-1) to (0,-.5) to (1.5,-1);
        \draw[very thick, in=180, out=180, looseness=2] (-1.5,1) to (-1.5,-1);
        \draw[very thick, in=0, out=0, looseness=2] (1.5,1) to (1.5,-1);
        \draw[very thick, bend right=30] (-2,0.05) to (-1,0.05);
        \draw[very thick, bend left=30] (-1.9,0) to (-1.1,0);
        \draw[very thick, bend left=30] (2,0.05) to (1,0.05);
        \draw[very thick, bend right=30] (1.9,0) to (1.1,0);
        \node at (-2.35,.3) {$P_0$};
        \node at (2.35,.3) {$P_0$};
        \node at (-1.5,.65) {$P_1$}; 
        \node at (-1.5,-.65) {$P_2$}; 
        \node at (1.5,.65) {$P_2$}; 
        \node at (1.5,-.65) {$P_1$}; 
        \node at (0,.27) {$P_3$};
        \end{scope}
    \end{tikzpicture}\!\! ,
\end{multline}
where the picture represents the ordinary conformal block. The only new ingredient is the normalization of the crosscap state given by $\Gamma(P_3)$. It is fully determined by requiring consistency with the bootstrap. It is in general given by \cite{Bianchi:1990yu}
\be 
\Gamma(P)=\frac{\mathbb{P}_{\id,P}}{\sqrt{\mathbb{S}_{\id,P}}} \times \sqrt{\rho_0(P)}=\mathbb{P}_{\id,P}\ .
\ee
Here, the first factor is the general result when the two-point function of the theory is canonically normalized. We then multiply by $\sqrt{\rho_0(P)}$ to account for our normalization of the two-point function. The $\mathbb{P}$-matrix describes the modular transformation of the M\"obius strip characters:
\be 
\mathbb{P}=\mathbb{T}^{\frac{1}{2}} \mathbb{S} \mathbb{T}^2 \mathbb{S} \mathbb{T}^{\frac{1}{2}}\ .
\ee
It is simple to work this out explicitly:
\be 
\mathbb{P}_{P_1,P_2}=8\int \d P\ \mathrm{e}^{\pi i (P_1^2+P_2^2+4P^2-\frac{1}{4})} \cos(4\pi P_1 P)\cos(4\pi P_2 P)=2 \cos(2\pi P_1 P_2)\ .
\ee
Thus we have
\be 
\Gamma(P)=\mathbb{P}_{\id,P}=\mathbb{P}_{P_1=\frac{i(b^2+1)}{2b},P}+\mathbb{P}_{P_1=\frac{i(b^2-1)}{2b},P}=4 \cosh(\pi b P) \cosh(\pi b^{-1} P)\ .
\ee
The $+$ sign comes from a careful treatment of the factor $\mathbb{T}^{\frac{1}{2}}$ in the definition of the $\mathbb{P}$-matrix; more physically, it comes because the orientifold projection acts by a factor $(-1)^N$ on a level $N$ descendant. This is the same result as obtained in \cite{Hikida:2002bt, Nakayama:2004vk} after translating to our conventions.
This fully specifies the Liouville partition function on any non-orientable surface and hence directly gives the value of $Z_\text{Vir}$ on any twisted $I$-bundle.

Finally, the gravity partition function is given by applying eq.~\eqref{eq:sum over topologies}, 
\be 
Z_\text{grav}(M_\Phi)=\sum_{\gamma \in \Map(\Sigma)/\Map(\widetilde{\Sigma})} |Z_\text{L}(\widetilde{\Sigma}^\gamma)|^2\ ,
\ee
where we used that the bulk mapping class group is the mapping class group of the non-orientable surface $\tilde{\Sigma}$ under which the Liouville partition function is invariant by crossing symmetry.

\section{The figure eight knot complement} \label{sec:figure eight knot}
In this section, we look at one particular hyperbolic 3-manifold in detail and illustrate some features of the theory at this example. The manifold in question is the figure eight knot complement, i.e.\ $\mathrm{S}^3$ with a Wilson line inside forming a figure eight knot. This manifold is known to admit a hyperbolic metric. The figure eight knot is the hyperbolic knot with the smallest possible volume and the only knot with the crossing number 4, as demonstrated 
\begin{figure}[ht]
	\centering
	 \begin{tikzpicture}        
        \draw[very thick, out = 90, in = 180, red] (-5/2,-3/2) to (-1,0);
        \draw[very thick, out = 0, in = 180, red] (-1,0) to (0,0);
        \draw[fill=white,draw=white] (-1,0) circle (1/10);
        \draw[very thick, out = 180, in = 90, red] (0,1) to (-1,0);

        \draw[very thick, out = 270, in = 150, red] (-1,0) to (0,-1);
        \draw[very thick, out = 330, in = 90, red] (0,-1) to (1,-2);
        \draw[fill=white,draw=white] (0,-1) circle (1/10);

        \draw[very thick, out = 90, in = 270, red] (-1,-2) to (1,0);
        \draw[very thick, out = 90, in = 0, red] (1,0) to (0,1);
        \draw[fill=white,draw=white] (1,0) circle (1/10);
        \draw[very thick, out = 0, in = 180, red] (0,0) to (1,0);
        \draw[very thick, out = 0, in = 90, red] (1,0) to (5/2,-3/2);
        \draw[very thick, out = 270, in = 330, red] (5/2,-3/2) to (0,-3);

        \draw[very thick, out = 150, in = 270, red] (0,-3) to (-1,-2);
        \draw[fill=white,draw=white] (0,-3) circle (1/5);

        \draw[very thick, out = 270, in= 0, red] (1,-2) to (0,-3);
        \draw[very thick, out = 180, in = 270, red] (0,-3) to (-5/2,-3/2);
\end{tikzpicture}
\caption{A visualization of the figure eight knot.}
\label{fig:figure-eight}
\end{figure}
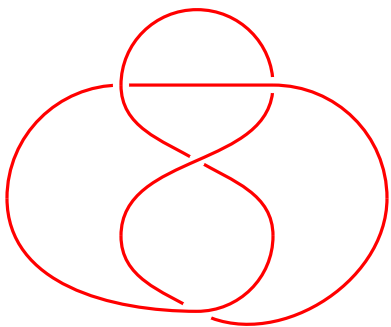
 in Figure \ref{fig:figure-eight}. Thus it is usually denoted as $4_1$. There are two approaches to calculate the Virasoro TQFT partition function of the figure eight knot complement. One way to compute the partition function is via the Heegaard splitting procedure. The other way is to consider the surface bundle construction of the figure eight knot, and to use the mapping torus technique introduced in \cite{Collier:2023fwi}. These two approaches will lead to different integral expressions as the final results. We check that these two expressions agree and both have the same semiclassical expansions as expected.

\subsection{Direct computation}
Let us first compute the partition function by successively undoing the over- and under-crossings in a particular projection of the knot.

 We start by computing the partition function via surgery. We embed the above knot configuration into a three-sphere to create the figure eight knot complement. In the TQFT setup, we consider the knot as a tangled Wilson loop with associated conformal weight $\Delta_0 = \frac{Q^2}{4}$, i.e. the cusp, although we will keep the label of the Wilson loop generic for most of the discussion. If we slice the above figure \ref{fig:figure-eight} into halves along the equatorial $\mathrm{S}^2$, we obtain two manifolds $M_1, M_2$ with boundaries as four-punctured sphere. The path integral over each half prepares a state in the Hilbert space $\mathcal{H}_{\Sigma_{0,4}}$, and the partition function is the inner product between these two sphere 4-point conformal blocks. Here the Wilson lines inside each component have nontrivial braidings. Before evaluating the inner product, we want to untangle the Wilson lines. For this purpose, we need to apply the crossing and braiding operations on the boundary surface $\Sigma_{0,4}$.
 
To make the crossing and braiding explicit, we firstly specify the intermediate channels in the figure eight knot.
 \begin{figure}[ht]
 \centering
 	\begin{tikzpicture}    
        \draw[thick, blue, out= 50, in = 130, densely dashed, looseness=.7] (-3/2,-2) to (3/2,-2);
        \draw[very thick, out = 90, in = 180, red] (-5/2,-3/2) to (-1,0);
        \draw[very thick, out = 0, in = 180, red] (-1,0) to (0,0);
        \draw[fill=white,draw=white] (-1,0) circle (1/10);
        \draw[very thick, out = 180, in = 90, red] (0,1) to (-1,0);

        \draw[very thick, out = 270, in = 150, red] (-1,0) to (0,-1);
        \draw[very thick, out = 330, in = 90, red] (0,-1) to (1,-2);
        \draw[fill=white,draw=white] (0,-1) circle (1/10);

        \draw[very thick, out = 90, in = 270, red] (-1,-2) to (1,0);
        \draw[very thick, out = 90, in = 0, red] (1,0) to (0,1);
        \draw[fill=white,draw=white] (1,0) circle (1/10);
        \draw[very thick, out = 0, in = 180, red] (0,0) to (1,0);

        \draw[very thick, out = 0, in = 90, red] (1,0) to (5/2,-3/2);
        \draw[very thick, out = 270, in = 330, red] (5/2,-3/2) to (0,-3);

        \draw[very thick, out = 150, in = 270, red] (0,-3) to (-1,-2);
        \draw[fill=white,draw=white] (0,-3) circle (1/5);

        \draw[very thick, out = 270, in= 0, red] (1,-2) to (0,-3);
        \draw[very thick, out = 180, in = 270, red] (0,-3) to (-5/2,-3/2);

        \draw[very thick, densely dashed, red] (0,1) to (0,0);
        \draw[very thick, densely dashed, red] (-1,-2) to (1,-2);
        \fill[red] (0,0) circle (.07);
        \fill[red] (0,1) circle (.07);
        \fill[red] (-1,-2) circle (.07);
        \fill[red] (1,-2) circle (.07);

        \node[left] at (0,.6) {$\mathds{1}$}; 
        \node[below] at (0,-2) {$\mathds{1}$};

        \draw[very thick,blue] (0,-2) circle (3/2);
        \draw[very thick, blue, out = -50, in = 230, looseness=.7] (-3/2,-2) to (3/2,-2);
\end{tikzpicture}
\caption{Heegaard splitting of the figure eight knot complement.}
\label{fig:figure-eight-intermediate}
 \end{figure}
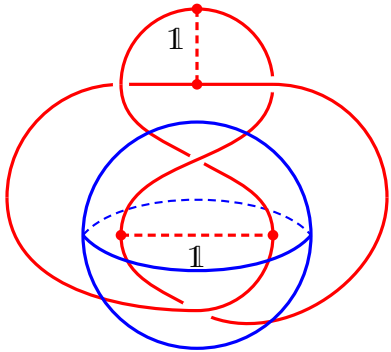
In the diagram, we have identity operators propagate in the intermediate channels corresponding to the contractible cycles in the bulk. We can use the fusion kernel $\mathbb{F}$ to transform the diagram \ref{fig:figure-eight-intermediate} into the other channel
\vspace{3.5ex}
\setlength{\jot}{-3ex}
    \begin{multline}
		Z_{\text{Vir}}\Biggl(\begin{tikzpicture}[baseline={([yshift=-.5ex]current bounding box.center)},scale=.6]        
        \draw[very thick, out = 90, in = 180, red] (-5/2,-3/2) to (-1,0);
        \draw[very thick, out = 0, in = 180, red] (-1,0) to (0,0);
        \draw[fill=white,draw=white] (-1,0) circle (1/10);
        \draw[very thick, out = 180, in = 90, red] (0,1) to (-1,0);
        \draw[very thick, out = 270, in = 150, red] (-1,0) to (0,-1);
        \draw[very thick, out = 330, in = 90, red] (0,-1) to (1,-2);
        \draw[fill=white,draw=white] (0,-1) circle (1/10);
        \draw[very thick, out = 90, in = 270, red] (-1,-2) to (1,0);
        \draw[very thick, out = 90, in = 0, red] (1,0) to (0,1);
        \draw[fill=white,draw=white] (1,0) circle (1/10);
        \draw[very thick, out = 0, in = 180, red] (0,0) to (1,0);
        \draw[very thick, out = 0, in = 90, red] (1,0) to (5/2,-3/2);
        \draw[very thick, out = 270, in = 330, red] (5/2,-3/2) to (0,-3);
        \draw[very thick, out = 150, in = 270, red] (0,-3) to (-1,-2);
        \draw[fill=white,draw=white] (0,-3) circle (1/5);
        \draw[very thick, out = 270, in= 0, red] (1,-2) to (0,-3);
        \draw[very thick, out = 180, in = 270, red] (0,-3) to (-5/2,-3/2);
        \draw[very thick, densely dashed, red] (0,1) to (0,0);
        \draw[very thick, densely dashed, red] (-1,-2) to (1,-2);
        \fill[red] (0,0) circle (.1);
        \fill[red] (0,1) circle (.1);
        \fill[red] (-1,-2) circle (.1);
        \fill[red] (1,-2) circle (.1);
        \node[left] at (0.2,1/2) {$\mathds{1}$}; 
        \node[below] at (0,-2) {$\mathds{1}$};
        \node[right] at (2.1,-1/2) {$P_0$};   
\end{tikzpicture}\Biggr) = \int \d P_s\,\d P_t\, \mathbb{F}_{\mathds{1},P_s}\begin{bmatrix}
	P_0 & P_0\\
	P_0 & P_0
\end{bmatrix}\,\mathbb{F}_{\mathds{1},P_t}\begin{bmatrix}
	P_0 & P_0\\
	P_0 & P_0
\end{bmatrix} \\
\times Z_{\text{Vir}}\Biggl(\begin{tikzpicture}[baseline={([yshift=-.5ex]current bounding box.center)}, scale=.6]   
        \draw[very thick, out = 270, in = 150, red] (-3/8,-5/2-1/8) to (0,-3);
        \draw[very thick, out = 90, in = 339, red] (3/8,-5/4-1/4+1/8) to (0,-1);
        \draw[very thick, red] (-1,0) to (-5/8,0);
        \draw[very thick, red] (1,3/8) to (1,0);
        \draw[very thick, out = 90, in = 180, red] (-5/2,-3/2) to (-1,0);
        \draw[fill=white,draw=white] (-1,0) circle (1/10);
        \draw[very thick, out = 270, in = 150, red] (-1,0) to (0,-1);
        \draw[fill=white,draw=white] (0,-1) circle (1/10);
        \draw[very thick, out= 30, in = 270, red] (0,-1) to (1,0);
        \draw[fill=white,draw=white] (1,0) circle (1/10); 
        \draw[very thick, out = 0, in = 90, red] (1,0) to (5/2,-3/2);
        \draw[very thick, out = 270, in = 330, red] (5/2,-3/2) to (0,-3);
        \draw[fill=white,draw=white] (0,-3) circle (1/8);
        \draw[very thick, out = 180, in = 270, red] (0,-3) to (-5/2,-3/2);
        \draw[very thick, out = 0, in = 270, red] (-5/8,0) to (-1/4,3/8);
        \draw[very thick, out = 90, in = 0, red] (-1/4,3/8) to (-5/8,3/4);
        \draw[very thick, out = 180, in = 90, red] (-5/8,3/4) to (-1,3/8);
        \draw[very thick, red] (-1,3/8) to (-1,0);       
        \draw[very thick, out = 180, in = 270, red] (5/8,0) to (1/4,3/8);
        \draw[very thick, out = 90, in = 180, red] (1/4,3/8) to (5/8,3/4);
        \draw[very thick, out = 0, in = 90, red] (5/8,3/4) to (1,3/8);
        \draw[very thick, red] (1,0) to (5/8,0);
        \draw[very thick, red] (-1/4,3/8) to (1/4,3/8);
        \node[above] at (0,3/8) {$P_t$};
        \draw[very thick,out = 210, in=90, red] (0,-1) to (-3/8,-3/2+1/8);
        \draw[very thick, out = 270, in = 180, red] (-3/8,-3/2+1/8) to (0,-15/8+1/8);
        \draw[very thick, out = 0, in = 270, red] (0,-15/8+1/8) to (3/8,-3/2+1/8);
        \draw[very thick, out = 0, in = 270, red] (0,-3) to (3/8,-5/2-1/8);
        \draw[very thick, out = 90, in = 0, red] (3/8,-5/2-1/8) to (0,-5/2+3/8-1/8);
        \draw[very thick, out = 180, in = 90, red] (0,-5/2+3/8-1/8) to (-3/8,-5/2-1/8);
        \draw[very thick, red] (0,-5/2+3/8-1/8) to (0,-15/8+1/8);
        \fill[red] (0,-5/2+3/8-1/8) circle (.1);
        \fill[red] (0,-15/8+1/8) circle (.1);
        \fill[red] (-1/4,3/8) circle (.1);
        \fill[red] (1/4,3/8) circle (.1);
        \node[right] at (0,-2) {$P_s$};
        \node[right] at (2.1,-1/2) {$P_0$};
\end{tikzpicture}\Biggr)\ ,
	\end{multline}
\setlength{\jot}{.5ex}%
where $P_0$ labels the conformal weight of the Wilson loop, i.e. $P_0 =0$ for $\Delta = Q^2/4$. After transforming the figure eight knot diagram into the other channel, we can untangle the knot at each trivalent node via the braiding move $\mathbb{B}$ as follows
\vspace{3.5ex}
\setlength{\jot}{-3ex}
\begin{multline}
		Z_{\text{Vir}}(4_1)
 = \int \d P_s\, \d P_t\,  \mathbb{F}_{\mathds{1},P_s}\begin{bmatrix}
	P_0 & P_0\\
	P_0 & P_0
\end{bmatrix}\,\mathbb{F}_{\mathds{1},P_t}\begin{bmatrix}
	P_0 & P_0\\
	P_0 & P_0
\end{bmatrix}(\mathbb{B}_{P_s}^{P_0,P_0})^2\\
\times(\mathbb{B}_{P_t}^{P_0,P_0})^{-2}
Z_\text{Vir}\Bigg(\begin{tikzpicture}[baseline={([yshift=-.5ex]current bounding box.center)}, scale=1.2]
        \draw[very thick, red] (0,0) to (0,-1);
        \draw[very thick, red] (0,0) to (.866,.5);
        \draw[very thick, red] (0,0) to (-.866,.5);
        \draw[very thick, red] (-.866,.5) to node[above, black] {$P_t$} (.866,.5);
        \draw[very thick, red] (0,-1) to node[right, black] {$P_0$} (.866,.5);
        \draw[very thick, red] (0,-1) to node[left, black] {$P_0$} (-.866,.5);
        \fill[red] (0,0) circle (0.06);
        \fill[red] (0,-1) circle (0.06);
        \fill[red] (.866,.5) circle (0.06);
        \fill[red] (-.866,.5) circle (0.06);
        \node at (.2,-.2) {$P_s$};
\end{tikzpicture}
\Bigg)\ .
\end{multline}
\setlength{\jot}{.5ex}
The fusion kernel $\mathbb{F}$ corresponding to the exchange of the identity operator can be written in terms of $\rho_0$ and $C_0$ as follows
\begin{equation}
	\mathbb{F}_{\mathds{1},P}\begin{bmatrix}
		P_0 & P_0\\
		P_0 & P_0
	\end{bmatrix} = \rho_0(P)C_0(P,P_0,P_0)\ .
\end{equation}
Meanwhile, we recognize the remaining contraction as the four-boundary wormhole discussed in Section~\ref{subsec:four-boundary wormhole} for which we can use the result \eqref{eq:four boundary wormhole partition function}, normalized by inverse structure constants to account for the normalization of the junctures. 

In the end, we obtain an integral expression of the figure eight knot partition function
\begin{equation}
\label{eq:figure-eight-heegaard}
	Z_{\text{Vir}}(4_1) = \int \d P_s\,\d P_t\, \rho_0(P_s) \rho_0(P_t)\,(\mathbb{B}_{P_s}^{P_0,P_0})^2(\mathbb{B}_{P_t}^{P_0,P_0})^{-2}\,\begin{Bmatrix} P_0 & P_0 & P_s \\ P_0 & P_0 & P_t \end{Bmatrix}\ .
\end{equation}

There are two momentum integrals in the above formula \eqref{eq:figure-eight-heegaard}, and we can reduce the number of integrals by one by using the relation \eqref{eq:FS relation} between the fusion kernel $\mathbb{F}$ and the modular S-matrix $\mathbb{S}$.  We hence get
\begin{align} 
Z_\text{Vir}(4_1)&=\int_0^\infty \d P\ \frac{\rho_0(P)}{\rho_0(P_0)}\, \mathrm{e}^{\frac{\pi i Q^2}{4}-3\pi i P^2} \, \mathbb{S}_{P_0,P_0}[P]\\
&=\int_0^\infty \d P\  \frac{\rho_0(P)\,\mathrm{e}^{\frac{3\pi i Q^2}{8}-\frac{5\pi i P^2}{2}}}{S_b(\frac{Q}{2}+i P)} \int_{-\infty}^\infty \d x \ \mathrm{e}^{-4\pi i x P_0}\, S_b(\tfrac{Q}{4}+\tfrac{i P}{2} \pm i P_0\pm i x)\ ,
\end{align}
where we inserted the explicit expression for the modular crossing kernel in the second line \cite{Teschner:2013tqy}.

Of course, this expression is dependent on the framing that we implicitly chose in this computation. For the figure eight knot complement, a nice way to fix the framing anomaly is by requiring that the partition function should be real. Indeed, complex conjugation corresponds to orientation reversal, but since the figure eight knot is invariant under orientation reversal (this property is called amphichirality), we can choose the partition function to be real.

One can easily check, for example numerically, that this is the case if we multiply the above expression with $\mathrm{e}^{4\pi i P_0^2}=\mathrm{e}^{4\pi i (\Delta_0-\frac{c}{24})}$, which is part of the ambiguity from framing. We hence have
\be 
Z_\text{Vir}(4_1)=\int_0^\infty \d P\  \frac{\rho_0(P)\,\mathrm{e}^{\frac{3\pi i Q^2}{8}+4\pi i P_0^2-\frac{5\pi i P^2}{2}}}{S_b(\frac{Q}{2}+i P)} \int_{-\infty}^\infty \d x \ \mathrm{e}^{-4\pi i x P_0}\, S_b(\tfrac{Q}{4}+\tfrac{i P}{2} \pm i P_0\pm i x)\ , \label{eq:figure eight knot partition function Virasoro TQFT}
\ee
which is the formula we will use from now on.

\paragraph{Choice of contour.} There is one additional subtlety with this formula. As it stands, the integral over $P$ is actually not convergent. Indeed, using the asymptotics of the double sine function, see e.g. \cite[eq.~(B.53)]{Eberhardt:2023mrq} and using that the integral over $x$ is dominated for small $x$, we see that
\be 
\int_{-\infty}^\infty \d x \ \mathrm{e}^{-4\pi i x P_0}\, S_b(\tfrac{Q}{4}+\tfrac{i P}{2} \pm i P_0\pm i x) \sim \frac{1}{\sqrt{2}}\, \mathrm{e}^{\frac{\pi i P^2}{2}-\frac{\pi P Q}{2}+\frac{\pi i Q^2}{24}-\frac{\pi i}{12}}\ .
\ee
Combining this with the asymptotics of the rest of the integrand, we see that the integrand behaves for large $\Re(P)$ as
\begin{align}
    \text{integrand}(P) \sim \mathrm{e}^{\frac{3 \pi Q P}{2}-\frac{5\pi i P^2}{2}}\times \mathcal{O}(\text{order 1 in $P$})\ .
\end{align}
Thus the integral in \eqref{eq:figure eight knot partition function Virasoro TQFT} doesn't converge for $P$ on the real axis. However, we see that we could have improved convergence by taking $P$ to run along a contour starting at $P=0$ and asymptoting for large $P$ the line $\RR-i a$, where the shift $a$ has to be at least $a>\frac{3Q}{10}$ to ensure convergence. Shifting the contour in this way doesn't cross any poles and is hence a generally harmless operation.
Thus it is understood that the integral over $P$ in \eqref{eq:figure eight knot partition function Virasoro TQFT} actually follows this modified contour.

\subsection{Comparison to Teichm\"uller TQFT}
The figure eight knot partition function can also be obtained in Teichm\"uller TQFT developed in \cite{Dimofte:2009yn, Dijkgraaf:2010ur, Dimofte:2011gm, EllegaardAndersen:2011vps}. Translating to our conventions, the expression for the Teichm\"uller TQFT partition function is\footnote{Teichm\"uller TQFT depends on a parameter $\hbar$, which, following the conventions of \cite{Dijkgraaf:2010ur}, we identify as $\hbar=-i \pi b^2$. This expression does not literally match the one given in \cite{Dimofte:2009yn, Dijkgraaf:2010ur, Dimofte:2011gm, EllegaardAndersen:2011vps}. We are unsure whether this is a typo in the previous literature. In any case, the semiclassical expansion that we discuss below \emph{does} match previous expressions, which gives us a lot of confidence in the correctness of \eqref{eq:figure eight knot partition function Teichmuller TQFT}. We thank Boris Post and Davide Saccardo for discussions about this.
}
\be 
Z_\text{Teich}(4_1)=\sqrt{2} \int_{\RR-i 0^+} \!\!\!\d x\ S_b(ix\pm 2i P_0)\ ,\label{eq:figure eight knot partition function Teichmuller TQFT}
\ee
Here the integral runs slightly below the real axis to avoid the poles at $x=\pm 2P_0$.
This formula can be obtained by realizing the figure eight knot complement as a gluing of two tetrahedra. Each tetrahedron gives rise to one double sine function and the gluing to the integral (modulo some constraints).

As we already conjectured in our previous paper \cite{Collier:2023fwi}, we expect that Virasoro TQFT is equivalent to Teichm\"uller TQFT and thus the two expressions should match,
\be 
Z_\text{Vir}(4_1) \overset{!}{=} Z_\text{Teich}(4_1)\ . \label{eq:figure eight knot Virasoro Teichmuller partition functions equality}
\ee
This equality turns out to be quite hard to prove analytically. However, we checked numerically for various values of $b$ and $P_0$ that the two expressions agree. 

\begin{figure}[ht]
    \centering
    \includegraphics[width=.9\textwidth]{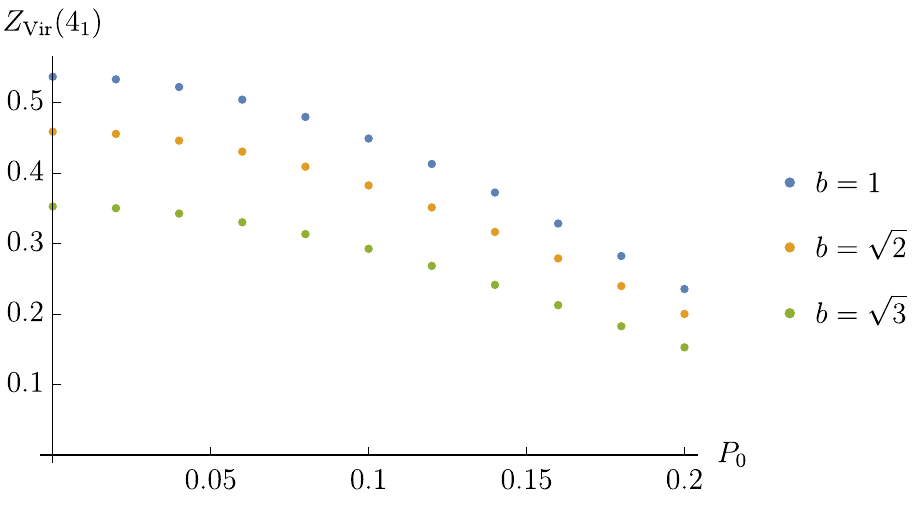}
    \caption{Numerical comparison of the Virasoro TQFT and the Teichm\"uller TQFT partition function of the figure eight knot complement. The plotted data points are for the Teichm\"uller expression \eqref{eq:figure eight knot partition function Teichmuller TQFT}, but are indistinguishable from the Virasoro data points.}
    \label{fig:numerical comparison Virasoro Teichmuller TQFT}
\end{figure}

The numerical evaluation is in principle straightforward. We restricted our attention to rational values of $b^2$, since in this case, there is a simple way to express the double sine function through the Barnes G-function for which we can use efficient implementations, for example in \texttt{Mathematica},
\be 
S_b(z)=(2\pi)^{\sqrt{mn}z-\frac{m+n}{2}}\prod_{k=0}^{m-1} \prod_{\ell=0}^{n-1} \frac{G\big(\frac{k+1}{m}+\frac{\ell+1}{n}-\frac{z}{\sqrt{mn}}\big)}{G\big(\frac{k}{m}+\frac{\ell}{n}+\frac{z}{\sqrt{mn}}\big)}\ .
\ee
It is then simple to compute the required integrals in \eqref{eq:figure eight knot partition function Virasoro TQFT} over a converging contour and compare with the simpler expression \eqref{eq:figure eight knot partition function Teichmuller TQFT}.
We computed the partition functions for $b=1$, $b=\sqrt{2}$ and $b=\sqrt{3}$ for $P_0=0,0.02,\dots,0.2$. To the precision we have computed, all values agree to seven decimal places, thus showing the equality \eqref{eq:figure eight knot Virasoro Teichmuller partition functions equality} beyond reasonable doubt. The data points are plotted in Figure~\ref{fig:numerical comparison Virasoro Teichmuller TQFT}.

From this discussion, it may seem that the Teichm\"uller TQFT always produces simpler expressions than Virasoro TQFT, but this is not the case. The expressions in Teichm\"uller TQFT become more complicated when the 3-manifold in question requires more tetrahedra to form a triangulation, while this is not necessarily so in Virasoro TQFT. It is in general quite hard to recognize when two integral representations of the partition function agree since there are an enormous number of non-trivial integral identities relating them. 

\subsection{Computation via the Seifert surface} \label{subsec:Seifert surface figure eight knot}
Let us explain a completely different way to compute the partition function that will lead to an inequivalent integral for the partition function.

\begin{figure}[ht]
    \centering
    \begin{tikzpicture}
        \fill[blue!30!white] (-.5,-1.25) to[out=-30, in=180] (0,-1.5) to[out=0, in=30] (.5,-1.25) to[out=150, in=0] (0,-1) to[out=180, in=30] (-.5,-1.25);
    \fill[red!30!white] (-1,1) to[out=180, in=90] (-2,0) to[out=-90, in=180] (-1,-1) to[out=0, in=150] (-.5,-1.25) to[out=210, in=0] (-1,-1.5) to[out=180, in=-90] (-2.5,0) to[out=90, in=180] (-1,1.5);
    \fill[red!30!white] (1,1) to[out=0, in=90] (2,0) to[out=-90, in=0] (1,-1) to[out=190, in=30] (.5,-1.25) to[out=-30, in=180] (1,-1.5) to[out=0, in=-90] (2.5,0) to[out=90, in=0] (1,1.5);
    \fill[red!30!white] (-1,1) to[out=0, in=60] (0,.5) to[out=240, in=-90] (-.5,.55) to[out=90, in=-45] (-.75,.9) to[out=-135, in=90] (-1,.55) to[out=-90, in=-120] (.5,.5) to[out=60, in=180] (1,1) to (1,1.5) to[out=180, in=-60] (.5,2) to[out=120, in=90] (-1,1.95) to[out=-90, in=135] (-.75,1.6) to[out=45, in=-90] (-.5,1.95) to[out=90, in=120] (0,2) to[out=-60, in=0] (-1,1.5);
        \draw[very thick] (0,.5) to[out=60, in=0] (-1,1) to[out=180, in=90] (-2,0) to[out=-90, in=180] (-1,-1) to[out=0, in=180] (0,-1.5);
        \fill[white] (-.5,-1.25) circle (.06);
        \draw[very thick] (0,2) to[out=-60, in=0] (-1,1.5) to[out=180, in=90] (-2.5,0) to[out=-90, in=180] (-1,-1.5) to[out=0, in=180] (0,-1);
        \draw[very thick] (0,-1) to[out=0, in=180] (1,-1.5) to[out=0, in=-90] (2.5,0) to[out=90, in=0] (1,1.5) to[out=180, in=-60] (.5,2);
        \fill[white] (.5,-1.25) circle (.06);
        \draw[very thick] (0,-1.5) to[out=0, in=180] (1,-1) to[out=0, in=-90] (2,0) to[out=90, in=0] (1,1) to[out=180, in=60] (.5,.5);
        \fill[blue!30!white] (-.75,.9) to[out=45, in=-90] (-.5,1.25) to[out=90, in=-45] (-.75,1.6) to[out=-135, in=90] (-1,1.25) to[out=-90, in=135] (-.75,.9);
        \draw[very thick] (0,.5) to[out=240, in=-90] (-.5,.55) to[out=90, in=-90] (-1,1.25);
        \fill[white] (-.75,.9) circle (.06);
        \draw[very thick] (.5,2) to[out=120, in=90] (-1,1.95) to[out=-90, in=90] (-.5,1.25);
        \fill[white] (-.75,1.6) circle (.06);
        \draw[very thick] (-1,1.25) to[out=90, in=-90] (-.5,1.95) to[out=90, in=120] (0,2);
        \draw[very thick] (-.5,1.25) to[out=-90, in=90] (-1,.55) to[out=-90, in=-120] (.5,.5);
    \end{tikzpicture}
    \caption{The Seifert surface of the figure eight knot. One can easily verify that the boundary of the Seifert surface traces out a figure eight knot.}
    \label{fig:figure eight knot Seifert surface}
\end{figure}
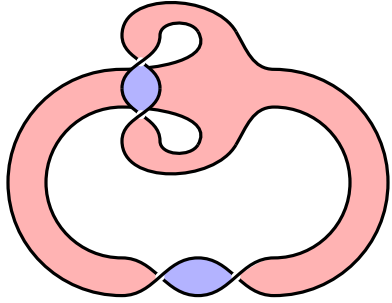

The figure eight knot admits a genus 1 Seifert surface. This means that we can realize the knot as the boundary of a one-holed torus embedded in $\mathrm{S}^3$, so that the boundary of the one-holed torus coincides with knot. This is depicted in Figure~\ref{fig:figure eight knot Seifert surface}. 
However, even more is true. One can slightly deform the Seifert surface and obtain a foliation of the knot complement in terms of one-holed tori. The figure eight knot complement is in fact a surface bundle over a circle, i.e.\ it is of the form
\be 
[0,1] \times \Sigma_{1,1}/\sim\ ,
\ee
where $\Sigma_{1,1}$ is the one-holed torus and we identify
\be 
(0,z) \sim (1,\phi(z))
\ee
with $\phi=S T^{3}$ being the corresponding mapping class group element in $\mathrm{SL}(2,\ZZ)$ generated by $S$ and $T$.

This might let one suspect that we can compute the partition function of the figure eight knot complement as
\be 
Z_\text{Vir}(4_1)\overset{?}{=}\tr_{\mathcal{H}_{1,1}}(\mathbb{S}[P_0]\,  \mathbb{T}^{3})\ ,
\ee
but this is not quite correct yet. Indeed, taking the trace over the Hilbert space $\mathcal{H}_{1,1}$ of conformal blocks on the once-punctured torus would lead to the partition function of the three-dimensional manifold where the Wilson line runs along the thermal circle $\mathrm{S}^1$. This is not what we want, since the Wilson line bounds the Seifert surface, which forms the meridian of the boundary torus of the manifold. This means that the correct expression is obtained by applying the S-modular transformation in the external parameter $P_0$. So we conclude that we should have
\be 
Z_\text{Vir}(4_1)=
\int_0^\infty \d P_0' \ \mathbb{S}_{P_0,P_0'}[\mathds{1}]\, \tr_{\mathcal{H}_{1,1}}(\mathbb{S}[P_0']\,  \mathbb{T}^{3})\ . \label{eq:figure eight knot Seifert surface partition function}
\ee
We can easily plug in the explicit expressions for the modular crossing kernel and get an alternative expression for the partition function of the figure eight knot complement.
This expression is even more unwieldy then the previous ones, since it involves three integrals, one from the definition of $\mathbb{S}$, one from the trace, and one from the integral over $P_0'$. This pushes our numerical capabilities a bit too far. Instead, we will check below that the first two terms in the semiclassical expansion agree with the semiclassical expansion of the previous expression.\footnote{We also checked that the corresponding expressions for the figure eight knot partition function in $\mathrm{SU}(2)_k$ Chern-Simons theory agree where all the integrals are just finite sums.}

We note that this expression makes reality of the partition function manifest, while it was obscured in the expression \eqref{eq:figure eight knot partition function Virasoro TQFT} that we discussed above. Indeed, one of the Moore-Seiberg relations states that $\mathbb{T}\mathbb{S}\mathbb{T}\mathbb{S}\mathbb{T}=\mathbb{S}$ as operators (see \cite[eq.~(A.5b)]{Collier:2023fwi}) and thus
\be 
\tr(\mathbb{S}  \mathbb{T}^{3})^*=\tr(\mathbb{T}^{-1}\mathbb{S}^{-1}  \mathbb{T}^{-2})=\tr(\mathbb{S} \mathbb{T}\mathbb{S}^{-1}\mathbb{T}^{-1})=\tr(\mathbb{S} \mathbb{T}^2\mathbb{S}\mathbb{T}\mathbb{S}^{-1})=\tr(\mathbb{S}\mathbb{T}^3)\ ,
\ee
and so after Fourier transformation we still get a real function.

\subsection{Semiclassical expansion}
We now write down the semiclassical expansion of the Virasoro TQFT partition function in the form \eqref{eq:figure eight knot partition function Teichmuller TQFT} and check the volume conjecture explicitly. This was already done before in the context of Teichm\"uller TQFT \cite{Dimofte:2009yn, Dijkgraaf:2010ur} and hence we shall be rather brief.

The key identity is the semiclassical expansion of the double sine function,
\begin{multline} 
\log S_b(x_0+x)=\sum_{n=0}^\infty \frac{(2\pi i b^2)^{n-1}}{2n!}\\
\times\Big(\Li_{2-n}\big(\mathrm{e}^{-2\pi i b x_0}\big)-(-1)^{n} \Li_{2-n}\big(\mathrm{e}^{2\pi i b x_0}\big)\Big)\, B_n\big(1-\tfrac{x}{b}\big)\ . \label{eq:double sine function semiclassical expansion}
\end{multline}
In this identity we think of $x_0$ as being of order $\mathcal{O}(\frac{1}{b})$, while $x$ is of order $\mathcal{O}(1)$.
This identity is standard for the quantum dilogarithm to which the double sine function is closely related, see e.g.~\cite[Proposition 6]{EllegaardAndersen:2011vps}. For completeness, we have included a short derivation in appendix~\ref{app:semiclassical expansion double sine}.

We now apply this expansion as follows. In the semiclassical limit, the argument of the double sine function in eq.~\eqref{eq:figure eight knot partition function Teichmuller TQFT} becomes large and we write $P_0=\frac{\eta_0}{b}$. We can then evaluate the integral via saddle point approximation. We write $x$ to leading order as $\frac{x_0}{b}$. Then the saddle-point equation is
\begin{align} 
0&=-\frac{1}{2\pi}\partial_{x_0} \sum_{\pm} \left(\Li_2(\mathrm{e}^{2\pi (x_0\pm 2\eta_0)})-\Li_2(\mathrm{e}^{-2\pi (x_0\pm 2\eta_0)})\right) \\
&=\log\prod_{\pm,\pm}\big(1-\mathrm{e}^{2\pi (\pm x_0 \pm 2\eta_0)}\big)\ .
\end{align}
The solution to this saddlepoint equation takes the form
\be 
x_0=\frac{1}{2\pi} \log \Big(\cosh(4\pi \eta_0)\pm_1\tfrac{1}{2}\pm_2\sqrt{\big(\cosh(4\pi \eta_0)\pm_1\tfrac{1}{2}\big)^2-1}\Big)+n i\ , \quad n \in \ZZ\ .
\ee
The steepest descent contour runs through the saddle point at
\be 
x_0=\frac{1}{2\pi} \log \Big(\cosh(4\pi \eta_0)-\tfrac{1}{2}-\sqrt{\big(\cosh(4\pi \eta_0)-\tfrac{1}{2}\big)^2-1}\Big)
\ee
and hence only that one is relevant for our analysis. For this to be valid, we should assume that
\be 
|\eta_0|<-\frac{1}{2\pi} \log\Big(\frac{\sqrt{5}-1}{2}\Big)\ ,
\ee
since otherwise $x_0$ becomes real and the saddle-point evaluation is different.
We obtain the semiclassical expansion
\begin{align}
\label{eq:figure eight expansion}
    Z_\text{Vir}(4_1)=\frac{\sqrt{2}\, \mathrm{e}^{-\frac{1}{2\pi b^2} \vol(4_1)}}{\Delta^\frac{1}{4}}\, \exp\Big(\sum_{n=1}^\infty S_n b^{2n}\Big)\ ,
\end{align}
where
\be 
\Delta=-X^2+2X+1+2X^{-1}-X^{-2}\ , \qquad X=\mathrm{e}^{4\pi \eta_0}\ ,
\ee
and
\begin{multline} 
\vol(4_1,\eta_0)=-\frac{i}{2} \Big(\Li_2\big(\tfrac{X-1+X^{-1}+i \sqrt{\Delta}}{2X}\big)+\Li_2\big(\tfrac{X-1+X^{-1}+i \sqrt{\Delta}}{2} X\big) \\
-\Li_2\big(\tfrac{X-1+X^{-1}-i \sqrt{\Delta}}{2X}\big)-\Li_2\big(\tfrac{X-1+X^{-1}-i \sqrt{\Delta}}{2} X\big)\Big)\ . \label{eq:figure eight knot complement volume nonzero eta}
\end{multline}
The first few orders for the higher loop corrections are given by
\begin{subequations}
\begin{align}
    S_1&=-\frac{\pi}{12\Delta^{\frac{3}{2}}}(X^3-X^2-2X^2+15-2X^{-1}-X^{-2}+X^{-3})\ , \\
    S_2&=\frac{2\pi^2}{\Delta^3}(X^3-X^2-2X+5-2X^{-1}-X^{-2}+X^{-3})\ , \\
    S_3&=\frac{\pi^3}{90\Delta^{\frac{9}{2}}}(X^{8}-4X^{7}-128X^{6}+36X^{5}+1074 X^{4}-5630 X^{3}+5782X^{2}\nonumber\\
    &\qquad+7484X^1-18311+7484X^{-1}+5782X^{-2}-5630X^{-3}+1074X^{-4}+36X^{-5}\nonumber\\
    &\qquad-128X^{-6}-4X^{-7}+X^{-8})\ .
\end{align}
\end{subequations}
Not surprisingly, this reproduces the semiclassical expansion given in \cite{Dimofte:2009yn, Dimofte:2011gm, Dijkgraaf:2010ur}. Also noticed there, the one-loop determinant equals the Reidemeister torsion of the figure eight knot, which can also be derived from computing the functional analytic one-loop determinants appearing in 3d gravity. Thus this shows the validity of the volume conjecture \eqref{eq:refined volume conjecture} for the figure eight knot. 
\paragraph{Expression from Seifert surface.} We now reproduce the semiclassical expansion from the expression that we got from the computation via the Seifert surface as described in section~\ref{subsec:Seifert surface figure eight knot}. This gives strong evidence that the expression \eqref{eq:figure eight knot Seifert surface partition function} is in fact equal to the simpler expression given by eq.~\eqref{eq:figure eight knot partition function Teichmuller TQFT}.

By using the explicit formula of the Virasoro crossing kernel shown in \cite{Collier:2023fwi}, we rewrite the integral formula \eqref{eq:figure eight knot Seifert surface partition function} in terms of double-sine functions
\begin{multline}
    Z_{\text{Vir}}(4_1)= 2\sqrt{2}\int_0^\infty \d P_0'\, \cos{4\pi P_0 P_0'}\int_0^\infty \d P\, \rho_0(P)\\
    \times  \int_{-\infty}^{\infty} \d\xi\ \frac{\mathrm{e}^{\frac{\pi i\Delta_0'}{2}-6\pi i (P^2-\frac{1}{24})-4\pi i \xi P}}{S_b(\tfrac{Q}{2}+i P_0')}S_b(\tfrac{Q}{4}+\tfrac{iP_0'}{2}\pm i P\pm i\xi)\ .
\end{multline}
For simplicity, in the following computation, we consider $P_0 = 0$ which sets the conformal weight of the knot to be $\Delta = \frac{Q^2}{4}$. As we will see later, the saddle-point equation in the semiclassical approximation will be simplified in this case. In general, we can also compute the partition function for the knot with a generic conformal weight, while the complexity of solving the saddle-point equations increases. Once we consider the semiclassical limit of this expression, we similarly rescale $P_0' = \frac{\eta_0}{b}$, $P = \frac{x}{b}$ and $\xi = \frac{\eta}{b}$. Then we apply the expansion formula of the double sine function to write the integrand into a expansion in $1/b^2$. 
\begin{equation}
    Z_{\text{Vir}}(4_1) = \int \frac{\d\eta_0\,\d x\, \d\eta}{b^3}\, \mathrm{e}^{\sum_{n=0}^\infty S^{(n)}b^{2(n-1)}}\ . \label{eq:Figure eight knot Seifert surface semiclassical expansion}
\end{equation}
In $b\rightarrow 0$ limit, we can approximate this integral by saddle-point. The leading order contribution is proportional to $1/b^2$ with the coefficient
\begin{multline}
     S^{(0)} =\frac{\pi i}{8}+2\pi x+\frac{\pi i \eta_0^2}{2}-6\pi i x^2-4\pi i \eta x-\frac{i}{4\pi}\, \big(\Li_2(\mathrm{e}^{-2\pi\eta_0+i\pi})-\Li_2(\mathrm{e}^{2\pi \eta_0-i\pi})\big)\\
     +\frac{i}{4\pi}\sum_{\pm,\pm} \big(\Li_2(\mathrm{e}^{-2\pi(\frac{\eta_0}{2}\pm x\pm\eta)+\frac{\pi i}{2}})-\Li_2(\mathrm{e}^{2\pi(\frac{\eta_0}{2}\pm x\pm\eta)-\frac{\pi i}{2}})\big)\ .
\end{multline}
This leads to three saddle-point equations.  

Since $\mathbb{S}[P_0']$ only depends on the conformal weight $\Delta_0'=P_0^{'2}+\frac{Q^2}{4}$, the function $\tr(\mathbb{S}[P_0']\mathbb{T}^3)$ is even in $P_0'$. This observation implies that $\eta_0 = 0$ will be a saddle-point and we can reduce one saddle-point equation with respect to $\eta_0$. When $\eta_0$ is set to be $0$, we have the saddle-point equations of $\eta$ and $x$ respectively as follow
\begin{subequations}
    \begin{align}
        0= &-4\pi i x+\frac{i}{2}\log{\left[\left(\frac{\cosh{(2\pi x)+i\sinh{(2\pi \eta)}}}{\cosh{(2\pi x)}-i\sinh{(2\pi \eta)}}\right)^2\right]}\ ,\\
        0=&-12 \pi i x+2 \pi-4\pi i\eta+\frac{i}{2}\log{\left[\left(\frac{\cosh{(2\pi\eta)+i\sinh{(2\pi x)}}}{\cosh{(2\pi\eta)}-i \sinh{(2\pi x)}}\right)^2\right]}\ .
    \end{align}
\end{subequations}
The first equation can be solved by taking $2\pi i\eta =\arcsin{(\sinh{(2\pi x)})}$. By plugging this relation between $\eta$ and $x$ into the second equation, we solve for $x$ and obtain the following saddle-point of $S^{(0)}$
\begin{equation}
    x = \frac{1}{4\pi}\log{\left(\frac{-1-3\sqrt{3}i-\sqrt{-42+6\sqrt{3}i}}{4}\right)}\ .
\end{equation}
We also explicitly check that $\frac{\partial S^{(0)}}{\partial \eta_0}$ is vanishing when $\eta_0 = 0$ and $x,\,\eta$ take the given saddle-point values. Therefore, $\eta_0 = 0$ is indeed the saddle-point along $\eta_0$ direction as we justified before. By evaluating the $S^{(0)}$ at the saddle point, we recover the hyperbolic volume of the figure eight knot as expected
\begin{equation}
    S^{(0)} = -\frac{\vol(4_1)}{2\pi}\ .
\end{equation}

In order to compare the semiclassical result with the refined volume conjecture \eqref{eq:refined volume conjecture}, we should also study the higher-loop corrections. Using the expansion of double-sine functions in \eqref{eq:double sine function semiclassical expansion}, we can compute the partition function to all orders perturbatively in $b^2$. Here we focus on the order one factor in the expansion 
\begin{equation}
    Z^{(1)} = \frac{1}{2}\sqrt{ -\frac{(2\pi)^3}{\det(\mathop{\text{Hess}}S^{(0)})}} \ \mathrm{e}^{S^{(1)}}\ , \label{eq:one-loop piece figure eight knot Seifert surface}
\end{equation}
since this factor is closely related to the one-loop determinant in the 3d gravity calculation. The prefactor comes from the Gaussian integral around the saddle point. The additional factor of $\frac{1}{2}$ appears because the integral is restricted to $P_0'>0$, while the minus sign inside the square root originates from the fact that the Gaussian integral has the form $\mathrm{e}^{\frac{1}{b^2} S_0}$. The three factors of $b$ get cancelled against the three $b$'s from the Jacobian in \eqref{eq:Figure eight knot Seifert surface semiclassical expansion}. We collect all order-one terms in the expansion
\begin{align}
    \mathrm{e}^{S^{(1)}}=\frac{4 i \sinh(2\pi x)}{\sinh(\pi (\frac{i}{4} -\frac{\eta_0}{2} \pm x \pm \eta))^{\frac{1}{4}}}\ ,
\end{align}
which upon inserting the saddlepoint value simplifies to
\be 
    \mathrm{e}^{S^{(1)}}=4 \sqrt{2} i \sinh(2\pi x)\ .
\ee
We then take the Gaussian integral contribution to \eqref{eq:one-loop piece figure eight knot Seifert surface} into account, we obtain the order-one correction to the partition function
\begin{equation}
    Z^{(1)} = \frac{2\sqrt{2} i \sinh(2\pi x)}{\sqrt{7i+5 \cosh(2\pi x) \sqrt{6-2\cosh(4\pi x)}-5i \cosh(4\pi x)}}=\frac{\sqrt{2}}{3^{\frac{1}{4}}}\ .
\end{equation}
This result matches with the order-one term in the expression \eqref{eq:figure eight expansion} with the Reidemeister torsion $\sqrt{\Delta} = \sqrt{3}$ at $P_0=0$.

Note that in the refined volume conjecture \eqref{eq:refined volume conjecture}, we write the semiclassical expansion of the partition function in terms of the central charge $c$, while we have the $b^2$ expansion in this part of calculation. The central charge $c$ is defined as $c =1+6(b+\frac{1}{b})^2 = 13+\frac{6}{b^2}+6b^2$. Therefore, strictly speaking, the one-loop determinant from the gravity calculation is not equal to $Z^{(1)}$. Instead, we need to renormalize $Z^{(1)}$ to obtain the one-loop determinant
\begin{equation}
    Z_{\text{one-loop}} = Z^{(1)}\, \mathrm{e}^{\frac{13}{12\pi}\vol(4_1)} \ ,
\end{equation}
which should be compared with the calculations performed in \cite{Giombi:2008vd}.\footnote{The computation in \cite{Giombi:2008vd} is not directly applicable to the figure eight knot case because of the presence of the cusp, in which case the relevant Kleinian groups has parabolic elements.}
\subsection{Dehn surgery}
As final application to the figure eight knot computation, we discuss an example of Dehn surgery.
Consider the figure eight knot and excise a small tubular neighborhood around the knot. We can then glue back a torus, but twisted by an $\SL(2,\ZZ)$ element. Such an element is specified by a two coprime integers $(p,q)$ specifying the slope of the meridian (the contractible curve).

The Virasoro TQFT partition function on a solid torus gives simply the vacuum character $\chi_\id$ in the appropriate channel, while it gives a generic Virasoro character $\chi_P$ with the inclusion of a Wilson line of momentum $P$. We can write\footnote{As explained in \cite{Collier:2023fwi}, the normalization of the inner product on the torus is somewhat ambiguous, but this ambiguity will cancel out of the calculation.}
\be 
Z_\text{Vir}(4_1,P_0)=\langle Z_\text{Vir}(4_1^\circ) \, | \, \chi_{P_0} \rangle\ ,
\ee
where $4_1^\circ$ is the figure eight knot complement with a tubular neighborhood around the knot removed and we emphasize the $P_0$-dependence of the Virasoro TQFT partition function.

Thus the partition function of a manifold obtained by Dehn surgery from the figure eight knot is given by
\begin{align} 
Z_\text{Vir}(4_1(p,q))&=\langle Z_\text{Vir}(4_1^\circ) \, | \, \mathbb{U}(p,q) \, |\,  \chi_{\text{vac}} \rangle
\\
&=\int_0^\infty \d P\  \mathbb{U}(p,q)_{\mathds{1},P}\, \langle Z_\text{Vir}(4_1^\circ) \, | \,  \chi_P \rangle\\
&=\int_0^\infty \d P\  \mathbb{U}(p,q)_{\mathds{1},P}\, Z_\text{Vir}(4_1,P)
\end{align}
where $\mathbb{U}(p,q)$ is the representation of the $\SL(2,\ZZ)$ modular transformation on the Virasoro characters. It takes the explicit form (see e.g. \cite{Benjamin:2020mfz})
\be 
\mathbb{U}(p,q)_{\mathds{1},P}=\varepsilon(p,q) \sqrt{\frac{8}{q}} \, \mathrm{e}^{-\frac{2\pi i}{q} (p^* \frac{Q^2}{4}-p P^2)}\bigg(\cosh\Big(\frac{2Q P \pi}{q}\Big)-\mathrm{e}^{\frac{2\pi i p^*}{q}}\cosh\Big(\frac{2\hat{Q} P \pi}{q}\Big)\bigg)\ .
\ee
Here $\hat{Q}=b-b^{-1}$, $\varepsilon(p,q)$ is a $P$-independent 24-th root of unity coming from the transformation behaviour of the Dedekind $\eta$-function and $p^*$ is the modular inverse of $p$, $pp^*\equiv 1 \bmod q$. This leaves an ambiguity in the expression which can be absorbed in the framing ambiguity. For the figure eight knot, we should also notice that because of amphichirality, the Dehn surgeries $(p,q)$ and $(-p,q)$ are equivalent and we can focus on $p,\, q \ge 0$.

It is in particular simple to evaluate the hyperbolic volume of this class of manifolds via saddle point approximation. Set $P=\frac{\eta}{b}$ as before. Then the action is
\begin{align}
    S=\vol(4_1,\eta)+\frac{\pi^2i}{q}(p^*-4p \eta^2)\pm \frac{4\pi^2\eta}{q}\ .
\end{align}
Since we focus on the volume, we can omit the purely imaginary part involving $p^*$. The sign choice of the last term is also immaterial, since we can send $\eta \to -\eta$. We hence find that
\be 
\vol(4_1(p,q))=\Re \bigg(\vol(4_1,\eta)+\frac{4\pi^2\eta(1-p i \eta)}{q}\bigg)\bigg|_{\eta=\eta^*}\ , \label{eq:volume figure eight knot Dehn surgeries}
\ee
where we plug in the saddle-point value $\eta^*$ and the volume is given by \eqref{eq:figure eight knot complement volume nonzero eta}. The saddle-point equation is transcendental and doesn't admit a closed form solution. However, it is straightforward to compute the volumes of various examples numerically, see Table~\ref{tab:Dehn surgery figure eight knot volumes}. We compared them to the volumes as computed by the program \texttt{SnapPy}.
\begin{table}[ht]
    \centering
    \begin{tabular}{c|cccccccccc}
\diagbox[width=1cm, height=.6cm, innerleftsep=.1cm,innerrightsep=.1cm]{$p$}{\raisebox{.1cm}{$q$}}    & 1 & 2 & 3 & 4 & 5 & 6 & 7 & 8 & 9 \\
\hline 
1& 0 & 1.3985 & 1.7320 & 1.8581 & 1.9186 & 1.9521 & 1.9725 & 1.9858 &  1.9950  \\
2&  0 &  & 1.7371 &  & 1.9195 &  & 1.9727 &  & 1.9951   \\
3&  0 & 1.4407 &  & 1.8634 & 1.9210 &  & 1.9732 & 1.9862 &   \\
4& 0 &  & 1.7571 &  & 1.9231 &  & 1.9738 &  & 1.9955   \\
5&  0.9813 & 1.5295 & 1.7714 & 1.8735 &  & 1.9557 & 1.9745 & 1.9870 & 1.9958   \\
6&  1.2845 &  &  &  & 1.9287 &  & 1.9754 &  &    \\
7&  1.4638 & 1.6496 & 1.8058 & 1.8871 & 1.9321 & 1.9591 &  & 1.9882 & 1.9965  \\
8&  1.5832 &  & 1.8243 &  & 1.9358 &  & 1.9776 &  & 1.9970   \\
9&  1.6678 & 1.7521 &  & 1.9027 & 1.9397 &  & 1.9789 & 1.9897 &   \\
        \end{tabular}
        \caption{The volumes of manifolds obtained from Dehn surgery from the figure eight knot. Zero entries indicate that the corresponding manifolds do not admit a hyperbolic metric. The other two exceptional cases that do not admit a hyperbolic metric are $(p,q)=(1,0)$ and $(0,1)$, see also \cite[Theorem 4.7]{Thurston}.}
        \label{tab:Dehn surgery figure eight knot volumes}
\end{table}
It is also simple to compute the  volumes in a large $p$ and $q$ expansion, since for large $p$ or $q$, the saddle point $\eta^* \to 0$ and the volume converges to the volume of the figure eight knot. We find to the first few orders
\begin{multline}
    \vol(4_1(p,q))=\vol(4_1)
-\frac{2\sqrt{3} \pi^2}{p^2+12q^2}+\frac{4\pi^4(p^4-72p^2q^2+144q^4)}{\sqrt{3}(p^2+12q^2)^4}\\
-\frac{8 \pi ^6 (23 p^8-8904 p^6 q^2+302400 p^4 q^4-1620864 p^2 q^6+767232 q^8)}{45 \sqrt{3} (p^2+12 q^2)^7}+\cdots
\end{multline}
The correction to the figure eight knot volume is always negative as required by general theorems about Dehn surgery \cite[Theorem 6.5.6.]{Thurston}. This expansion is a known result, see \cite{NeumannZagier}.
This case of Dehn surgery exemplifies the existence of accumulation points in the spectrum of three-manifolds. We discussed their implications for the gravitational path integral in our previous paper \cite{Collier:2023fwi}.

\section*{Acknowledgements} We would like to thank Alex Belin, Jeevan Chandra, Tom Hartman, Daniel Jafferis, Diego Li\v{s}ka, Alex Maloney, Baur Mukhametzhanov, Boris Post, Sahand Seifnashri, Steve Shenker, Julian Sonner, J\"org Tesch\-ner and Ka Ho Wong for useful discussions. 
While at the IAS, L.E.\ was supported by the grant DE-SC0009988 from the U.S.\ Department of Energy. This material is based upon work supported by the U.S. Department of Energy, Office of Science, Office of High Energy Physics of U.S. Department of Energy under grant Contract Number  DE-SC0012567 (High Energy Theory research), DOE Early Career Award  DE-SC0021886  and the Packard Foundation Award in Quantum Black Holes and Quantum Computation.

\appendix

\section{Semiclassical expansion of the double sine function} \label{app:semiclassical expansion double sine}
In this appendix, we will derive the semiclassical expansion of the double sine function \eqref{eq:double sine function semiclassical expansion}. We start from the integral representation
\begin{align}
    \log S_b(x_0+x)&=\frac{1}{4}\int_{\RR+i 0^+} \frac{\d t}{t} \, \frac{\sinh\big((\frac{Q}{2}-x-x_0)t\big)}{\sinh(\frac{bt}{2})\sinh(\frac{t}{2b})} \\
    &=\frac{1}{4} \int_{\RR+i 0^+} \frac{\d t}{t} \, \frac{\mathrm{e}^{(\frac{b^2+1}{2}-bx-bx_0)t}-\mathrm{e}^{-(\frac{b^2+1}{2}-bx-bx_0)t}}{(\mathrm{e}^{\frac{b^2t}{2}}-\mathrm{e}^{-\frac{b^2t}{2}})\sinh(\frac{t}{2})} \\
    &=\frac{1}{4} \int_{(\RR+i 0^+)\cup (\RR+i 0^-)} \frac{\d t}{t} \, \frac{\mathrm{e}^{(b^2+\frac{1}{2}-bx-bx_0)t}}{(\mathrm{e}^{b^2t}-1)\sinh(\frac{t}{2})} \ .
\end{align}
Here we rescaled $t$ and put $t \to -t$ in the second expression to have the integrand have the same form. We can now use the definition of the Bernoulli polynomials and get as formal expansion
\begin{align}
    \log S_b(x_0+x)&=\sum_{n=0}^\infty \frac{b^{2n-2}}{4n!}\, B_n(1-\tfrac{x}{b}) \int_{(\RR+i 0^+)\cup (\RR+i 0^-)}\d t \, \frac{t^{n-2}\, \mathrm{e}^{(\frac{1}{2}-bx_0)t}}{\sinh(\frac{t}{2})}\ .
\end{align}
The remaining integral can be computed for example by pulling off the contour off and summing over the residues at $t=2\pi i m$. This gives
\begin{align}
    \int_{\RR+i 0^+}\d t \, \frac{t^{n-2}\, \mathrm{e}^{(\frac{1}{2}-bx_0)t}}{\sinh(\frac{t}{2})}&=\sum_{m=1}^\infty 2(2\pi i m)^{n-2}\, \mathrm{e}^{-2\pi i m b x_0}=2(2\pi i)^{n-2} \Li_{2-n}(\mathrm{e}^{-2\pi i b x_0})\ .
\end{align}
We similarly evaluate the contribution from the other contour $\RR+i0^-$ which then recovers \eqref{eq:double sine function semiclassical expansion}.
\bibliographystyle{JHEP}
\bibliography{bib}

\providecommand{\href}[2]{#2}\begingroup\raggedright\begin{thebibliography}{10}

\bibitem{Collier:2023fwi}
S.~Collier, L.~Eberhardt and M.~Zhang, \emph{{Solving 3d Gravity with Virasoro
  TQFT}},  \href{https://arxiv.org/abs/2304.13650}{{\ttfamily 2304.13650}}.

\bibitem{Yin:2007gv}
X.~Yin, \emph{{Partition Functions of Three-Dimensional Pure Gravity}},
  \href{https://doi.org/10.4310/CNTP.2008.v2.n2.a1}{\emph{Commun. Num. Theor.
  Phys.} {\bfseries 2} (2008) 285}
  [\href{https://arxiv.org/abs/0710.2129}{{\ttfamily 0710.2129}}].

\bibitem{Giombi:2008vd}
S.~Giombi, A.~Maloney and X.~Yin, \emph{{One-loop Partition Functions of 3D
  Gravity}}, \href{https://doi.org/10.1088/1126-6708/2008/08/007}{\emph{JHEP}
  {\bfseries 08} (2008) 007} [\href{https://arxiv.org/abs/0804.1773}{{\ttfamily
  0804.1773}}].

\bibitem{Cotler:2018zff}
J.~Cotler and K.~Jensen, \emph{{A theory of reparameterizations for
  $\mathrm{AdS}_3$ gravity}},
  \href{https://doi.org/10.1007/JHEP02(2019)079}{\emph{JHEP} {\bfseries 02}
  (2019) 079} [\href{https://arxiv.org/abs/1808.03263}{{\ttfamily
  1808.03263}}].

\bibitem{Maxfield:2020ale}
H.~Maxfield and G.~J. Turiaci, \emph{{The path integral of 3D gravity near
  extremality; or, JT gravity with defects as a matrix integral}},
  \href{https://doi.org/10.1007/JHEP01(2021)118}{\emph{JHEP} {\bfseries 01}
  (2021) 118} [\href{https://arxiv.org/abs/2006.11317}{{\ttfamily
  2006.11317}}].

\bibitem{Cotler:2020ugk}
J.~Cotler and K.~Jensen, \emph{{$\mathrm{AdS}_{3}$ gravity and random CFT}},
  \href{https://doi.org/10.1007/JHEP04(2021)033}{\emph{JHEP} {\bfseries 04}
  (2021) 033} [\href{https://arxiv.org/abs/2006.08648}{{\ttfamily
  2006.08648}}].

\bibitem{Eberhardt:2022wlc}
L.~Eberhardt, \emph{{Off-shell Partition Functions in 3d Gravity}},
  \href{https://arxiv.org/abs/2204.09789}{{\ttfamily 2204.09789}}.

\bibitem{Belin:2020hea}
A.~Belin and J.~de~Boer, \emph{{Random statistics of OPE coefficients and
  Euclidean wormholes}},
  \href{https://doi.org/10.1088/1361-6382/ac1082}{\emph{Class. Quant. Grav.}
  {\bfseries 38} (2021) 164001}
  [\href{https://arxiv.org/abs/2006.05499}{{\ttfamily 2006.05499}}].

\bibitem{Schlenker:2022dyo}
J.-M. Schlenker and E.~Witten, \emph{{No ensemble averaging below the black
  hole threshold}}, \href{https://doi.org/10.1007/JHEP07(2022)143}{\emph{JHEP}
  {\bfseries 07} (2022) 143}
  [\href{https://arxiv.org/abs/2202.01372}{{\ttfamily 2202.01372}}].

\bibitem{Chandra:2022bqq}
J.~Chandra, S.~Collier, T.~Hartman and A.~Maloney, \emph{{Semiclassical 3D
  gravity as an average of large-c CFTs}},
  \href{https://arxiv.org/abs/2203.06511}{{\ttfamily 2203.06511}}.

\bibitem{Belin:2023efa}
A.~Belin, J.~de~Boer, D.~L. Jafferis, P.~Nayak and J.~Sonner,
  \emph{{Approximate CFTs and Random Tensor Models}},
  \href{https://arxiv.org/abs/2308.03829}{{\ttfamily 2308.03829}}.

\bibitem{DiUbaldo:2023qli}
G.~Di~Ubaldo and E.~Perlmutter, \emph{{AdS$_3$/RMT$_2$ Duality}},
  \href{https://arxiv.org/abs/2307.03707}{{\ttfamily 2307.03707}}.

\bibitem{Collier:2023cyw}
S.~Collier, L.~Eberhardt, B.~M\"uhlmann and V.~A. Rodriguez, \emph{{The
  Virasoro Minimal String}},
  \href{https://arxiv.org/abs/2309.10846}{{\ttfamily 2309.10846}}.

\bibitem{deBoer:2023vsm}
J.~de~Boer, D.~Liska, B.~Post and M.~Sasieta, \emph{{A principle of maximum
  ignorance for semiclassical gravity}},
  \href{https://arxiv.org/abs/2311.08132}{{\ttfamily 2311.08132}}.

\bibitem{Ponsot:1999uf}
B.~Ponsot and J.~Teschner, \emph{{Liouville bootstrap via harmonic analysis on
  a noncompact quantum group}},
  \href{https://arxiv.org/abs/hep-th/9911110}{{\ttfamily hep-th/9911110}}.

\bibitem{Ponsot:2000mt}
B.~Ponsot and J.~Teschner, \emph{{Clebsch-Gordan and Racah-Wigner coefficients
  for a continuous series of representations of
  $U_q(\mathfrak{sl}(2,\mathbb{R}))$}},
  \href{https://doi.org/10.1007/PL00005590}{\emph{Commun. Math. Phys.}
  {\bfseries 224} (2001) 613}
  [\href{https://arxiv.org/abs/math/0007097}{{\ttfamily math/0007097}}].

\bibitem{Teschner:2012em}
J.~Teschner and G.~Vartanov, \emph{{6j symbols for the modular double, quantum
  hyperbolic geometry, and supersymmetric gauge theories}},
  \href{https://doi.org/10.1007/s11005-014-0684-3}{\emph{Lett. Math. Phys.}
  {\bfseries 104} (2014) 527}
  [\href{https://arxiv.org/abs/1202.4698}{{\ttfamily 1202.4698}}].

\bibitem{Teschner:2013tqy}
J.~Teschner and G.~S. Vartanov, \emph{{Supersymmetric gauge theories,
  quantization of $\mathcal{M}_{\mathrm{flat}}$, and conformal field theory}},
  \href{https://doi.org/10.4310/ATMP.2015.v19.n1.a1}{\emph{Adv. Theor. Math.
  Phys.} {\bfseries 19} (2015) 1}
  [\href{https://arxiv.org/abs/1302.3778}{{\ttfamily 1302.3778}}].

\bibitem{Mertens:2017mtv}
T.~G. Mertens, G.~J. Turiaci and H.~L. Verlinde, \emph{{Solving the Schwarzian
  via the Conformal Bootstrap}},
  \href{https://doi.org/10.1007/JHEP08(2017)136}{\emph{JHEP} {\bfseries 08}
  (2017) 136} [\href{https://arxiv.org/abs/1705.08408}{{\ttfamily
  1705.08408}}].

\bibitem{Lam:2018pvp}
H.~T. Lam, T.~G. Mertens, G.~J. Turiaci and H.~Verlinde, \emph{{Shockwave
  S-matrix from Schwarzian Quantum Mechanics}},
  \href{https://doi.org/10.1007/JHEP11(2018)182}{\emph{JHEP} {\bfseries 11}
  (2018) 182} [\href{https://arxiv.org/abs/1804.09834}{{\ttfamily
  1804.09834}}].

\bibitem{Jafferis:2022wez}
D.~L. Jafferis, D.~K. Kolchmeyer, B.~Mukhametzhanov and J.~Sonner, \emph{{JT
  gravity with matter, generalized ETH, and Random Matrices}},
  \href{https://arxiv.org/abs/2209.02131}{{\ttfamily 2209.02131}}.

\bibitem{Kashaev:1996kc}
R.~M. Kashaev, \emph{{The Hyperbolic volume of knots from quantum
  dilogarithm}}, \href{https://doi.org/10.1023/A:1007364912784}{\emph{Lett.
  Math. Phys.} {\bfseries 39} (1997) 269}.

\bibitem{Henningson:1998gx}
M.~Henningson and K.~Skenderis, \emph{{The Holographic Weyl anomaly}},
  \href{https://doi.org/10.1088/1126-6708/1998/07/023}{\emph{JHEP} {\bfseries
  07} (1998) 023} [\href{https://arxiv.org/abs/hep-th/9806087}{{\ttfamily
  hep-th/9806087}}].

\bibitem{Teschner:2003em}
J.~Teschner, \emph{{On the relation between quantum Liouville theory and the
  quantized Teichmuller spaces}},
  \href{https://doi.org/10.1142/S0217751X04020579}{\emph{Int. J. Mod. Phys. A}
  {\bfseries 19S2} (2004) 459}
  [\href{https://arxiv.org/abs/hep-th/0303149}{{\ttfamily hep-th/0303149}}].

\bibitem{Apresyan:2022erh}
E.~Apresyan, G.~Sarkissian and V.~P. Spiridonov, \emph{{A parafermionic
  hypergeometric function and supersymmetric 6j-symbols}},
  \href{https://doi.org/10.1016/j.nuclphysb.2023.116170}{\emph{Nucl. Phys. B}
  {\bfseries 990} (2023) 116170}
  [\href{https://arxiv.org/abs/2205.10276}{{\ttfamily 2205.10276}}].

\bibitem{Eberhardt:2023mrq}
L.~Eberhardt, \emph{{Notes on crossing transformations of Virasoro conformal
  blocks}},  \href{https://arxiv.org/abs/2309.11540}{{\ttfamily 2309.11540}}.

\bibitem{Belavin:1984vu}
A.~A. Belavin, A.~M. Polyakov and A.~B. Zamolodchikov, \emph{{Infinite
  Conformal Symmetry in Two-Dimensional Quantum Field Theory}},
  \href{https://doi.org/10.1016/0550-3213(84)90052-X}{\emph{Nucl. Phys. B}
  {\bfseries 241} (1984) 333}.

\bibitem{Zamolodchikov:1984eqp}
A.~B. Zamolodchikov, \emph{{Conformal symmetry in two dimensions: an explicit
  recurrence formula for the conformal partial wave amplitude}},
  \href{https://doi.org/10.1007/BF01214585}{\emph{Commun. Math. Phys.}
  {\bfseries 96} (1984) 419}.

\bibitem{Besken:2019jyw}
M.~Be\c{s}ken, S.~Datta and P.~Kraus, \emph{{Semi-classical Virasoro blocks:
  proof of exponentiation}},
  \href{https://doi.org/10.1007/JHEP01(2020)109}{\emph{JHEP} {\bfseries 01}
  (2020) 109} [\href{https://arxiv.org/abs/1910.04169}{{\ttfamily
  1910.04169}}].

\bibitem{Krasnov:2000zq}
K.~Krasnov, \emph{{Holography and Riemann surfaces}},
  \href{https://doi.org/10.4310/ATMP.2000.v4.n4.a5}{\emph{Adv. Theor. Math.
  Phys.} {\bfseries 4} (2000) 929}
  [\href{https://arxiv.org/abs/hep-th/0005106}{{\ttfamily hep-th/0005106}}].

\bibitem{ZografTakhtajan}
P.~G. Zograf and L.~A. Takhtadzhyan, \emph{{On uniformization of Riemann
  surfaces and the Weil-Petersson metric on Teichm{\"u}ller and Schottky
  spaces}},
  \href{https://doi.org/10.1070/SM1988v060n02ABEH003170}{\emph{Mathematics of
  the USSR-Sbornik} {\bfseries 60} (1988) 297}.

\bibitem{McIntyre:2004xs}
A.~McIntyre and L.~A. Takhtajan, \emph{{Holomorphic factorization of
  determinants of Laplacians on Riemann surfaces and a higher genus
  generalization of Kronecker's first limit formula}}, {\emph{Analysis}
  {\bfseries 16} (2006) 1291}
  [\href{https://arxiv.org/abs/math/0410294}{{\ttfamily math/0410294}}].

\bibitem{Cho:2017oxl}
M.~Cho, S.~Collier and X.~Yin, \emph{{Recursive Representations of Arbitrary
  Virasoro Conformal Blocks}},
  \href{https://doi.org/10.1007/JHEP04(2019)018}{\emph{JHEP} {\bfseries 04}
  (2019) 018} [\href{https://arxiv.org/abs/1703.09805}{{\ttfamily
  1703.09805}}].

\bibitem{Mutation_volume}
D.~Ruberman, \emph{Mutation and volumes of knots in $\mathrm{S}^3$},
  \href{https://doi.org/10.1007/BF01389038}{\emph{Inventiones mathematicae}
  {\bfseries 90} (1987) 189}.

\bibitem{SnapPy}
M.~Culler, N.~M. Dunfield, M.~Goerner and J.~R. Weeks, ``Snap{P}y, a computer
  program for studying the geometry and topology of $3$-manifolds.'' Available
  at \url{http://snappy.computop.org} (06/04/2023).

\bibitem{Moore:1988qv}
G.~W. Moore and N.~Seiberg, \emph{{Classical and Quantum Conformal Field
  Theory}}, \href{https://doi.org/10.1007/BF01238857}{\emph{Commun. Math.
  Phys.} {\bfseries 123} (1989) 177}.

\bibitem{Kitaev:2005hzj}
A.~Kitaev, \emph{{Anyons in an exactly solved model and beyond}},
  \href{https://doi.org/10.1016/j.aop.2005.10.005}{\emph{Annals Phys.}
  {\bfseries 321} (2006) 2}
  [\href{https://arxiv.org/abs/cond-mat/0506438}{{\ttfamily
  cond-mat/0506438}}].

\bibitem{Belin:2021ryy}
A.~Belin, J.~de~Boer and D.~Liska, \emph{{Non-Gaussianities in the statistical
  distribution of heavy OPE coefficients and wormholes}},
  \href{https://doi.org/10.1007/JHEP06(2022)116}{\emph{JHEP} {\bfseries 06}
  (2022) 116} [\href{https://arxiv.org/abs/2110.14649}{{\ttfamily
  2110.14649}}].

\bibitem{Chandra:2023dgq}
J.~Chandra and T.~Hartman, \emph{{Toward random tensor networks and holographic
  codes in CFT}},  \href{https://arxiv.org/abs/2302.02446}{{\ttfamily
  2302.02446}}.

\bibitem{Mukhametzhanov:2023notes}
B.~Mukhametzhanov, ``{Unpublished notes}.''

\bibitem{Ghosh:2019rcj}
A.~Ghosh, H.~Maxfield and G.~J. Turiaci, \emph{{A universal Schwarzian sector
  in two-dimensional conformal field theories}},
  \href{https://doi.org/10.1007/JHEP05(2020)104}{\emph{JHEP} {\bfseries 05}
  (2020) 104} [\href{https://arxiv.org/abs/1912.07654}{{\ttfamily
  1912.07654}}].

\bibitem{Coleman:1988cy}
S.~R. Coleman, \emph{{Black Holes as Red Herrings: Topological Fluctuations and
  the Loss of Quantum Coherence}},
  \href{https://doi.org/10.1016/0550-3213(88)90110-1}{\emph{Nucl. Phys. B}
  {\bfseries 307} (1988) 867}.

\bibitem{Giddings:1987cg}
S.~B. Giddings and A.~Strominger, \emph{{Axion Induced Topology Change in
  Quantum Gravity and String Theory}},
  \href{https://doi.org/10.1016/0550-3213(88)90446-4}{\emph{Nucl. Phys. B}
  {\bfseries 306} (1988) 890}.

\bibitem{Giddings:1988cx}
S.~B. Giddings and A.~Strominger, \emph{{Loss of Incoherence and Determination
  of Coupling Constants in Quantum Gravity}},
  \href{https://doi.org/10.1016/0550-3213(88)90109-5}{\emph{Nucl. Phys. B}
  {\bfseries 307} (1988) 854}.

\bibitem{Yin:2007at}
X.~Yin, \emph{{On Non-handlebody Instantons in 3D Gravity}},
  \href{https://doi.org/10.1088/1126-6708/2008/09/120}{\emph{JHEP} {\bfseries
  09} (2008) 120} [\href{https://arxiv.org/abs/0711.2803}{{\ttfamily
  0711.2803}}].

\bibitem{Bianchi:1990yu}
M.~Bianchi and A.~Sagnotti, \emph{{On the systematics of open string
  theories}}, \href{https://doi.org/10.1016/0370-2693(90)91894-H}{\emph{Phys.
  Lett. B} {\bfseries 247} (1990) 517}.

\bibitem{Hikida:2002bt}
Y.~Hikida, \emph{{Liouville field theory on a unoriented surface}},
  \href{https://doi.org/10.1088/1126-6708/2003/05/002}{\emph{JHEP} {\bfseries
  05} (2003) 002} [\href{https://arxiv.org/abs/hep-th/0210305}{{\ttfamily
  hep-th/0210305}}].

\bibitem{Nakayama:2004vk}
Y.~Nakayama, \emph{{Liouville field theory: A Decade after the revolution}},
  \href{https://doi.org/10.1142/S0217751X04019500}{\emph{Int. J. Mod. Phys. A}
  {\bfseries 19} (2004) 2771}
  [\href{https://arxiv.org/abs/hep-th/0402009}{{\ttfamily hep-th/0402009}}].

\bibitem{Dimofte:2009yn}
T.~Dimofte, S.~Gukov, J.~Lenells and D.~Zagier, \emph{{Exact Results for
  Perturbative Chern-Simons Theory with Complex Gauge Group}},
  \href{https://doi.org/10.4310/CNTP.2009.v3.n2.a4}{\emph{Commun. Num. Theor.
  Phys.} {\bfseries 3} (2009) 363}
  [\href{https://arxiv.org/abs/0903.2472}{{\ttfamily 0903.2472}}].

\bibitem{Dijkgraaf:2010ur}
R.~Dijkgraaf, H.~Fuji and M.~Manabe, \emph{{The Volume Conjecture, Perturbative
  Knot Invariants, and Recursion Relations for Topological Strings}},
  \href{https://doi.org/10.1016/j.nuclphysb.2011.03.014}{\emph{Nucl. Phys. B}
  {\bfseries 849} (2011) 166}
  [\href{https://arxiv.org/abs/1010.4542}{{\ttfamily 1010.4542}}].

\bibitem{Dimofte:2011gm}
T.~Dimofte, \emph{{Quantum Riemann Surfaces in Chern-Simons Theory}},
  \href{https://doi.org/10.4310/ATMP.2013.v17.n3.a1}{\emph{Adv. Theor. Math.
  Phys.} {\bfseries 17} (2013) 479}
  [\href{https://arxiv.org/abs/1102.4847}{{\ttfamily 1102.4847}}].

\bibitem{EllegaardAndersen:2011vps}
J.~Ellegaard~Andersen and R.~Kashaev, \emph{{A TQFT from Quantum Teichm\"uller
  Theory}}, \href{https://doi.org/10.1007/s00220-014-2073-2}{\emph{Commun.
  Math. Phys.} {\bfseries 330} (2014) 887}
  [\href{https://arxiv.org/abs/1109.6295}{{\ttfamily 1109.6295}}].

\bibitem{Benjamin:2020mfz}
N.~Benjamin, S.~Collier and A.~Maloney, \emph{{Pure Gravity and Conical
  Defects}}, \href{https://doi.org/10.1007/JHEP09(2020)034}{\emph{JHEP}
  {\bfseries 09} (2020) 034}
  [\href{https://arxiv.org/abs/2004.14428}{{\ttfamily 2004.14428}}].

\bibitem{Thurston}
W.~P. Thurston, \emph{{The Geometry and Topology of Three-Manifolds: With a
  Preface by Steven P. Kerckhoff}},  2022.

\bibitem{NeumannZagier}
W.~D. Neumann and D.~Zagier, \emph{Volumes of hyperbolic three-manifolds},
  \href{https://doi.org/https://doi.org/10.1016/0040-9383(85)90004-7}{\emph{Topology}
  {\bfseries 24} (1985) 307}.

\end{thebibliography}\endgroup
\end{document}